\documentclass[twocolumn]{aastex631}

\newcommand\fakesection[1]{%
  \refstepcounter{section}%
  \addcontentsline{toc}{section}{\protect\numberline{\thesection}#1}%
  \sectionmark{#1}}

\usepackage{tabularx}

\shorttitle{New density for the LHS~1140 planets}
\shortauthors{Cadieux et al.}

\graphicspath{{./}{figures/}}

\begin{document}

\defcitealias{Dittmann_2017}{D17}
\defcitealias{Ment_2019}{M19}
\defcitealias{Lillo-Box_2020}{LB20}

\title{New Mass and Radius Constraints on the LHS~1140 Planets -- LHS~1140\,b is Either a Temperate Mini-Neptune or a Water World}

\correspondingauthor{Charles Cadieux}
\email{charles.cadieux.1@umontreal.ca}

\author[0000-0001-9291-5555]{Charles Cadieux} 
\affiliation{Institut Trottier de recherche sur les exoplanètes, Université de Montréal, 1375 Ave Thérèse-Lavoie-Roux, Montréal, QC, H2V 0B3, Canada}

\author[0000-0002-9479-2744]{Mykhaylo Plotnykov}
\affiliation{Department of Physics, University of Toronto, Toronto, ON M5S 3H4, Canada}

\author[0000-0001-5485-4675]{Ren\'e Doyon} 
\affiliation{Institut Trottier de recherche sur les exoplanètes, Université de Montréal, 1375 Ave Thérèse-Lavoie-Roux, Montréal, QC, H2V 0B3, Canada}
\affiliation{Observatoire du Mont-M\'egantic, Universit\'e de Montr\'eal, Montr\'eal H3C 3J7, Canada}

\author[0000-0003-3993-4030]{Diana Valencia}
\affiliation{Department of Physical \& Environmental Sciences, University of Toronto at Scarborough, Toronto, ON M1C 1A4, Canada}
\affiliation{David A. Dunlap Dept.\ of Astronomy \& Astrophysics, University of Toronto, 50 St. George Street, Toronto, Ontario, M5S 3H4, Canada}

\author[0000-0003-0029-2835]{Farbod Jahandar}
\affiliation{Institut Trottier de recherche sur les exoplanètes, Université de Montréal, 1375 Ave Thérèse-Lavoie-Roux, Montréal, QC, H2V 0B3, Canada}

\author[0000-0003-4987-6591]{Lisa Dang}
\affiliation{Institut Trottier de recherche sur les exoplanètes, Université de Montréal, 1375 Ave Thérèse-Lavoie-Roux, Montréal, QC, H2V 0B3, Canada}

\author[0000-0003-2260-9856]{Martin Turbet}
\affiliation{Laboratoire de M\'et\'eorologie Dynamique/IPSL, CNRS, Sorbonne Universit\'e, Ecole Normale Sup\'erieure, PSL Research University, Ecole Polytechnique, 75005 Paris, France}
\affiliation{Laboratoire d'astrophysique de Bordeaux, Univ. Bordeaux, CNRS, B18N, allée Geoffroy Saint-Hilaire, 33615 Pessac, France}

\author[0000-0002-5967-9631]{Thomas J. Fauchez}
\affiliation{NASA Goddard Space Flight Center
8800 Greenbelt Road Greenbelt, MD 20771, USA}
\affiliation{Integrated Space Science and Technology Institute, Department of Physics, American University, Washington DC}
\affiliation{NASA GSFC Sellers Exoplanet Environments Collaboration}

\author[0000-0001-5383-9393]{Ryan Cloutier}
\affiliation{Department of Physics \& Astronomy, McMaster University, 1280 Main St W, Hamilton, ON, L8S 4L8, Canada}

\author[0000-0002-8466-5469]{Collin Cherubim}
\affiliation{Earth and Planetary Science, Harvard University, 20 Oxford St, Cambridge, MA 02138, USA}
\affiliation{Center for Astrophysics | Harvard \& Smithsonian, 60 Garden Street, Cambridge, MA 02138, USA}

\author[0000-0003-3506-5667]{\'Etienne Artigau}
\affiliation{Institut Trottier de recherche sur les exoplanètes, Université de Montréal, 1375 Ave Thérèse-Lavoie-Roux, Montréal, QC, H2V 0B3, Canada}
\affiliation{Observatoire du Mont-M\'egantic, Universit\'e de Montr\'eal, Montr\'eal H3C 3J7, Canada}

\author[0000-0003-4166-4121]{Neil J. Cook}
\affiliation{Institut Trottier de recherche sur les exoplanètes, Université de Montréal, 1375 Ave Thérèse-Lavoie-Roux, Montréal, QC, H2V 0B3, Canada}

\author[0000-0002-5494-3237]{Billy Edwards}
\affiliation{SRON, Netherlands Institute for Space Research, Niels Bohrweg 4, NL-2333 CA, Leiden, The Netherlands}
\affiliation{Department of Physics and Astronomy, University College London, London, United Kingdom}

\author[0000-0003-4992-8427]{Tim Hallatt}
\affiliation{Department of Physics and Trottier Space Institute, McGill University, Montréal, Québec, H3A 2T8, Canada}
\affiliation{Institut Trottier de recherche sur les exoplanètes, Université de Montréal, 1375 Ave Thérèse-Lavoie-Roux, Montréal, QC, H2V 0B3, Canada}

\author[0000-0003-0977-6545]{Benjamin Charnay}
\affiliation{LESIA, Observatoire de Paris, Université PSL, CNRS, Sorbonne Université, Université Paris-Cité, 5 place Jules Janssen, 92195 Meudon, France}

\author[0000-0002-7613-393X]{Fran\c cois Bouchy}
\affiliation{Departement d’astronomie, Universit\'e de Gen\`eve, Chemin Pegasi, 51, CH-1290 Versoix, Switzerland}

\author[0000-0002-1199-9759]{Romain Allart}
\affiliation{Institut Trottier de recherche sur les exoplanètes, Université de Montréal, 1375 Ave Thérèse-Lavoie-Roux, Montréal, QC, H2V 0B3, Canada}

\author[0000-0002-5407-3905]{Lucile Mignon}
\affiliation{Departement d’astronomie, Universit\'e de Gen\`eve, Chemin Pegasi, 51, CH-1290 Versoix, Switzerland}

\author[0000-0002-5074-1128]{Frédérique Baron} 
\affiliation{Institut Trottier de recherche sur les exoplanètes, Université de Montréal, 1375 Ave Thérèse-Lavoie-Roux, Montréal, QC, H2V 0B3, Canada}

\author[0000-0003-2434-3625]{Susana C. C. Barros}
\affiliation{Instituto de Astrof\'isica e Ci\^encias do Espa\c{c}o, Universidade do Porto, CAUP, Rua das Estrelas, PT4150-762 Porto, Portugal}
\affiliation{Departamento\,de\,Fisica\,e\,Astronomia,\,Faculdade\,de\,Ciencias,\,Universidade\,do\,Porto,\,Rua\,Campo\,Alegre,\,4169-007\,Porto,\,Portugal}

\author[0000-0001-5578-1498]{Bj\"orn Benneke}
\affiliation{Institut Trottier de recherche sur les exoplanètes, Université de Montréal, 1375 Ave Thérèse-Lavoie-Roux, Montréal, QC, H2V 0B3, Canada}

\author[0000-0001-5578-7400]{B. L. Canto Martins}
\affiliation{Departamento de Física Teórica e Experimental, Universidade Federal do Rio Grande do Norte, Campus Universitário, Natal, RN, 59072-970, Brazil}

\author[0000-0001-6129-5699]{Nicolas B. Cowan}
\affiliation{Department of Earth \& Planetary Sciences, McGill University, 3450 rue University, Montréal, QC H3A 0E8, Canada}
\affiliation{Department of Physics and Trottier Space Institute, McGill University, Montréal, Québec, H3A 2T8, Canada}

\author[0000-0001-8218-1586]{J. R. De Medeiros}
\affiliation{Departamento de Física Teórica e Experimental, Universidade Federal do Rio Grande do Norte, Campus Universitário, Natal, RN, 59072-970, Brazil}

\author[0000-0001-5099-7978]{Xavier Delfosse}
\affiliation{Univ.\ Grenoble Alpes, CNRS, IPAG, 38000 Grenoble, France}

\author[0000-0003-4434-2195]{Elisa Delgado-Mena}
\affiliation{Instituto de Astrof\'isica e Ci\^encias do Espa\c{c}o, Universidade do Porto, CAUP, Rua das Estrelas, PT4150-762 Porto, Portugal}

\author[0000-0002-9332-2011]{Xavier Dumusque}
\affiliation{Departement d’astronomie, Universit\'e de Gen\`eve, Chemin Pegasi, 51, CH-1290 Versoix, Switzerland}

\author[0000-0001-9704-5405]{David Ehrenreich}
\affiliation{Departement d’astronomie, Universit\'e de Gen\`eve, Chemin Pegasi, 51, CH-1290 Versoix, Switzerland}

\author[0000-0003-4009-0330]{Yolanda G. C. Frensch}
\affiliation{Departement d’astronomie, Universit\'e de Gen\`eve, Chemin Pegasi, 51, CH-1290 Versoix, Switzerland}

\author[0000-0002-0264-7356]{J. I. Gonz\'alez Hern\'andez }
\affiliation{Instituto de Astrofísica de Canarias, 38205 La Laguna, Tenerife, Spain}
\affiliation{Departamento de Astrofísica, Universidad de La Laguna, 38206 La Laguna, Tenerife, Spain}

\author[0000-0001-9232-3314]{Nathan C. Hara}
\affiliation{Departement d’astronomie, Universit\'e de Gen\`eve, Chemin Pegasi, 51, CH-1290 Versoix, Switzerland}

\author[0000-0002-6780-4252]{David Lafreni\`ere}
\affiliation{Institut Trottier de recherche sur les exoplanètes, Université de Montréal, 1375 Ave Thérèse-Lavoie-Roux, Montréal, QC, H2V 0B3, Canada}

\author[0000-0002-1158-9354]{Gaspare Lo Curto}
\affiliation{European Southern Observatory, Karl-Schwarzschild-Strasse 2, 85748 Garching, Germany}

\author[0000-0002-8786-8499]{Lison Malo}
\affiliation{Institut Trottier de recherche sur les exoplanètes, Université de Montréal, 1375 Ave Thérèse-Lavoie-Roux, Montréal, QC, H2V 0B3, Canada}
\affiliation{Observatoire du Mont-M\'egantic, Universit\'e de Montr\'eal, Montr\'eal H3C 3J7, Canada}

\author[0000-0002-6090-8446]{Claudio Melo}
\affiliation{European Southern Observatory, Karl-Schwarzschild-Strasse 2, 85748 Garching, Germany}

\author[0000-0002-8070-2058]{Dany Mounzer}
\affiliation{Departement d’astronomie, Universit\'e de Gen\`eve, Chemin Pegasi, 51, CH-1290 Versoix, Switzerland}

\author[0000-0002-8569-7243]{Vera Maria Passeger}
\affiliation{Instituto de Astrofísica de Canarias, 38205 La Laguna, Tenerife, Spain}
\affiliation{Departamento de Astrofísica, Universidad de La Laguna, 38206 La Laguna, Tenerife, Spain}
\affiliation{Hamburger Sternwarte, Gojenbergsweg 112, 21029 Hamburg, Germany}

\author[0000-0002-9815-773X]{Francesco Pepe}
\affiliation{Departement d’astronomie, Universit\'e de Gen\`eve, Chemin Pegasi, 51, CH-1290 Versoix, Switzerland}

\author{Anne-Sophie Poulin-Girard}
\affiliation{ABB, Québec, G1P 0B2, Canada}

\author[0000-0003-4422-2919]{Nuno C. Santos}
\affiliation{Instituto de Astrof\'isica e Ci\^encias do Espa\c{c}o, Universidade do Porto, CAUP, Rua das Estrelas, PT4150-762 Porto, Portugal}
\affiliation{Departamento\,de\,Fisica\,e\,Astronomia,\,Faculdade\,de\,Ciencias,\,Universidade\,do\,Porto,\,Rua\,Campo\,Alegre,\,4169-007\,Porto,\,Portugal}

\author{Danuta Sosnowska}
\affiliation{Departement d’astronomie, Universit\'e de Gen\`eve, Chemin Pegasi, 51, CH-1290 Versoix, Switzerland}

\author[0000-0002-3814-5323]{Alejandro Su\'arez Mascare\~{n}o}
\affiliation{Instituto de Astrofísica de Canarias, 38205 La Laguna, Tenerife, Spain}
\affiliation{Departamento de Astrofísica, Universidad de La Laguna, 38206 La Laguna, Tenerife, Spain}
 
\author[0000-0002-2791-0595]{Simon Thibault} 
\affiliation{Centre d’optique, photonique et lasers, Université Laval, Québec, G1V 0A6, Canada}

\author[0000-0001-7329-3471]{Valentina Vaulato}
\affiliation{Departement d’astronomie, Universit\'e de Gen\`eve, Chemin Pegasi, 51, CH-1290 Versoix, Switzerland}

\author[0000-0002-1854-0131]{Gregg A. Wade}
\affiliation{Department of Physics and Space Science, Royal Military College of Canada, PO Box 17000, Station Forces, Kingston, ON, Canada}

\author[0000-0002-9216-4402]{François Wildi} 
\affiliation{Departement d’astronomie, Universit\'e de Gen\`eve, Chemin Pegasi, 51, CH-1290 Versoix, Switzerland}


\begin{abstract}
The two-planet transiting system LHS~1140 has been extensively observed since its discovery in 2017, notably with \textit{Spitzer}, HST, TESS, and ESPRESSO, placing strong constraints on the parameters of the M4.5 host star and its small temperate exoplanets, LHS~1140~b~and~c. Here, we reanalyse the ESPRESSO observations of LHS~1140 with the novel line-by-line framework designed to fully exploit the radial velocity content of a stellar spectrum while being resilient to outlier measurements. The improved radial velocities, combined with updated stellar parameters, consolidate our knowledge on the mass of LHS~1140\,b (5.60$\pm$0.19\,M$_{\oplus}$) and LHS~1140\,c (1.91$\pm$0.06\,M$_{\oplus}$) with unprecedented precision of 3\%. Transits from \textit{Spitzer}, HST, and TESS are jointly analysed for the first time, allowing us to refine the planetary radii of b (1.730$\pm$0.025\,R$_{\oplus}$) and c (1.272$\pm$0.026\,R$_{\oplus}$). Stellar abundance measurements of refractory elements (Fe, Mg and Si) obtained with NIRPS are used to constrain the internal structure of LHS~1140\,b. This planet is unlikely to be a rocky super-Earth as previously reported, but rather a mini-Neptune with a $\sim$0.1\% H/He envelope by mass or a water world with a water-mass fraction between 9 and 19\% depending on the atmospheric composition and relative abundance of Fe and Mg. While the mini-Neptune case would not be habitable, a water-abundant LHS~1140\,b potentially has habitable surface conditions according to 3D global climate models, suggesting liquid water at the substellar point for atmospheres with relatively low CO$_2$ concentration, from Earth-like to a few bars.
\end{abstract}


\section{Introduction}\label{sec:intro}

The last few years have been fruitful in the quest to uncover exoplanets transiting nearby low-mass stars. Unlike their solar counterparts, M dwarfs represent optimal targets for detailed studies of their planetary systems. They have smaller sizes (0.1--0.6\,R$_{\odot}$) and masses (0.1--0.6\,M$_{\odot}$), facilitating the characterization of exoplanets through transit and radial velocity (RV) observations. As they are less luminous, their Habitable Zone (HZ) is more compact than in our solar system, corresponding to orbital periods usually well sampled by current surveys (typically 60 days for an M0, 3 days for an M9). M dwarfs have at least twice as many small exoplanets with $M_{\rm p} \sin i < 10$\,M$_{\oplus}$ than G-type stars \citep{Sabotta_2021} and make up the majority of systems in the vicinity of the Sun (\citealt{Reyle_2021}, \citeyear{Reyle_2022}). We thus expect the nearest HZ planets to orbit such kind of stars. This is exemplified with Proxima Centauri (M5.5V), our closest neighbour (1.3\,pc; \citealt{Gaia_Collaboration_2021}), hosting a non-transiting terrestrial planet in the HZ \citep{Anglada_2016, Faria_2022}. The M5.5V dwarf GJ 1002 at 4.84\,pc also has two non-transiting Earth-mass companions in the HZ recently discovered by \cite{Suarez_2023}. The TRAPPIST-1 system \citep{Gillon_2017} at 12.5\,pc has seven terrestrial planets, including three in the HZ, all transiting the M8V ultracool host. These transiting systems are extremely valuable because the radius of exoplanets\,---\,sometimes even the mass through transit timing variations as for TRAPPIST-1 (e.g., \citealt{Agol_2021})\,---\,is only accessible via the transit method. Combined with dynamical mass constraints from Doppler spectroscopy, the bulk density of exoplanets can be obtained, revealing whether their interior is mostly rocky, gaseous, or perhaps even water-rich \citep{Luque_2022}, but evidence of water worlds remains elusive \citep{Rogers_2023}.

The M4.5 dwarf LHS~1140 located at 15.0\,pc is another intriguing system, currently the second closest to a transiting HZ exoplanet after TRAPPIST-1. A super-Earth on a 24.7-day temperate orbit was detected in 2017 (\citealt{Dittmann_2017}, hereafter \citetalias{Dittmann_2017}) from MEarth photometry \citep{Irwin_2009}, followed by the discovery of a second rocky planet with a shorter 3.8-day period (\citealt{Ment_2019}, hereafter \citetalias{Ment_2019}). The follow-up study of \citetalias{Ment_2019} presents a transit visit of LHS~1140~b~and~c with the Spitzer Space Telescope, largely improving the radius constraints of the two planets. Their masses were derived from HARPS \citep{Pepe_2002} radial velocities, initially for b only by \citetalias{Dittmann_2017}, and subsequently for b and c by \citetalias{Ment_2019} using an extended data set. This planetary system was revisited in 2020 (\citealt{Lillo-Box_2020}, hereafter \citetalias{Lillo-Box_2020}) with the ESPRESSO spectrograph \citep{Pepe_2021} and the Transiting Exoplanet Survey Satellite (TESS; \citealt{Ricker_2015}), offering an update of the bulk densities of LHS~1140~b~and~c, while also hinting at a possible third non-transiting planet on a longer orbit ($P_{\rm d} = 78.9$\,days). Lastly, transit spectroscopy of LHS~1140\,b has been obtained from the ground \citep{Diamond-Lowe_2020} and from space \citep{Edwards_2021} with the Wide Field Camera 3 (WFC3) on the Hubble Space Telescope (HST). \cite{Edwards_2021} reported a tentative detection of a hydrogen-dominated atmosphere with H$_2$O on LHS~1140\,b, but the signal ($\sim$100\,ppm) could also be explained by stellar contamination.

In this letter, we present a new analysis of archival data of LHS~1140 and derive stellar abundances from near-infrared spectroscopy with NIRPS \citep{Bouchy_2017}. The ESPRESSO radial velocities were significantly improved using a line-by-line extraction \citep{Artigau_2022} and, for the first time, transit data sets from \textit{Spitzer}, HST, and TESS were jointly analysed. We describe the observations in Section~\ref{sec:obs} and characterize the host star in Section~\ref{sec:stellar}. We present our revision of the mass and radius of the LHS~1140 planets and discuss their plausible internal structures in Section~\ref{sec:discussion}. Concluding remarks follow in Section~\ref{sec:conclusion}.

\section{Observations} \label{sec:obs}

\subsection{Spitzer photometry} \label{sec:Spitzer}

We recovered four \textit{Spitzer} transits of LHS~1140\,b taken with the Infrared Array Camera \citep{Fazio_2004} at 4.5\,$\mu$m from the NASA/IPAC Infrared Science Archive\footnote{\href{https://irsa.ipac.caltech.edu/}{\texttt{irsa.ipac.caltech.edu/}}}. These data include the double transit of LHS~1140~b~and~c analysed by \citetalias{Ment_2019}, with three additional unpublished transits of LHS~1140\,b (PI: J.\,A.\,Dittmann) from the same program. The observations were taken on UT2017-10-02, 2017-10-27, 2018-03-24, and 2018-04-18, hereafter referred to as Transit 1 to 4 (Transit 4 is presented in \citetalias{Ment_2019}).

\begin{figure}[t!]
\centering
\includegraphics[width=0.95\linewidth]{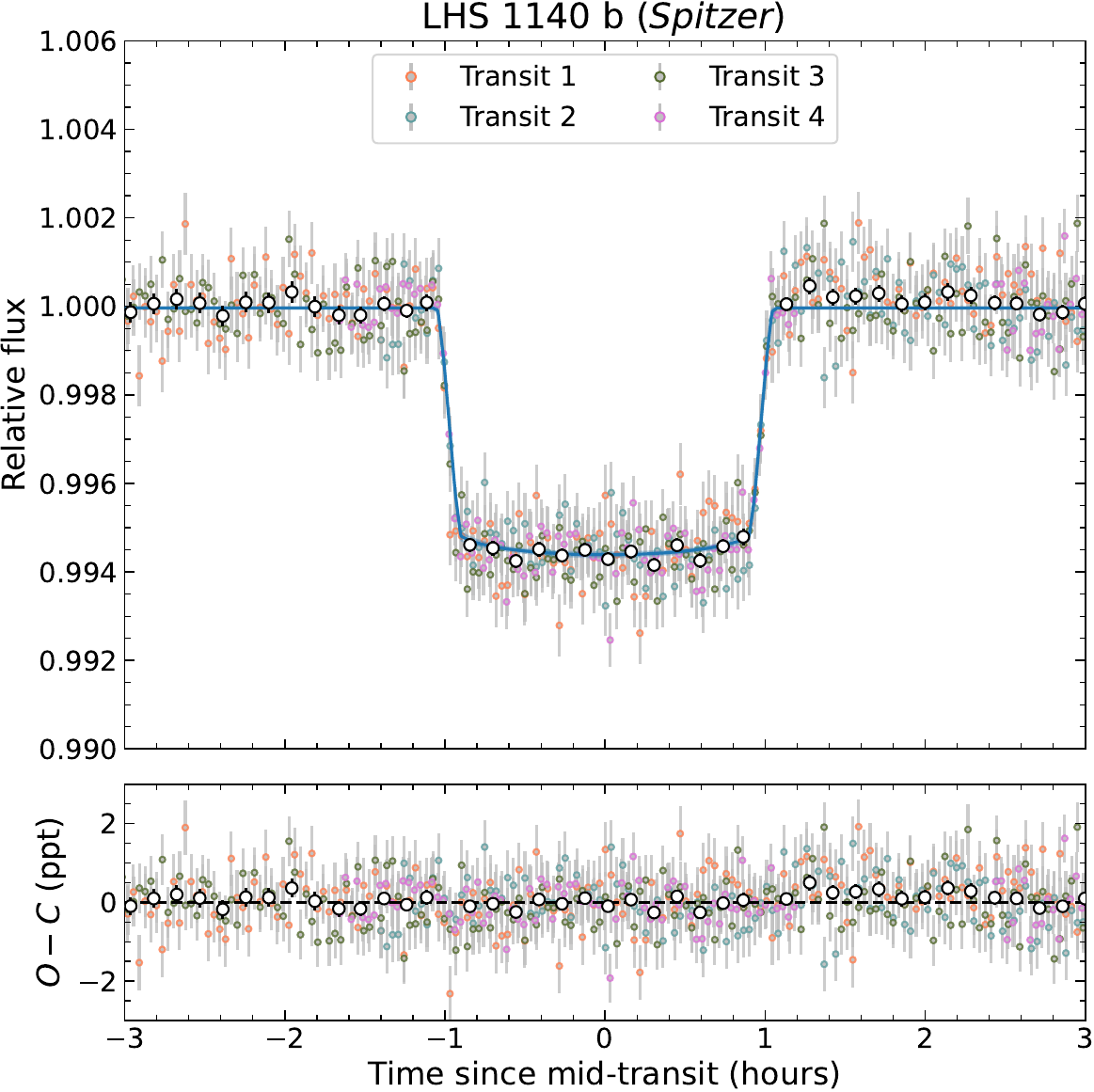}
  \caption{Phase-folded transit of LHS~1140\,b observed with \textit{Spitzer} Infrared Array Camera (at 4.5\,$\mu$m). The transit of LHS~1140\,c from Transit~4 is masked (see Fig.~\ref{fig:spitzer_individual} for individual transits). The open circles represent 8-minute binned photometry and the blue curve depicts the best-fit transit model. The residuals of this fit are shown below.}
  \label{fig:spitzer_transit}
\end{figure}

The observations were acquired using the subarray mode with a 2\,s exposure producing datacubes of 64 subarray images of 32$\times$32 pixels. We used the Spitzer Phase Curve Analysis (SPCA) pipeline \citep{Dang2018, Bell2021} to extract the photometry and decorrelate against instrumental systematics. For Transits 1 and 3, we use a 3$\times$3 pixel area to extract the target's intensity for each subarray frame. We then median-binned each datacube to mitigate the known subarray instrumental systematics and use Pixel Level Decorrelation \citep[PLD;][]{Deming2015} to detrend against detector systematics. Similarly to \citetalias{Ment_2019}, we elect to discard the first 78 minutes of Transit 4 during which the target's centroid had not yet settled on the detector. For Transit 2 and 4, we opted for a different detrending strategy as the shorter baseline before transit tends to bias the retrieved eclipse depth with PLD. Instead, we find that extracting the target photometry with an exact circular aperture with a radius of 3.0 pixels centered on the target's centroid yields the optimal photometric scheme. We then binned each datacube and detrended the instrumental systematics using a 2D polynomial as a function of centroid. The stacked (phase-folded) transit of LHS~1140\,b is presented in Figure~\ref{fig:spitzer_transit}. The individual detrended transits are presented Figure~\ref{fig:spitzer_individual}. 

\subsection{HST WFC3 white light curve} \label{sec:HST}

Two transits of LHS~1140\,b were observed with HST WFC3 on UT2017-01-28 and UT2017-12-15 (PN: 14888; PI: J.\ A.\ Dittmann). Unfortunately, due to large shifts in the position of the spectrum on the detector, the observation from UT2017-01-28 could not be reliably analysed \citep{Edwards_2021}. The transit on UT2017-12-15 was successfully analysed by \citet{Edwards_2021} to constrain the transmission spectrum of LHS~1140\,b near 1.4\,$\mu$m. The observations were conducted in the G141 grism configuration with the GRISM256 aperture (256$\times$256 subarray) and 103.13\,s integration time. Readers are referred to \citet{Edwards_2021} for a complete description of the HST data reduction that made use of the Iraclis software \citep{Tsiaras_2016}. Here, we include the extracted white light curve from HST (Fig.~\ref{fig:hst_transit}) for our transit analysis.

\subsection{TESS photometry} \label{sec:TESS}

LHS~1140 (TIC 92226327, TOI-256) was observed with TESS at a 2-minute cadence during its primary mission from September 20 to October 17, 2018 (Sector 3) and during its first extended mission from September 23 to October 20, 2020 (Sector 30). We used the Presearch Data Conditioning Simple Aperture Photometry (\texttt{PDCSAP}; \citealt{Smith_2012}; \citealt{Stumpe_2012}, \citeyear{Stumpe_2014}) data product issued by the TESS Science Processing Operations Center (SPOC, \citealt{Jenkins_2016}) at NASA Ames Research Center and available on the Mikulski Archive for Space Telescopes\footnote{\href{https://archive.stsci.edu/tess/}{\texttt{archive.stsci.edu/tess/}}}. The \texttt{PDCSAP} data includes corrections for instrumental systematics and for flux dilution from known \textit{Gaia} sources within several TESS pixels (21$\arcsec$ per pixel). The light curve from Sector 3 was reprocessed with a more recent release of SPOC (version 5.0.20), which applies the new background correction implemented for the extended mission to the first sectors of TESS. This new correction typically reduces the inferred transit depths by less than 2\%. Using the same pipeline (SPOC v5.0) for Sectors 3 and 30 ensures consistent transit depths for LHS~1140~b~and~c between primary and extended mission data. The full TESS light curve of LHS~1140 shown in Figure~\ref{fig:tess_transit} captures 2 and 11 transits of planet b and c, respectively, twice as many as in \citetalias{Lillo-Box_2020} based on Sector 3 data only. The phase-folded transits from TESS are also shown in the same Figure~\ref{fig:tess_transit}.

\subsection{ESPRESSO radial velocity} \label{sec:ESPRESSO}

We retrieved publicly available ESPRESSO data of LHS~1140 from the European Southern Observatory (ESO) science archive\footnote{\href{http://archive.eso.org/}{\texttt{archive.eso.org/}}} \citep{Delmotte_2006}. 
These data consist of the same 117 spectra analysed in \citetalias{Lillo-Box_2020} and taken with the SINGLEHR21 mode between October 2018 and December 2019. We used the bias \& dark subtracted, extracted and flat-fielded spectra reduced with the ESPRESSO pipeline (version 2.2.1)\footnote{\href{https://www.eso.org/sci/software/pipelines/espresso/}{\texttt{eso.org/sci/software/pipelines/espresso/}}}. The RV extraction from the reduced data was performed with the line-by-line (LBL, version 0.52) method of \cite{Artigau_2022} available as an open source package\footnote{\href{https://github.com/njcuk9999/lbl}{\texttt{github.com/njcuk9999/lbl}}}. A simple telluric correction is first performed inside the LBL code by fitting a TAPAS \citep{Bertaux_2014} atmospheric model. This correction step, comparable to the approach of \cite{Allart_2022}, has been demonstrated to improve the RV precision of ESPRESSO particularly for M-type stars.

At the core of the LBL method, first explored by \cite{Dumusque_2018}, Doppler shifts are measured on the smallest spectral range possible, i.e., a spectral line, from a high signal-to-noise ratio (SNR) template spectrum of the star and its first derivative \citep{Bouchy_2001}. This template is constructed by coadding all our 117 telluric-corrected spectra. Then, the statistical consistency between all per-line velocities ($\sim$38\,000 for LHS~1140) is verified using a simple mixture model (Appendix B of \citealt{Artigau_2022}) that effectively remove high-sigma outliers to produce a final error-weighted average of valid lines. This approach fully exploits the RV content of a stellar spectrum and is conceptually similar to widely employed template-matching algorithms (e.g., \citealt{Anglada-Escude_2012}, \citealt{Astudillo-Defru_2017}, \citealt{Zechmeister_2018}, \citealt{Silva_2022}) while being more resilient to outlying spectral features (e.g., telluric residuals, cosmic rays, detector defects).

\begin{figure}[t!]
\centering
  \includegraphics[width=0.90\linewidth]{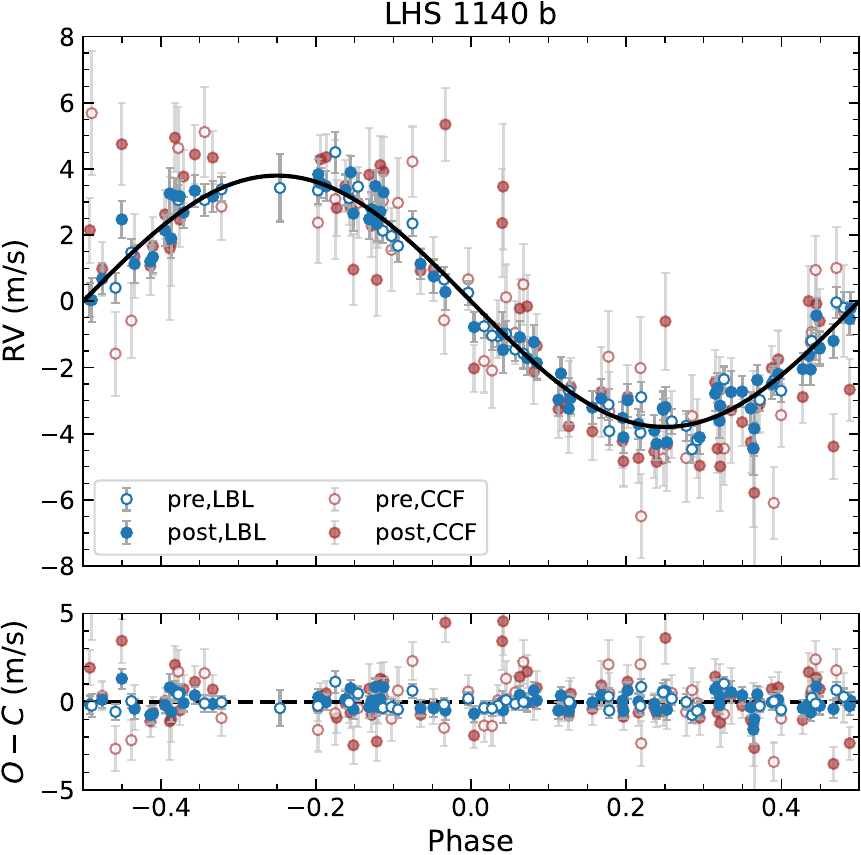}\\[0.25cm]
  \includegraphics[width=0.90\linewidth]{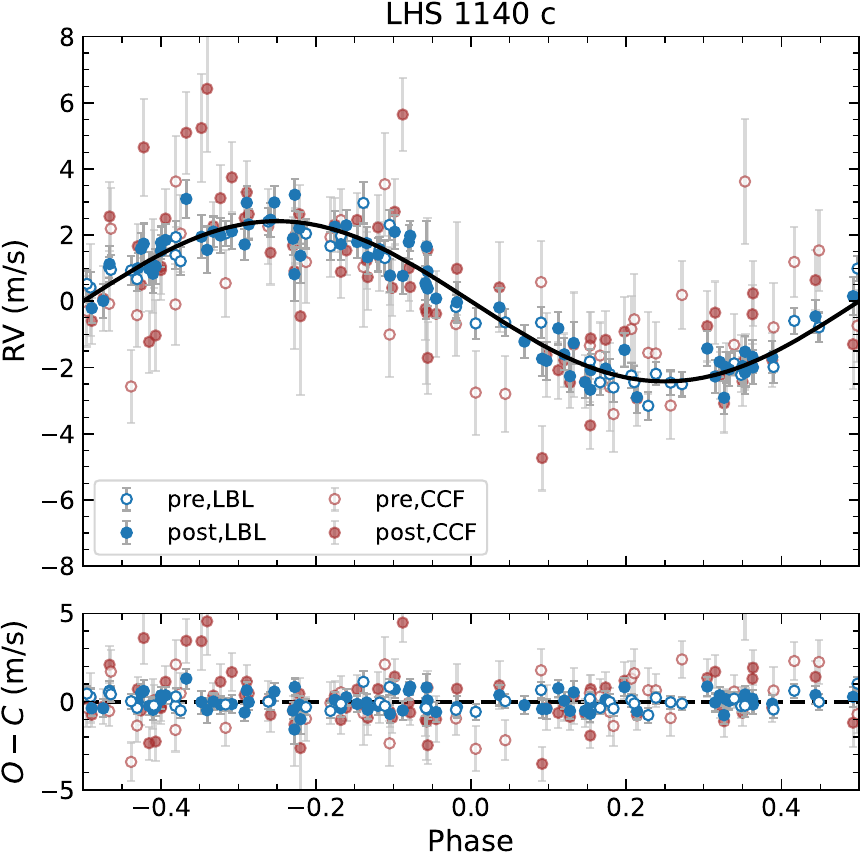}
  \caption{Phase-folded Keplerian signals of LHS~1140\,b (top) and c (bottom) with ESPRESSO from RVs produced with the line-by-line (LBL) method in blue and from cross-correlation function (CCF) in red \citep{Lillo-Box_2020}. Open and solid data points represent pre- and post-fiber upgrade of ESPRESSO (June 2019). The best-fit circular orbital solutions (black curves) of LHS~1140~b ($K_{\rm b} = 3.80 \pm 0.11$\,m\,s$^{-1}$) and c ($K_{\rm c} = 2.42 \pm 0.07$\,m\,s$^{-1}$) are respectively improved by 50\% and 70\% compared with previous estimates with CCF. The residuals RMS is equivalent to 0.50\,m\,s$^{-1}$ for LBL and 1.28\,m\,s$^{-1}$ for CCF.}
  \label{fig:RV}
\end{figure}

Following \citetalias{Lillo-Box_2020}, we separately analysed the data taken before and after the fiber link change of ESPRESSO in June 2019 \citep{Pepe_2021}, hereafter designated as ``pre" and ``post" velocities. Using an iterative sigma-clipping algorithm, we removed two epochs (BJD~=~2458703.787210 and 2458766.704332) flagged as $>$3$\sigma$ outliers. The final radial velocities are given in Table~\ref{table:rv} and show a median uncertainty of 0.36\,m\,s$^{-1}$ and dispersion (RMS) of 4.07\,m\,s$^{-1}$. As a comparison, the published values of \citetalias{Lillo-Box_2020} derived from the cross-correlation function (CCF) technique have a median precision of 0.99\,m\,s$^{-1}$, almost three times larger than LBL, and a 4.76\,m\,s$^{-1}$ scatter. Figure~\ref{fig:RV} presents a comparison between LBL and CCF for the best-fit orbits of LHS~1140~b~and~c. The full LBL RV sequence is shown in Figure~\ref{fig:RV_raw}. Given the extreme precision of ESPRESSO with LBL, a joint analysis with the HARPS data also available through the ESO archive has resulted in identical semi-amplitudes for LHS~1140~b~and~c. For this reason and to simplify the analysis, we opted to only use ESPRESSO in this work. The ESPRESSO observations span approximately 400 days, a long enough baseline to characterize signals at longer periods, such as the candidate LHS~1140\,d ($P_{\rm d} = 78.9$\,days) or the rotation of the star ($P_{\rm rot} = 131$\,days; \citetalias{Dittmann_2017}). As discussed in Appendices~\ref{sec:jointfit} and \ref{sec:validating}, we find no evidence for LHS~1140\,d and attribute this 80-day signal most likely to stellar activity.

\subsection{NIRPS high-resolution spectroscopy} \label{sec:NIRPS}

We acquired 29 high-resolution spectra of LHS~1140 with the Near-InfraRed Planet Searcher (NIRPS; \citealt{Bouchy_2017}; \citealt{Wildi_2022}) during one of its commissioning phases (Prog-ID 60.A-9109) from 2022-11-26 to 2022-12-06. NIRPS is a new echelle spectrograph designed for precision RV at the ESO 3.6-m telescope in La Silla, Chile covering the $YJH$ bands (980--1800\,nm). The instrument is equipped with a high-order Adaptive Optics (AO) system and two observing modes, High Accuracy (HA, $R\approx85\,000$, 0.4$^{\prime\prime}$ fiber) and High Efficiency (HE, $R\approx70\,000$, 0.9$^{\prime\prime}$ fiber), that can both be utilized simultaneously with HARPS. LHS~1140 was observed in HE mode as an RV standard star to test the stability of the instrument preceding the official start of NIRPS operation in April 2023. The observations were reduced with \texttt{APERO} v0.7.271 \citep{Cook_2022}, the standard data reduction software for the SPIRou near-infrared spectrograph \citep{Donati_2020}, fully compatible with NIRPS. We built a template spectrum of LHS~1140 from the telluric-corrected data product from \texttt{APERO} to derive independent stellar parameters and the abundances of several elements (Sect.~\ref{sec:abundances}). This template spectrum combines 29 individual spectra each with a SNR per pixel of about 70 in the middle of $H$ band.

\section{Stellar Characterization} \label{sec:stellar}

The star LHS~1140 was characterized in previous studies (\citetalias{Dittmann_2017}; \citetalias{Ment_2019}; \citetalias{Lillo-Box_2020}). In this section, we summarize our work to revise the stellar mass and radius and to measure the effective temperature and stellar abundances with NIRPS. An analysis of the stellar kinematics confirming the age ($>$5\,Gyr, \citetalias{Dittmann_2017}) and galactic thin disk membership of LHS~1140 is presented in Appendix~\ref{sec:age}.

\subsection{Stellar mass and radius update}
We pulled the $K_{\rm s}$ magnitude (8.82$\pm$0.02) of LHS~1140 from 2MASS \citep{Skrutskie_2006} and its distance $d$ from Gaia DR3 (14.96$\pm$0.01\,pc; \citealt{Gaia_Collaboration_2021}).
Then, from the absolute $K_{\rm s}$ magnitude ($M_{K_{\rm s}}$) to $M_{\star}$ empirical relation of \cite{Mann_2019}, we obtain a stellar mass of $0.1844 \pm 0.0046$\,M$_{\odot}$, with uncertainty propagating the errors on $K_{\rm s}$, $d$, and the scatter of this relation. This revised and more precise stellar mass is consistent with that of \citetalias{Ment_2019} ($M_{\star} = 0.179 \pm 0.014$\,M$_{\odot}$) obtained from a similar mass–luminosity calibration \citep{Benedict_2016}, but using a smaller sample of nearby binaries. In a similar way, but using the $M_{K_{\rm s}}$--$R_{\star}$ relationship of \cite{Mann_2015}, we obtain a radius of $0.2153 \pm 0.0065$\,R$_{\odot}$ for LHS~1140. \citetalias{Ment_2019} determine the stellar radius from an analysis of the transits, yielding a slightly more precise $R_{\star} = 0.2139 \pm 0.0041$\,R$_{\odot}$. To make sure that our results are completely independent of previous analyses, we first adopt the value derived from \cite{Mann_2015} as a prior, then further constrained the radius from the stellar density inferred from transits. This Bayesian method is detailed in Appendix~\ref{sec:bayes_stellar} and results in a new stellar mass and radius of $M_{\star} = 0.1844 \pm 0.0045$\,R$_{\odot}$ and $R_{\star} = 0.2159 \pm 0.0030$\,R$_{\odot}$. The stellar parameters of LHS~1140 are listed in Table~\ref{table:stellarparams}.

\subsection{Stellar abundances from NIRPS}\label{sec:abundances}

We follow the methodology of \cite{Jahandar_2023}, also applied in \cite{Cadieux_2022} for TOI-1452 (M4) and in \cite{Gan_2023} for TOI-4201 (M0.5), to derive the effective temperature as well as the abundances of several chemical species in LHS~1140 from the NIRPS template spectrum. A global fit ($\chi^2$ minimization) to a selection of strong spectral lines using ACES stellar models (\citealt{allard2012models}; \citealt{husser2013new}) convolved to match NIRPS resolution resulted in a  $T_{\rm eff} = 3096\pm48$\,K and a [M/H] = $0.01\pm0.04$ for LHS~1140. Note we fixed $\log g = 5$ (cgs) for our grid of models in accordance to LHS~1140 (Table~\ref{table:stellarparams}). This method was empirically calibrated for log~$g$ of 5.0$\pm$0.2\,dex \citep{Jahandar_2023}.

We then performed a series of fits for a fixed $T_{\rm eff} = 3100$\,K on individual spectral lines of known chemical species to derive their elemental abundances (again following \citealt{Jahandar_2023}). We show an example of this method in Figure~\ref{fig:nirps_met} for the Al I line at 1675.514\,nm from which we measure [Al/H] = $0.0\pm0.1$\,dex from this single line. The average abundances for all chemical species detected in LHS~1140 are given in Table~\ref{table:abundances} including the refractory elements Fe, Mg, and Si that form the bulk material of planetary cores and mantles. 
As shown in Table~\ref{table:ratios}, LHS~1140 features a relatively low Fe/Mg weight ratio (1.03$^{+0.40}_{-0.29}$) compared to the Sun (1.87$\pm0.22$) and other solar neighbourhood M dwarfs, with a C/O measurement consistent with the solar value. The measured Fe/Mg abundance ratio is used later as input to planetary internal structure models (Sect.~\ref{sec:nature_of_lhs1140b}).

\section{Results \& Discussion} \label{sec:discussion}

\subsection{New density measurements} \label{sec:composition}

We measure the physical and orbital parameters of LHS~1140~b~and~c by jointly fitting transit and Keplerian models to the photometric (\textit{Spitzer}, HST, and TESS) and RV (ESPRESSO) observations. The details of this joint transit RV fit are presented in Appendix~\ref{sec:jointfit}. Notably, the best-fit solution is two planets on circular orbits (Table~\ref{table:derivedparams}) with no evidence of candidate LHS~1140\,d. We adopt the average radius measured by \textit{Spitzer} and TESS for LHS~1140\,c but discuss an important discrepancy between the two measurements in Appendix~\ref{sec:tension}.

We infer a mass of 5.60$\pm$0.19\,M$_{\oplus}$ for LHS~1140\,b and 1.91$\pm$0.06\,M$_{\oplus}$ for LHS~1140\,c, as well as planetary radii of 1.730$\pm$0.025\,R$_{\oplus}$ for b and 1.272$\pm$0.026\,R$_{\oplus}$ for c. The LHS~1140 planets are among the best-characterized exoplanets to date, with relative uncertainties of only 3\% for the mass and 2\% for the radius, reaching a similar precision to the TRAPPIST-1 planets \citep{Agol_2021}. These measurements correspond to bulk densities of $5.9 \pm 0.3$\,g\,cm$^{-3}$ and $5.1 \pm 0.4$\,g\,cm$^{-3}$ for planet~b~and~c, respectively. The results of previous studies for the semi-amplitudes ($K$) and scaled radii ($R_{\rm p}/R_{\star}$) of the planets are shown in Table~\ref{table:comparison}. Since \citetalias{Lillo-Box_2020}, our updated $R_{\rm p}/R_{\star}$ ratios have increased back to \citetalias{Ment_2019} values. This change results from incorporating additional transits with \textit{Spitzer} and HST, as we retrieve the same $R_{\rm p}/R_{\star}$ as \citetalias{Lillo-Box_2020} to 1$\sigma$ when fitting the TESS data only (see Fig.~\ref{fig:rpr*}). Note our revision of the mass and radius of LHS~1140~b~and~c is dominated (by more than 80\%) by $K$ and $R_{\rm p}/R_{\star}$ changes, not by the update of stellar parameters.

\begin{deluxetable}{lcccc}
\tablecaption{Comparison of the semi-amplitudes and scaled radii of the LHS~1140 planets from different studies}
\tablehead{
    \colhead{Parameter} & \colhead{\citetalias{Dittmann_2017}} & \colhead{\citetalias{Ment_2019}} & \colhead{\citetalias{Lillo-Box_2020}} & \colhead{This work}
}
\startdata
\multicolumn{1}{c}{\textit{LHS~1140\,b}}\\
$K$ (m\,s$^{-1}$) & 5.34$\pm$1.10 & 4.85$\pm$0.55 & 4.21$\pm$0.24 & 3.80$\pm$0.11\\
$R_{\rm p}/R_{\star}$ (\%) & 7.08$\pm$0.13 & 7.39$\pm$0.01 & 7.02$\pm$0.19 & 7.33$\pm$0.04 \\[0.1cm]
\hline
\multicolumn{1}{c}{\textit{LHS~1140\,c}}\\
$K$ (m\,s$^{-1}$) & -- & 2.35$\pm$0.49 & 2.22$\pm$0.20 & 2.42$\pm$0.07\\
$R_{\rm p}/R_{\star}$ (\%) & -- &  5.49$\pm$0.01 & 5.02$\pm$0.16 & 5.39$\pm$0.08 
\enddata
\tablecomments{\citetalias{Dittmann_2017} with MEarth and HARPS. \citetalias{Ment_2019} with MEarth (extended), \textit{Spitzer} and HARPS (extended). \citetalias{Lillo-Box_2020} with TESS, HARPS (\citetalias{Ment_2019}), and ESPRESSO. This work with \textit{Spitzer} (extended), HST, TESS (extended), and ESPRESSO (\citetalias{Lillo-Box_2020}).}
\label{table:comparison}
\end{deluxetable}

\begin{figure*}[t!]
\centering
\includegraphics[width=1\linewidth]{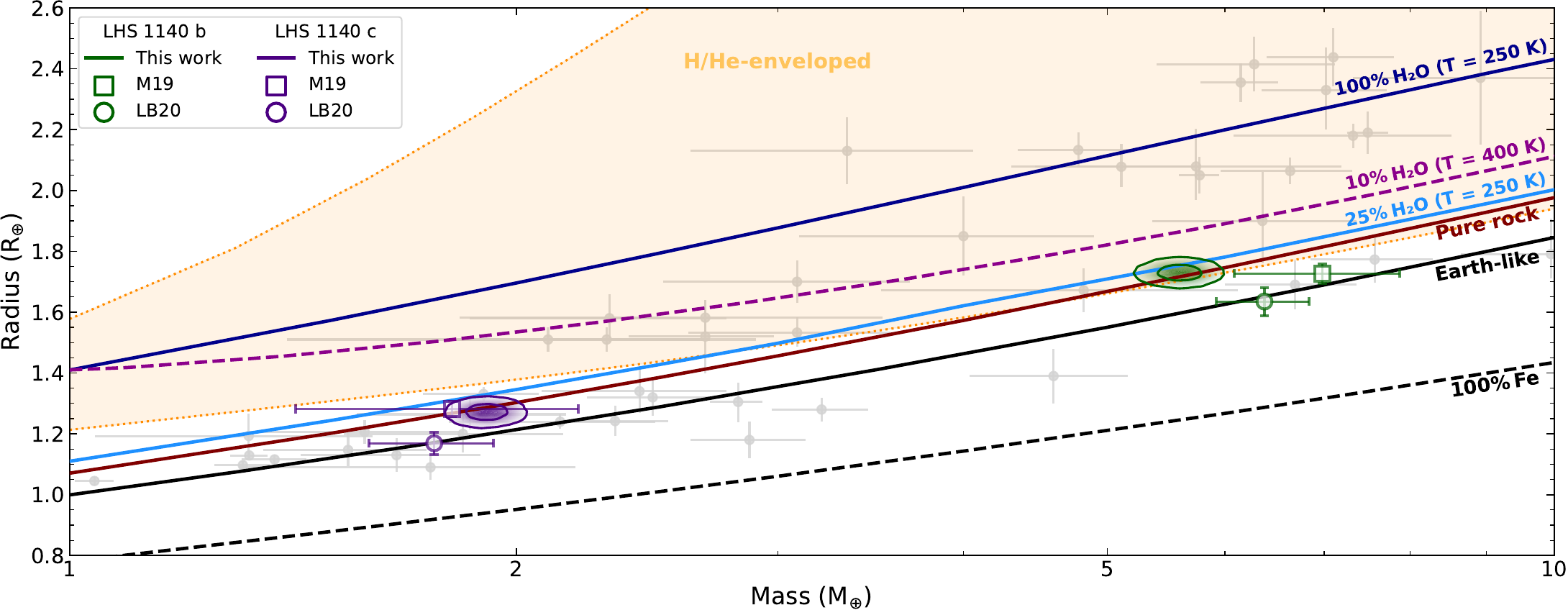}
  \caption{Mass--radius diagram of transiting exoplanets around M dwarfs (gray points) with well-established mass and radius (relative uncertainty $<30$\%) from the NASA Exoplanet Archive \citep{Akeson_2013}. Various theoretical composition curves from \cite{Guillot_1995, Valencia_2007,Valencia_2013,Aguichine_2021} are plotted with the water-rich models assuming an Earth-like interior (33\% iron by mass). The synthetic population of terrestrials enveloped in H/He (0.1--30\% initial mass fraction) of \citealt{Rogers_2023} is highlighted in orange. This region also highlights interior model degeneracy where the nature (H/He-rich or water-rich) of sub-Neptunes cannot be reliably determined from mass and radius alone. The mass and radius posteriors (1- and 2-$\sigma$ contours) of LHS~1140\,b (green) and LHS~1140\,c (purple) from our joint transit RV analysis are drawn. Published values from \citealt{Lillo-Box_2020} (LB20) derived from HARPS+ESPRESSO and TESS (Sector 3) are shown with open circles. The results of \citealt{Ment_2019} (M19) from HARPS and MEarth+\textit{Spitzer} are presented with open squares.}
  \label{fig:MR}
\end{figure*}

We compare our updated mass and radius to the M dwarf exoplanet population in Figure~\ref{fig:MR} with various pure composition curves included in the same figure. A detailed analysis of the internal structure of LHS~1140\,b is presented in Section~\ref{sec:nature_of_lhs1140b}. First examining LHS~1140\,b in Figure~\ref{fig:MR}, we see that our revised mass and radius are off the Earth-like track, contradicting previous results from \citetalias{Ment_2019} and \citetalias{Lillo-Box_2020} that this planet is a rocky, larger version of Earth (super-Earth). The mode of the distribution lies above the pure Mg--Si rock sequence, a region of the mass--radius diagram requiring an additional input of light elements, i.e., gas (H$_2$, He) or ices (e.g., H$_2$O, CH$_4$, NH$_3$), to explain the observed radius. LHS~1140\,b is on the lower limit of the synthetic sub-Neptune population around M dwarfs of \citealt{Rogers_2023} (orange region in Fig.~\ref{fig:MR}) that have undergone thermal evolution and atmospheric mass loss (photoevaporation) over 5\,Gyr assuming initial H/He mass fraction between 0.1\% and 30\%. With a small $T_{\rm eq}$ of 226\,K, \citealt{Rogers_2023} predict a $\sim$0.1\% H/He mass fraction for LHS~1140\,b which we confirm through simulation in Section~\ref{sec:interior_minineptune}. Alternatively, the planet could be purely rocky or with a water-mass fraction of 10--20\%, clearly not as high as the water-rich (50\% H$_2$O) population suggested by \cite{Luque_2022}. Note this latter scenario would involve a different formation mechanism, as suggested by recent studies \citep{Cloutier-Menou_2020, Luque_2022, Piaulet_2023, Cherubim_2023}, where small planets around M dwarfs could directly accrete icy materials outside the water snow line before migrating inwards, in which case sub-Neptunes would actually be water worlds rather than H/He-enveloped planets.

For LHS~1140\,c, our mass measurement agrees with previous studies, but we show in Appendix~\ref{sec:tension} that \textit{Spitzer} and TESS radii are in 4$\sigma$ disagreement, complicating the determination of its internal structure. In Figure~\ref{fig:MR}, we present the average \textit{Spitzer}+TESS radius of LHS~1140\,c most compatible with a rocky interior depleted in iron relative to Earth, but should the planet be smaller, as measured by TESS, an Earth-like interior remains plausible. However, given the planet size and higher $T_{\rm eq}$ of 422\,K, a hydrogen-dominated atmosphere or an important water content ($\gtrsim10\%$ by mass) similar to LHS~1140\,b are formally rejected.

\subsection{Nature of LHS~1140\,b} \label{sec:nature_of_lhs1140b}

The unprecedented precisions of both the mass and radius of the LHS~1140 planets combined with stellar abundance measurements provide a unique opportunity to better constrain the nature of these planets. Because of the radius uncertainty associated with LHS~1140\,c discussed above, we focus our analysis on LHS~1140\,b. The mass--radius diagram of Figure \ref{fig:MR} suggests three potential scenarios: (1) a mini-Neptune depleted in hydrogen, (2) a pure rocky (and airless) planet and (3) a water world. All three possibilities are discussed below.

To address scenarios 2 and 3, we follow the method of \cite{Plotnykov_2020} also applied to the water-world candidate TOI-1452\,b \citep{Cadieux_2022} to constrain both the core-mass fraction (CMF) and water-mass fraction (WMF) of the planet. The adaptation of this method to LHS~1140\,b is further detailed in Appendix~\ref{sec:interior_modeling} with the posterior distributions available in Appendix~\ref{posterior_dist}. In brief, this interior analysis treats the Fe/Mg ratio either as a free output parameter completely constrained by the mass-radius data (the no prior case) or as direct input informed by the measured stellar value (the stellar prior case) with $\mathrm{Fe/Mg_{planet}} \sim \mathcal{N}(\mathrm{Fe/Mg_{star}},\sigma_{\rm star}^2$). This latter case assumes that stellar abundances are a good proxy of planetary abundances as suggested by planet formation studies (e.g., \citealt{Bond_2010}, \citealt{Thiabaud_2015}, \citealt{Unterborn_2016}) and empirically (e.g., \citealt{Dorn2017}, \citealt{Bonsor_2021}) though this correlation is not necessarily 1:1 \citep{Adibekyan_2021}.

\subsubsection{Hydrogen-poor mini-Neptune}
\label{sec:interior_minineptune}

An Earth-like interior (CMF~=~33\%, WMF~$\sim$~0) overlain by a solar mixture of hydrogen-helium contributing $\sim$0.1\% of the mass, $\sim$10\% of the radius could explain the density of LHS~1140\,b. 
Here, we simulate the photoevaporation history of LHS~1140\,b for 10\,Gyr using the method of \cite{Cherubim_2023} to verify whether such hydrogen-rich envelope could survive at present day. These simulations take into account thermal evolution, photoevaporation (e.g., \citealt{Owen_2017}) from stellar extreme ultraviolet (XUV; 10--130\,nm) and core-powered atmospheric escape (e.g., \citealt{Ginzburg_2018}). 

\begin{figure}[t!]
\centering
\includegraphics[width=1\linewidth]{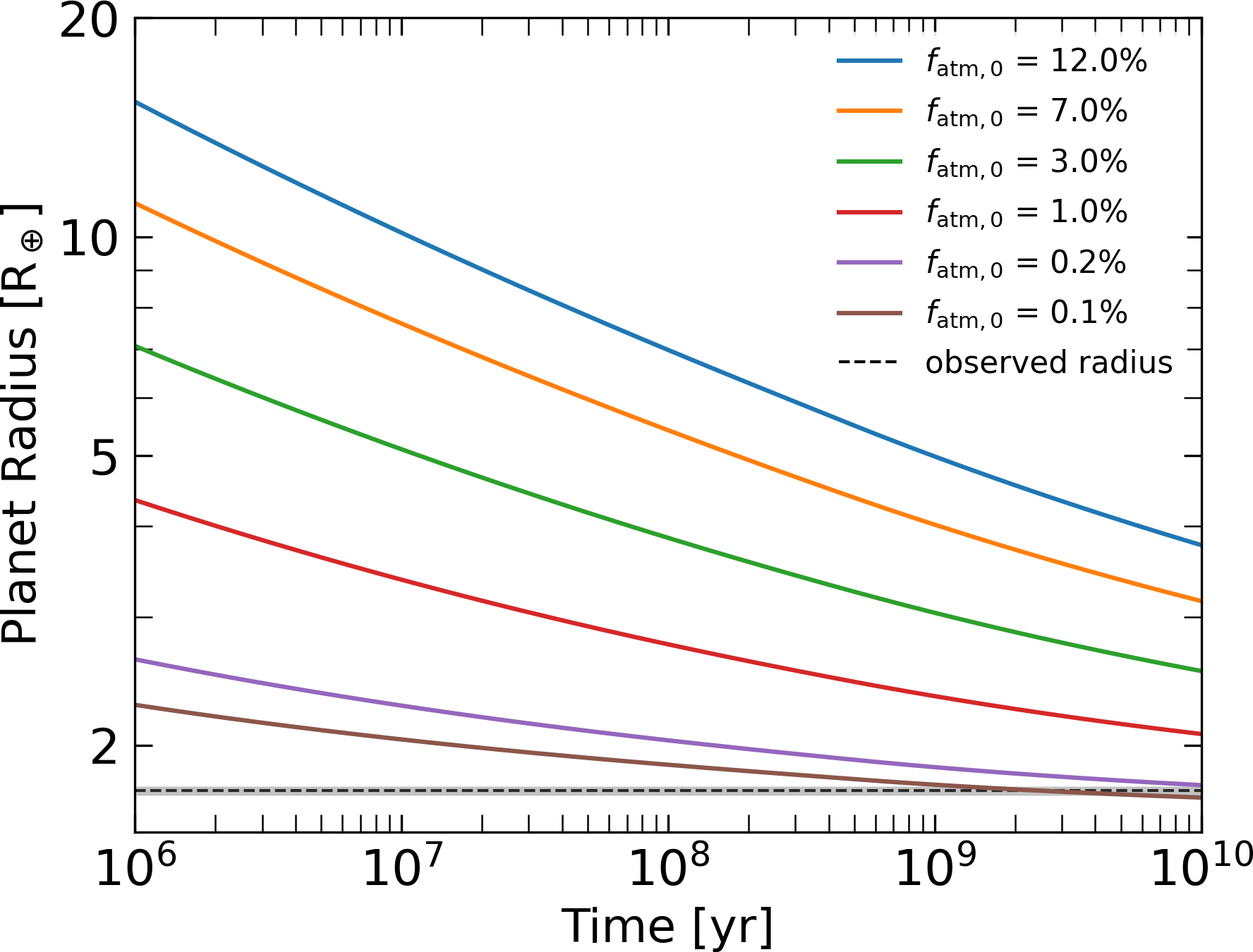}
  \caption{Primordial H/He atmosphere photoevaporation simulation of LHS~1140\,b. Each curve corresponds to different initial envelope mass fractions $f_{\rm atm,0}$. These simulations model thermal contraction, XUV-driven atmospheric escape, and core-powered mass loss using the methodology of \cite{Cherubim_2023}. A $f_{\rm atm,0}$~=~0.1--0.2\% is consistent with the observed radius given the age of the system ($>$5\,Gyr).}  
  \label{fig:LHS1140b_photoevaporation}
\end{figure}

The results for a range of initial envelope mass fractions ($f_{\rm atm,0}$ between 0.1\% and 12\%) are shown in Figure~\ref{fig:LHS1140b_photoevaporation}. Assuming that LHS~1140\,b did not undergo important migration after formation, it appears that a $f_{\rm atm,0}=0.1$--0.2\% (brown and purple curves in Fig.~\ref{fig:LHS1140b_photoevaporation}) is in agreement with the observed radius after 5\,Gyr, the minimum age estimate of the system. Since gas accretion models typically predict initial envelopes with mass fractions $\gtrsim$1\% \citep{Ginzburg_2016}, such a small $f_{\rm atm,0}$ for LHS~1140\,b would imply a formation in a gas-poor environment in less than 0.1\,Myr \citep{Lee_2021} or that it lost part of its atmosphere during giant impacts \citep{Inamdar_2016}. The relatively large semi-major axis of LHS~1140\,b ($\sim$0.1\,au) is just beyond the instellation needed to strip the atmosphere: the final $f_{\rm atm}$ after 10\,Gyr is close to the initial $f_{\rm atm,0}$. In other words, the radius evolution over the simulation timescale is dominated by the cooling/contraction of the atmosphere.

\subsubsection{Pure rocky planet}\label{sec:interior_barerock}

This is the special case of a rocky planet with no water envelope, i.e., the WMF forced to zero. With no prior on the Fe/Mg ratio, our model converges to a very small CMF (${4.3}_{-2.7}^{+4.1}$\%) essentially consistent with a coreless planet with a predicted Fe/Mg ratio (0.37$^{+0.12}_{-0.07}$) significantly lower (2.3$\sigma$) than observed for the host star (1.03$^{+0.40}_{-0.29}$) and much smaller than the lowest value ever measured in M dwarfs of the solar neighbourhood (see Fig.~\ref{fig:LHS1140b_Fe2Mg}). Moreover, this model converges to a larger mass and smaller radius than observed (2.3$\sigma$ offset in density). We argue that the inconsistency with the observations coupled with the challenge of forming highly iron depleted (coreless) planets (\citealt{Carter2015,Scora2020,Spaargaren2023}) make this scenario implausible for LHS~1140\,b. This conclusion is in line with exoplanet demographics \citep{Rogers_2015} and the empirical rocky-to-gaseous transition around M dwarfs \citep{Cloutier-Menou_2020} that most $\sim$1.6\,R$_{\oplus}$ exoplanets are not rocky.

\subsubsection{Water world}\label{sec:interior_waterworld}

For this scenario, we also include an atmospheric layer, essentially a (small) fixed radius correction associated with a potential atmosphere. For the general case of a water world receiving more irradiation than the runaway greenhouse threshold ($T_{\rm eq} \gtrsim 300$\,K; \citealt{Turbet_2020}; \citealt{Aguichine_2021}), the outer layer is likely to be supercritical, which would significantly inflate the radius, requiring a proper joint modeling of the warm water layer in vapor/supercritical state on top of a core+mantle interior. While we defer this general case to a future publication (Plotnykov et al.\ in prep.), LHS~1140\,b with a small $T_{\rm eq} = 226\pm6$\,K does not warrant such a detailed treatment since a potential outer water layer on this planet is most likely to be either in frozen or liquid state as the planet resides in the Water Condensation Zone \citep{Turbet_2023}. We thus assume the atmospheric layer of LHS~1140\,b to be an Earth-like atmosphere with a surface pressure of 1~bar and surface temperature equal to $T_{\rm eq}$. The radius correction (a few tens of kilometers) for such a thin atmosphere is negligible.

Two prior cases for the Fe/Mg abundance ratio are considered. The unconstrained case  (no prior) yields CMF~$={28.3}_{-15.9}^{+14.3}$\% and a WMF~$={18.7}_{-9.8}^{+12.6}$\% while assuming that the planet shares the same Fe/Mg ratio as the host star (stellar prior), we obtain CMF~$={20.5}_{-5.8}^{+5.5}$\% and WMF~$={13.7}_{-4.5}^{+5.4}$\%. As shown in Table~\ref{table:interiorparams}, adopting a solar Fe/Mg ratio instead of those measured on LHS~1140 yields an even larger WMF. We tested changing the surface temperature and pressure (using climate predictions from Sect.~\ref{sec:GCM}) to generate fully solid/liquid water surface. The effect of phase change is small: both CMF and WMF remained within the reported uncertainty for all models.

A variant of this model is the Hycean world \citep{Madhusudhan_2021}, i.e., a water world surrounded by a thin H/He-rich layer as was recently proposed for the temperate mini-Neptune K2-18\,b \citep{Madhusudhan_2023}. This new result opens the possibility that LHS~1140\,b may be a lower-mass version of such a Hycean planet in the middle of the radius valley \citep{Fulton_2017}. In this scenario, the lower mean molecular weight of the atmosphere yields a higher radius correction (up to 250 km for $\mu$\,$\sim$\,2) corresponding to $\sim$2\% of the planet radius. This case with stellar prior on Fe/Mg still yields a significant WMF~$={9.3}_{-3.9}^{+4.6}$\% with CMF~$={21.7}_{-6.6}^{+5.4}$\%.

The main conclusion from this modeling exercise summarized in Table~\ref{table:interiorparams} is that LHS~1140\,b is unlikely to be a rocky super-Earth. The planet is either a unique mini-Neptune with a thin $\sim$0.1\% H/He atmosphere or a water world with a WMF in the 9--19\% range depending on the atmospheric composition and the Fe/Mg ratio of the planetary interior. Transmission spectroscopic observations with JWST \citep{Gardner_2023} will be key to discriminate between these scenarios.

\begin{deluxetable}{lccc}
\tablecaption{Summary of the interior models of LHS~1140\,b}
\tablehead{
    \colhead{Model} & \colhead{CMF\,(\%)} & \colhead{WMF\,(\%)} & \colhead{Fe/Mg [w]}
}
\startdata
\multicolumn{1}{c}{\textit{LHS~1140}}\\
\footnotesize{Host star (reference)} & -- & -- & 1.0$^{+0.4}_{-0.3}$\\
\hline
\multicolumn{1}{c}{\textit{LHS~1140\,b}}\\
\footnotesize{Purely rocky (no prior)} & ${4.3}_{-2.7}^{+4.2}$ & -- & ${0.4}_{-0.1}^{+0.1}$\\
\footnotesize{Water world (no prior)}  & ${28.3}_{-15.9}^{+14.3}$ & ${18.7}_{-9.8}^{+12.6}$ & ${1.6}_{-0.9}^{+2.5}$\\
\footnotesize{Water world} & & &\\[-0.15cm]
\hspace{0.5cm} \footnotesize{(stellar prior)} & ${20.5}_{-5.8}^{+5.5}$ & ${13.7}_{-4.5}^{+5.4}$ & ${1.1}_{-0.3}^{+0.3}$\\
\footnotesize{Water world} & & &\\[-0.15cm]
\hspace{0.5cm} \footnotesize{(solar prior)} & ${27.8}_{-4.2}^{+4.6}$ & ${17.1}_{-5.0}^{+5.2}$ & ${1.8}_{-0.2}^{+0.2}$\\
\footnotesize{Hycean world$^{*}$} & & &\\[-0.15cm]
\hspace{0.5cm} \footnotesize{(stellar prior)} & ${21.7}_{-6.6}^{+5.4}$ & ${9.3}_{-3.9}^{+4.6}$ & ${1.1}_{-0.3}^{+0.3}$
\enddata
\tablecomments{$^{*}$The Hycean model is derived from the Water world model (stellar prior) by subtracting 250\,km to the planetary radius corresponding to 5 atmospheric scale heights of H$_2$.}
\label{table:interiorparams}
\end{deluxetable}

\subsection{3D GCM of LHS~1140\,b and prospects for atmospheric characterization} \label{sec:GCM}

\begin{figure*}[t!]
\centering
  \includegraphics[width=1.0\linewidth]{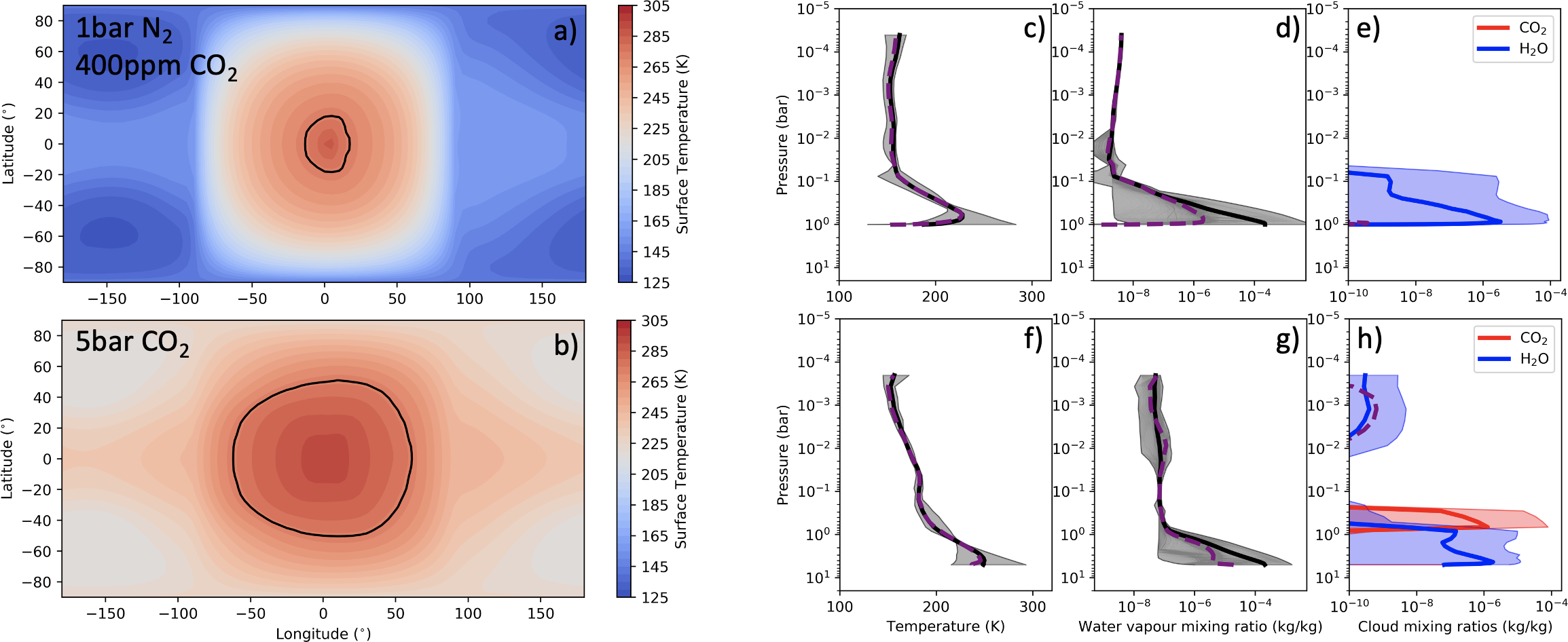}\\[0.3cm]
    \caption{Results of Global Climate Model (GCM) simulations of LHS~1140\,b assuming an Earth-like atmosphere (1 bar N$_2$, 400\,ppm CO$_2$; panels a, c, d, e) and a thick CO$_2$-dominated atmosphere (5\,bar CO$_2$; panels b, f, g, h). \textit{Panels a, b}: Surface temperature maps of LHS~1140\,b. The black line indicates the line of stability between liquid water and sea ice. The extent of the ice-free ocean strongly depends on the amount of CO$_2$ in the atmosphere. \textit{Panels c, d, e, f, g, and h}: Vertical profiles of the atmospheric temperatures, water vapor, and cloud mixing ratios (both H$_2$O and CO$_2$ ice clouds). The solid bold lines indicate the global mean vertical profiles and the dotted lines indicate terminator vertical profiles (impacting transit spectra).}
    \label{fig:GCM}
\end{figure*}

Here we consider the case of a water world with a thin atmosphere for LHS~1140\,b as this scenario presents broader implications for habitability. Planets with large amounts of water are likely to have CO$_2$+N$_2$+H$_2$O atmospheres \citep{Forget_2014,Kite_2018}, possibly with high amounts of CO$_2$ \citep{Marounina_2022}. While the diversity of water-world atmospheres has yet to be explored, we adopt here two distinct atmospheric compositions as a first working hypothesis and proof of concept to predict the transmission spectrum of LHS~1140\,b: an Earth-like atmosphere (1\,bar N$_2$, 400\,ppm CO$_2$) and a CO$_2$-dominated atmosphere (5\,bar CO$_2$). 

The simulations were computed with the Generic Planetary Climate Model (hereafter simply called GCM), a state-of-the-art 3D climate model \citep{Wordsworth_2011} historically known as the LMD (\textit{Laboratoire de
Météorologie Dynamique}, Paris, France) Generic Global Climate Model. The model has been widely applied to simulate all types of exoplanets, ranging from terrestrial planets like the TRAPPIST-1 planets \citep{Turbet_2018,Fauchez_2019} to mini-Neptunes like GJ\,1214\,b \citep{Charnay_2015} or K2-18\,b \citep{Charnay_2021}. The GCM uses an up-to-date generalized radiative transfer and can simulate a wide range of atmospheric compositions (N$_2$, H$_2$O, CO$_2$, etc.) including clouds self-consistently.

The model was used to make realistic predictions of LHS~1140\,b's atmospheric properties (temperature--pressure profile, water vapor, and cloud mixing ratios), summarized in Figure~\ref{fig:GCM}. We find that no matter how much CO$_2$ is included in the model, the planet has a patch of liquid water at the substellar point. The extent of the ice-free ocean grows with increasing atmospheric CO$_2$, due to the greenhouse effect. This result is similar to that shown by \cite{Turbet_2016,Boutle_2017,DelGenio_2018} for Proxima b and \cite{Wolf2017, Turbet_2018,Fauchez_2019} for TRAPPIST-1 planets, that water worlds synchronously rotating in the Habitable Zone of low-mass stars almost always have surface liquid water, at least at the substellar point. 

Note that the GCM simulations presented here do not include dynamic ocean and sea ice transport, which could slightly change the extent of the substellar liquid water ocean depending on the amount of CO$_2$ \citep{DelGenio_2019,Yang_2020}. Note also that we assumed that LHS~1140\,b is in synchronous rotation. This is the most likely rotation mode given LHS~1140\,b's proximity to its star and low eccentricity \citep{Ribas_2016}, producing gravitational tides which are expected to dominate atmospheric tides for this planet \citep{Leconte_2015}.

\begin{figure}[ht!]
\centering
  \includegraphics[width=1\linewidth]{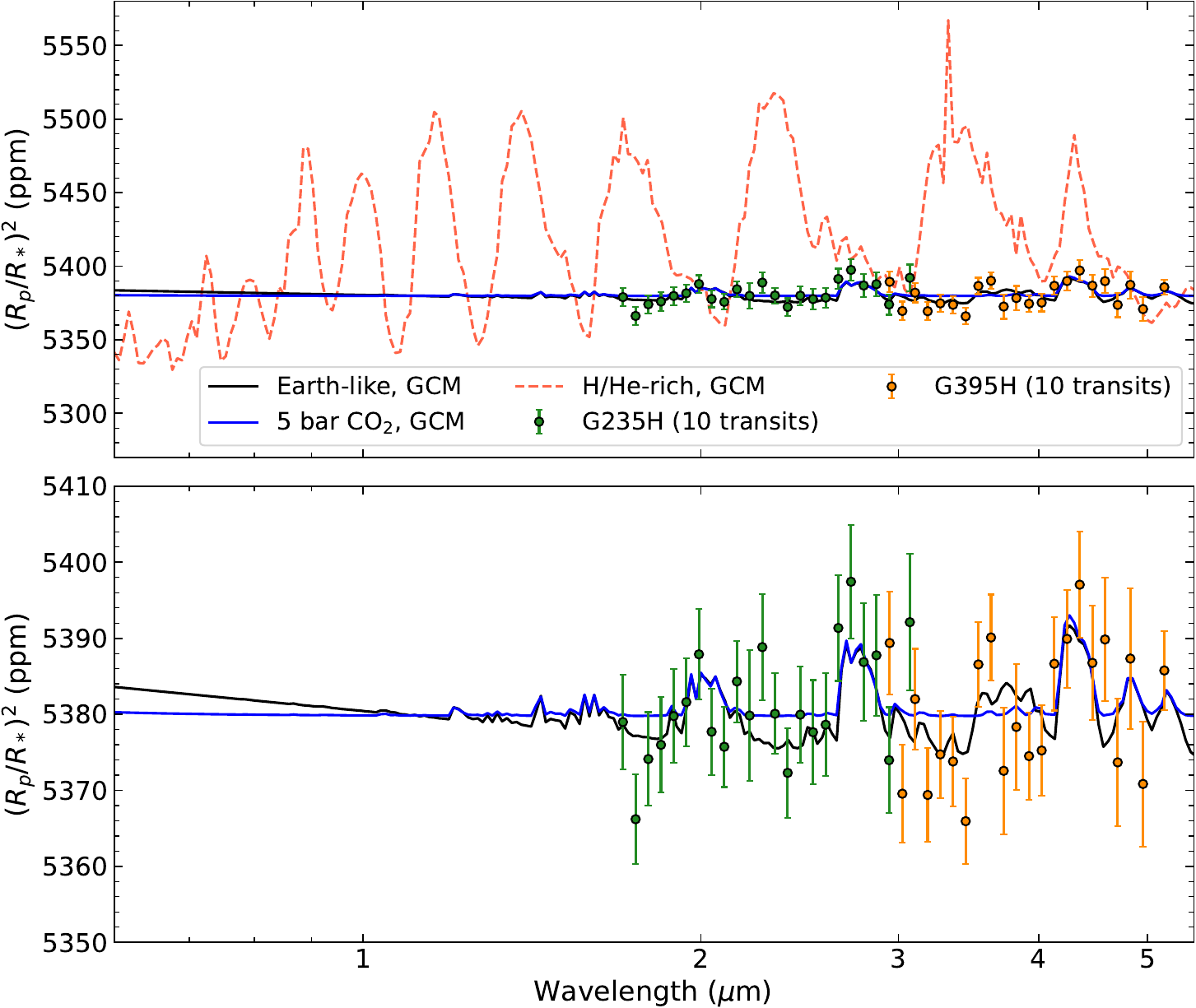}
  \caption{Predicted transmission spectrum of LHS~1140\,b from 0.6 to 5.5\,$\mu$m for the Earth-like (black) and 5\,bar CO$_2$ (blue) global climate models (GCM, Fig.~\ref{fig:GCM}). The bottom panel is a zoom of the top panel. Simulated JWST/NIRSpec data with the G235H (green) and G395H (orange) dispersers are shown. We estimate that 20 transits equally divided between G235H and G395H are required to detect a secondary atmosphere around LHS~1140\,b at 4$\sigma$. The red dashed curve is a GCM-based representation of a mini-Neptune (H/He-rich) that can be easily detected with a single visit.}  
  \label{fig:transit_spectrum}
\end{figure}

The outputs of the GCM were used to compute realistic transmission spectra that take into account the 3D nature of the atmosphere. We used the Planetary Spectrum Generator (PSG; \citealt{Villanueva_2018}, \citeyear{Villanueva_2022}) to produce transmission spectrum models (following the methodology of \citealt{Fauchez_2019}) of LHS~1140\,b based on the output of the two GCM simulations including the effects of H$_2$O/CO$_2$ clouds. As shown in Figure~\ref{fig:transit_spectrum}, the strongest atmospheric feature predicted by these models is CO$_2$ near 4.3\,$\mu$m with a strength of about 15\,ppm. Changing the partial pressure of CO$_2$ from 400~ppm to 5 bars produces a similar CO$_2$ bump as this molecule essentially condenses at higher concentration to form ice particles at low altitude (see Fig.~\ref{fig:GCM}, panel h). 

LHS~1140\,b was observed twice in transit with JWST in July 2023 during Cycle 1 (PID: 2334, PI: M.\ Damiano, unpublished). The two visits were made with NIRSpec \citep{Boker_2023} using the Bright Object Time Series mode, one transit with the G235H disperser (1.66--3.05\,$\mu$m) and the other with G395H (2.87--5.14\,$\mu$m). We used \texttt{PandExo} \citep{Batalha_2017} to simulate NIRSpec observations, keeping the same observing strategy of alternating between G235H and G395H. For a degraded spectral resolution of $R \simeq 20$ and assuming a conservative 5\,ppm noise floor \citep{Coulombe_2023}, these simulations predict sensitivity of about {20}\,ppm for G235H/G395H per spectral bin and per transit. As shown in Figure~\ref{fig:transit_spectrum}, the Cycle 1 data has the capability to identify the mini-Neptune scenario, but should LHS~1140\,b be a water world, we predict that an additional 18 transits (20 in total) would be required to detect its atmosphere (at 4$\sigma$). This estimation is made by simulating 1000 transmission spectra and comparing the log Bayesian evidence ($\ln Z$) between a flat spectrum and the Earth-like model of Figure~\ref{fig:transit_spectrum} with the median simulation shown in the same figure yielding a $\Delta \ln Z = 6.5$ or 4$\sigma$ \citep{Benneke_2013}. Note, the optimal observing strategy has yet to be fine-tuned depending on the level of stellar contamination yet to be characterized. This may warrant some visits to be obtained regularly with NIRISS SOSS (0.6--2.8\,$\mu$m; \citealt{Doyon_2023}, \citealt{Albert_2023}).

It should be noted that only four transits are observable every year with JWST as a result of the 24.7-day period of LHS~1140\,b and the fact that the system is near the ecliptic. An in-depth atmospheric characterization will realistically require that {\it all } future transit events of this planet be observed with JWST over several years (at least 3 years to detect a 15\,ppm CO$_2$ signal at 3$\sigma$). Irrespective of the nature of LHS~1140\,b, we advocate initiating an extensive campaign as soon as possible considering the uniqueness of this temperate world.

\section{Summary \& Conclusion} \label{sec:conclusion}
In this letter, we revisited the M4.5 LHS~1140 system hosting two transiting small exoplanets including LHS~1140\,b in the Habitable Zone. We applied the novel line-by-line precision radial velocity method of \cite{Artigau_2022} to publicly available ESPRESSO data, previously analysed by \cite{Lillo-Box_2020} using the cross-correlation function technique. The improvement on the radial velocities is significant: the errors are reduced by almost a factor three, the residual dispersion is halved, and no important excess white noise is detected. By jointly fitting the RVs with transits from \textit{Spitzer} (3 new archival transits), HST, and TESS (new Sector 30), we update the planetary mass and radius to 5.60$\pm$0.19\,M$_{\oplus}$ and 1.730$\pm$0.025\,R$_{\oplus}$ for LHS~1140\,b, 1.91$\pm$0.06\,M$_{\oplus}$ and 1.272$\pm$0.026\,R$_{\oplus}$ for LHS~1140\,c. The improved radial velocity data do not support the existence of the non-transiting candidate LHS~1140\,d announced by \cite{Lillo-Box_2020}.

Our revised mass and radius measurements reveal that LHS~1140\,b is unlikely to be a rocky super-Earth as previously reported, as it would require: (1) A density larger (by 2.3$\sigma$) than observed, (2) A core-mass fraction consistent with a coreless planet (CMF~$={4.3}_{-2.7}^{+4.2}$\%), (3) A planetary Fe/Mg weight ratio smaller (2.3$\sigma$) than measured on its host star and never measured in any solar neighbourhood M dwarfs. Instead, our analysis shows that LHS~1140\,b could either be one of the smallest known mini-Neptune ($\sim$0.1\% H/He by mass) with an atmosphere stable to mass loss over the lifetime of the system or a water world with a significant water-mass fraction of ${13.7}_{-4.5}^{+5.5}$\% when the iron-to-magnesium weight ratio of the planet is informed by those measured with NIRPS on the host star (Fe/Mg = 1.03$^{+0.40}_{-0.29}$). For LHS~1140\,c, our updated density is consistent with a rocky world depleted in iron relative to Earth's, but this result is highly influenced by radius measurement currently discrepant between \textit{Spitzer} and TESS to 4$\sigma$.

We recommend in-depth transit spectroscopy with JWST to characterize the atmosphere on LHS~1140\,b which we predict would create a CO$_2$ feature at 4.3\,$\mu$m, potentially as small as $\sim$15\,ppm according to self-consistent 3D global climate modeling of the planet in the water world hypothesis. These simulations show that the atmospheric CO$_2$ concentration controls the surface temperature and the extent of a liquid water ocean. For an Earth-like case (1\,bar N$_2$, 400\,ppm CO$_2$), liquid water is limited to a small patch at the substellar point, while for a CO$_2$-dominated atmosphere (5\,bar CO$_2$), almost a whole hemisphere is covered. These future observations could reveal the first exoplanet with a potentially habitable atmosphere and surface.

\fakesection{Acknowledgments}
\vspace{1cm}
We thank the anonymous referee for constructive comments and suggestions that improved the presentation of this letter.

This study uses public ESPRESSO data under program IDs 0102.C-0294(A), 0103.C-0219(A), and 0104.C-0316(A) (PI: J.\ Lillo-Box).

This work is partly supported by the Natural Science and Engineering Research Council of Canada and the Trottier Institute for Research on Exoplanets through the Trottier Family Foundation.

We acknowledge the use of public TESS Alert data from pipelines at the TESS Science Office and at the TESS Science Processing Operations Center. Resources supporting this work were provided by the NASA High-End Computing (HEC) Program through the NASA Advanced Supercomputing (NAS) Division at Ames Research Center for the production of the SPOC data products. This letter includes data collected by the TESS mission that are publicly available from the Mikulski Archive for Space Telescopes (MAST). 

This work has been carried out within the framework of the NCCR PlanetS supported by the Swiss National Science Foundation (SNSF) under grants 51NF40\_182901 and 51NF40\_205606. This project has received funding from the SNSF for project 200021\_200726. F.P. would like to acknowledge the SNSF for supporting research with ESPRESSO and NIRPS through grants nr. 140649, 152721, 166227, 184618 and 215190.

This project has received funding from the European Research Council (ERC) under the European Union’s Horizon 2020 research and innovation program (grant agreement SCORE No 851555)

Co-funded by the European Union (ERC, FIERCE, 101052347). Views and opinions expressed are however those of the author(s) only and do not necessarily reflect those of the European Union or the European Research Council. Neither the European Union nor the granting authority can be held responsible for them.

M.T. thanks the Gruber Foundation for its generous support to this research, support from the Tremplin 2022 program of the Faculty of Science and Engineering of Sorbonne University, and the Generic PCM team for the teamwork development and improvement of the model. This work was performed using the High-Performance Computing (HPC) resources of Centre Informatique National de l'Enseignement Supérieur (CINES) under the allocations No. A0100110391 and A0120110391 made by Grand Équipement National de Calcul Intensif (GENCI).

T.J.F. acknowledges support from the GSFC Sellers Exoplanet Environments Collaboration (SEEC), which is funded in part by the NASA Planetary Science Divisions Internal Scientist Funding Model.

R.A. is a Trottier Postdoctoral Fellow and acknowledges support from the Trottier Family Foundation. This work was supported in part through a grant from the Fonds de Recherche du Qu\'ebec - Nature et Technologies (FRQNT).

J.I.G.H., V.M.P., and A.S.M. acknowledge financial support from the Spanish Ministry of Science and Innovation (MICINN) project PID2020-117493GB-I00 and A.S.M. from the Government of the Canary Islands project ProID2020010129.

This work was supported by Funda\c{c}\~ao para a Ci\^encia e a Tecnologia (FCT) through national funds and by FEDER through COMPETE2020 - Programa Operacional Competitividade e Internacionalização by these grants: UIDB/04434/2020; UIDP/04434/2020. The research leading to these results has received funding from the European Research Council through the grant agreement 101052347 (FIERCE). E.D.M. acknowledges the support from through Stimulus FCT contract 2021.01294.CEECIND and by the following grants: UIDB/04434/2020 \& UIDP/04434/2020 and 2022.04416.PTDC.

B.L.C.M. and J.R.M. acknowledge continuous grants from the Brazilian funding agencies CNPq and Print/CAPES/UFRN. This study was financed in part by the Coordenação de Aperfeiçoamento de Pessoal de Nível Superior - Brasil (CAPES) - Finance Code 001.

T.H. acknowledges support from an NSERC Alexander Graham Bell CGS-D scholarship.

X.D. acknowledges support by the French National Research Agency in the framework of the Investissements d'Avenir program (ANR-15-IDEX-02), through the funding of the ``Origin of Life" project of the Grenoble-Alpes University.

\facilities{\textit{Spitzer}, HST, TESS, ESO-VLT/ESPRESSO, ESO-La Silla/NIRPS}

\software{\texttt{emcee} \citep{Foreman-Mackey_2013}; \texttt{Astropy} \citep{Astropy_2018}; \texttt{radvel} \citep{Fulton_2018}; \texttt{matplotlib} \citep{Hunter_2007}; \texttt{juliet} \citep{Espinoza_2019}; \texttt{batman} \citep{Kreidberg_2015}; \texttt{SciPy} \citep{Virtanen_2020}; \texttt{NumPy} \citep{Harris_2020} \texttt{zeus} \citep{Karamanis_2021}; \texttt{seaborn} \citep{Waskom_2021}; \texttt{exofile} (\href{https://github.com/AntoineDarveau/exofile}{github.com/AntoineDarveau/exofile}); \texttt{corner} \citep{Foreman-Mackey_2016}; \texttt{PSG} \citep{Villanueva_2022}}

\bibliography{LHS1140}{}

\begin{thebibliography}{}
\expandafter\ifx\csname natexlab\endcsname\relax\def\natexlab#1{#1}\fi
\providecommand{\url}[1]{\href{#1}{#1}}
\providecommand{\dodoi}[1]{doi:~\href{http://doi.org/#1}{\nolinkurl{#1}}}
\providecommand{\doeprint}[1]{\href{http://ascl.net/#1}{\nolinkurl{http://ascl.net/#1}}}
\providecommand{\doarXiv}[1]{\href{https://arxiv.org/abs/#1}{\nolinkurl{https://arxiv.org/abs/#1}}}

\bibitem[{{Adibekyan} {et~al.}(2021){Adibekyan}, {Dorn}, {Sousa}, {Santos},
  {Bitsch}, {Israelian}, {Mordasini}, {Barros}, {Delgado Mena}, {Demangeon},
  {Faria}, {Figueira}, {Hakobyan}, {Oshagh}, {Soares}, {Kunitomo}, {Takeda},
  {Jofr{\'e}}, {Petrucci}, \& {Martioli}}]{Adibekyan_2021}
{Adibekyan}, V., {Dorn}, C., {Sousa}, S.~G., {et~al.} 2021, Science, 374, 330,
  \dodoi{10.1126/science.abg8794}

\bibitem[{{Agol} {et~al.}(2021){Agol}, {Dorn}, {Grimm}, {Turbet}, {Ducrot},
  {Delrez}, {Gillon}, {Demory}, {Burdanov}, {Barkaoui}, {Benkhaldoun},
  {Bolmont}, {Burgasser}, {Carey}, {de Wit}, {Fabrycky}, {Foreman-Mackey},
  {Haldemann}, {Hernandez}, {Ingalls}, {Jehin}, {Langford}, {Leconte},
  {Lederer}, {Luger}, {Malhotra}, {Meadows}, {Morris}, {Pozuelos}, {Queloz},
  {Raymond}, {Selsis}, {Sestovic}, {Triaud}, \& {Van Grootel}}]{Agol_2021}
{Agol}, E., {Dorn}, C., {Grimm}, S.~L., {et~al.} 2021, Planet. Sci. J., 2, 1,
  \dodoi{10.3847/PSJ/abd022}

\bibitem[{{Aguichine} {et~al.}(2021){Aguichine}, {Mousis}, {Deleuil}, \&
  {Marcq}}]{Aguichine_2021}
{Aguichine}, A., {Mousis}, O., {Deleuil}, M., \& {Marcq}, E. 2021, \apj, 914,
  84, \dodoi{10.3847/1538-4357/abfa99}

\bibitem[{{Ahumada} {et~al.}(2020){Ahumada}, {Prieto}, {Almeida}, {Anders},
  {Anderson}, {Andrews}, {Anguiano}, {Arcodia}, {Armengaud}, {Aubert}, {Avila},
  {Avila-Reese}, {Badenes}, {Balland}, {Barger}, {Barrera-Ballesteros}, {Basu},
  {Bautista}, {Beaton}, {Beers}, {Benavides}, {Bender}, {Bernardi}, {Bershady},
  {Beutler}, {Bidin}, {Bird}, {Bizyaev}, {Blanc}, {Blanton}, {Boquien},
  {Borissova}, {Bovy}, {Brandt}, {Brinkmann}, {Brownstein}, {Bundy}, {Bureau},
  {Burgasser}, {Burtin}, {Cano-D{\'\i}az}, {Capasso}, {Cappellari}, {Carrera},
  {Chabanier}, {Chaplin}, {Chapman}, {Cherinka}, {Chiappini}, {Doohyun Choi},
  {Chojnowski}, {Chung}, {Clerc}, {Coffey}, {Comerford}, {Comparat}, {da
  Costa}, {Cousinou}, {Covey}, {Crane}, {Cunha}, {Ilha}, {Dai}, {Damsted},
  {Darling}, {Davidson}, {Davies}, {Dawson}, {De}, {de la Macorra}, {De Lee},
  {Queiroz}, {Deconto Machado}, {de la Torre}, {Dell'Agli}, {du Mas des
  Bourboux}, {Diamond-Stanic}, {Dillon}, {Donor}, {Drory}, {Duckworth},
  {Dwelly}, {Ebelke}, {Eftekharzadeh}, {Davis Eigenbrot}, {Elsworth},
  {Eracleous}, {Erfanianfar}, {Escoffier}, {Fan}, {Farr},
  {Fern{\'a}ndez-Trincado}, {Feuillet}, {Finoguenov}, {Fofie},
  {Fraser-McKelvie}, {Frinchaboy}, {Fromenteau}, {Fu}, {Galbany}, {Garcia},
  {Garc{\'\i}a-Hern{\'a}ndez}, {Oehmichen}, {Ge}, {Maia}, {Geisler}, {Gelfand},
  {Goddy}, {Gonzalez-Perez}, {Grabowski}, {Green}, {Grier}, {Guo}, {Guy},
  {Harding}, {Hasselquist}, {Hawken}, {Hayes}, {Hearty}, {Hekker}, {Hogg},
  {Holtzman}, {Horta}, {Hou}, {Hsieh}, {Huber}, {Hunt}, {Chitham}, {Imig},
  {Jaber}, {Angel}, {Johnson}, {Jones}, {J{\"o}nsson}, {Jullo}, {Kim},
  {Kinemuchi}, {Kirkpatrick}, {Kite}, {Klaene}, {Kneib}, {Kollmeier}, {Kong},
  {Kounkel}, {Krishnarao}, {Lacerna}, {Lan}, {Lane}, {Law}, {Le Goff}, {Leung},
  {Lewis}, {Li}, {Lian}, {Lin}, {Long}, {Longa-Pe{\~n}a}, {Lundgren}, {Lyke},
  {Ted Mackereth}, {MacLeod}, {Majewski}, {Manchado}, {Maraston}, {Martini},
  {Masseron}, {Masters}, {Mathur}, {McDermid}, {Merloni}, {Merrifield},
  {M{\'e}sz{\'a}ros}, {Miglio}, {Minniti}, {Minsley}, {Miyaji}, {Mohammad},
  {Mosser}, {Mueller}, {Muna}, {Mu{\~n}oz-Guti{\'e}rrez}, {Myers}, {Nadathur},
  {Nair}, {Nandra}, {do Nascimento}, {Nevin}, {Newman}, {Nidever}, {Nitschelm},
  {Noterdaeme}, {O'Connell}, {Olmstead}, {Oravetz}, {Oravetz}, {Osorio},
  {Pace}, {Padilla}, {Palanque-Delabrouille}, {Palicio}, {Pan}, {Pan},
  {Parker}, {Paviot}, {Peirani}, {Ram{\'r}ez}, {Penny}, {Percival},
  {Perez-Fournon}, {P{\'e}rez-R{\`a}fols}, {Petitjean}, {Pieri},
  {Pinsonneault}, {Poovelil}, {Povick}, {Prakash}, {Price-Whelan}, {Raddick},
  {Raichoor}, {Ray}, {Rembold}, {Rezaie}, {Riffel}, {Riffel}, {Rix}, {Robin},
  {Roman-Lopes}, {Rom{\'a}n-Z{\'u}{\~n}iga}, {Rose}, {Ross}, {Rossi},
  {Rowlands}, {Rubin}, {Salvato}, {S{\'a}nchez}, {S{\'a}nchez-Menguiano},
  {S{\'a}nchez-Gallego}, {Sayres}, {Schaefer}, {Schiavon}, {Schimoia},
  {Schlafly}, {Schlegel}, {Schneider}, {Schultheis}, {Schwope}, {Seo},
  {Serenelli}, {Shafieloo}, {Shamsi}, {Shao}, {Shen}, {Shetrone}, {Shirley},
  {Aguirre}, {Simon}, {Skrutskie}, {Slosar}, {Smethurst}, {Sobeck}, {Sodi},
  {Souto}, {Stark}, {Stassun}, {Steinmetz}, {Stello}, {Stermer},
  {Storchi-Bergmann}, {Streblyanska}, {Stringfellow}, {Stutz}, {Su{\'a}rez},
  {Sun}, {Taghizadeh-Popp}, {Talbot}, {Tayar}, {Thakar}, {Theriault}, {Thomas},
  {Thomas}, {Tinker}, {Tojeiro}, {Toledo}, {Tremonti}, {Troup}, {Tuttle},
  {Unda-Sanzana}, {Valentini}, {Vargas-Gonz{\'a}lez}, {Vargas-Maga{\~n}a},
  {V{\'a}zquez-Mata}, {Vivek}, {Wake}, {Wang}, {Weaver}, {Weijmans}, {Wild},
  {Wilson}, {Wilson}, {Wolthuis}, {Wood-Vasey}, {Yan}, {Yang}, {Y{\`e}che},
  {Zamora}, {Zarrouk}, {Zasowski}, {Zhang}, {Zhao}, {Zhao}, {Zheng}, {Zheng},
  {Zhu}, \& {Zou}}]{Ahumada_2020}
{Ahumada}, R., {Prieto}, C.~A., {Almeida}, A., {et~al.} 2020, \apjs, 249, 3,
  \dodoi{10.3847/1538-4365/ab929e}

\bibitem[{{Akeson} {et~al.}(2013){Akeson}, {Chen}, {Ciardi}, {Crane}, {Good},
  {Harbut}, {Jackson}, {Kane}, {Laity}, {Leifer}, {Lynn}, {McElroy}, {Papin},
  {Plavchan}, {Ram{\'\i}rez}, {Rey}, {von Braun}, {Wittman}, {Abajian}, {Ali},
  {Beichman}, {Beekley}, {Berriman}, {Berukoff}, {Bryden}, {Chan}, {Groom},
  {Lau}, {Payne}, {Regelson}, {Saucedo}, {Schmitz}, {Stauffer}, {Wyatt}, \&
  {Zhang}}]{Akeson_2013}
{Akeson}, R.~L., {Chen}, X., {Ciardi}, D., {et~al.} 2013, \pasp, 125, 989,
  \dodoi{10.1086/672273}

\bibitem[{Albert {et~al.}(2023)Albert, Lafrenière, Doyon, Artigau, Volk,
  Goudfrooij, Martel, Radica, Rowe, Espinoza, Roy, Filippazzo, Darveau-Bernier,
  Talens, Sivaramakrishnan, Willott, Fullerton, LaMassa, Hutchings, Rowlands,
  Begoña~Vila, Zhou, Aldridge, Maszkiewicz, Beaulieu, Cook, Piaulet, Roy,
  Lamontagne, Morel, Frost, Salhi, Coulombe, Benneke, MacDonald, Johnstone,
  Turner, Fournier-Tondreau, Allart, \& Kaltenegger}]{Albert_2023}
Albert, L., Lafrenière, D., Doyon, R., {et~al.} 2023, Publications of the
  Astronomical Society of the Pacific, 135, 075001,
  \dodoi{10.1088/1538-3873/acd7a3}

\bibitem[{{Allard} {et~al.}(2012){Allard}, {Homeier}, \&
  {Freytag}}]{allard2012models}
{Allard}, F., {Homeier}, D., \& {Freytag}, B. 2012, Philosophical Transactions
  of the Royal Society of London Series A, 370, 2765,
  \dodoi{10.1098/rsta.2011.0269}

\bibitem[{{Allart} {et~al.}(2022){Allart}, {Lovis}, {Faria}, {Dumusque},
  {Sosnowska}, {Figueira}, {Silva}, {Mehner}, {Pepe}, {Cristiani}, {Rebolo},
  {Santos}, {Adibekyan}, {Cupani}, {Di Marcantonio}, {D'Odorico}, {Gonz{\'a}lez
  Hern{\'a}ndez}, {Martins}, {Milakovi{\'c}}, {Nunes}, {Sozzetti}, {Su{\'a}rez
  Mascare{\~n}o}, {Tabernero}, \& {Zapatero Osorio}}]{Allart_2022}
{Allart}, R., {Lovis}, C., {Faria}, J., {et~al.} 2022, \aap, 666, A196,
  \dodoi{10.1051/0004-6361/202243629}

\bibitem[{{Ambikasaran} {et~al.}(2015){Ambikasaran}, {Foreman-Mackey},
  {Greengard}, {Hogg}, \& {O'Neil}}]{Ambikasaran_2015}
{Ambikasaran}, S., {Foreman-Mackey}, D., {Greengard}, L., {Hogg}, D.~W., \&
  {O'Neil}, M. 2015, IEEE Transactions on Pattern Analysis and Machine
  Intelligence, 38, 252, \dodoi{10.1109/TPAMI.2015.2448083}

\bibitem[{{Anglada-Escud{\'e}} \& {Butler}(2012)}]{Anglada-Escude_2012}
{Anglada-Escud{\'e}}, G., \& {Butler}, R.~P. 2012, \apjs, 200, 15,
  \dodoi{10.1088/0067-0049/200/2/15}

\bibitem[{{Anglada-Escud{\'e}} {et~al.}(2016){Anglada-Escud{\'e}}, {Amado},
  {Barnes}, {Berdi{\~n}as}, {Butler}, {Coleman}, {de La Cueva}, {Dreizler},
  {Endl}, {Giesers}, {Jeffers}, {Jenkins}, {Jones}, {Kiraga}, {K{\"u}rster},
  {L{\'o}pez-Gonz{\'a}lez}, {Marvin}, {Morales}, {Morin}, {Nelson}, {Ortiz},
  {Ofir}, {Paardekooper}, {Reiners}, {Rodr{\'\i}guez},
  {Rodr{\'\i}guez-L{\'o}pez}, {Sarmiento}, {Strachan}, {Tsapras}, {Tuomi}, \&
  {Zechmeister}}]{Anglada_2016}
{Anglada-Escud{\'e}}, G., {Amado}, P.~J., {Barnes}, J., {et~al.} 2016, \nat,
  536, 437, \dodoi{10.1038/nature19106}

\bibitem[{{Artigau} {et~al.}(2022){Artigau}, {Cadieux}, {Cook}, {Doyon},
  {Vandal}, {Donati}, {Moutou}, {Delfosse}, {Fouqu{\'e}}, {Martioli}, {Bouchy},
  {Parsons}, {Carmona}, {Dumusque}, {Astudillo-Defru}, {Bonfils}, \&
  {Mignon}}]{Artigau_2022}
{Artigau}, {\'E}., {Cadieux}, C., {Cook}, N.~J., {et~al.} 2022, \aj, 164, 84,
  \dodoi{10.3847/1538-3881/ac7ce6}

\bibitem[{{Asplund} {et~al.}(2021){Asplund}, {Amarsi}, \&
  {Grevesse}}]{Asplund_2021}
{Asplund}, M., {Amarsi}, A.~M., \& {Grevesse}, N. 2021, \aap, 653, A141,
  \dodoi{10.1051/0004-6361/202140445}

\bibitem[{{Astropy Collaboration} {et~al.}(2018){Astropy Collaboration},
  {Price-Whelan}, {Sip{\H{o}}cz}, {G{\"u}nther}, {Lim}, {Crawford}, {Conseil},
  {Shupe}, {Craig}, {Dencheva}, {Ginsburg}, {VanderPlas}, {Bradley},
  {P{\'e}rez-Su{\'a}rez}, {de Val-Borro}, {Aldcroft}, {Cruz}, {Robitaille},
  {Tollerud}, {Ardelean}, {Babej}, {Bach}, {Bachetti}, {Bakanov}, {Bamford},
  {Barentsen}, {Barmby}, {Baumbach}, {Berry}, {Biscani}, {Boquien}, {Bostroem},
  {Bouma}, {Brammer}, {Bray}, {Breytenbach}, {Buddelmeijer}, {Burke},
  {Calderone}, {Cano Rodr{\'\i}guez}, {Cara}, {Cardoso}, {Cheedella}, {Copin},
  {Corrales}, {Crichton}, {D'Avella}, {Deil}, {Depagne}, {Dietrich}, {Donath},
  {Droettboom}, {Earl}, {Erben}, {Fabbro}, {Ferreira}, {Finethy}, {Fox},
  {Garrison}, {Gibbons}, {Goldstein}, {Gommers}, {Greco}, {Greenfield},
  {Groener}, {Grollier}, {Hagen}, {Hirst}, {Homeier}, {Horton}, {Hosseinzadeh},
  {Hu}, {Hunkeler}, {Ivezi{\'c}}, {Jain}, {Jenness}, {Kanarek}, {Kendrew},
  {Kern}, {Kerzendorf}, {Khvalko}, {King}, {Kirkby}, {Kulkarni}, {Kumar},
  {Lee}, {Lenz}, {Littlefair}, {Ma}, {Macleod}, {Mastropietro}, {McCully},
  {Montagnac}, {Morris}, {Mueller}, {Mumford}, {Muna}, {Murphy}, {Nelson},
  {Nguyen}, {Ninan}, {N{\"o}the}, {Ogaz}, {Oh}, {Parejko}, {Parley}, {Pascual},
  {Patil}, {Patil}, {Plunkett}, {Prochaska}, {Rastogi}, {Reddy Janga},
  {Sabater}, {Sakurikar}, {Seifert}, {Sherbert}, {Sherwood-Taylor}, {Shih},
  {Sick}, {Silbiger}, {Singanamalla}, {Singer}, {Sladen}, {Sooley},
  {Sornarajah}, {Streicher}, {Teuben}, {Thomas}, {Tremblay}, {Turner},
  {Terr{\'o}n}, {van Kerkwijk}, {de la Vega}, {Watkins}, {Weaver}, {Whitmore},
  {Woillez}, {Zabalza}, \& {Astropy Contributors}}]{Astropy_2018}
{Astropy Collaboration}, {Price-Whelan}, A.~M., {Sip{\H{o}}cz}, B.~M., {et~al.}
  2018, \aj, 156, 123, \dodoi{10.3847/1538-3881/aabc4f}

\bibitem[{{Astudillo-Defru} {et~al.}(2017){Astudillo-Defru}, {D{\'\i}az},
  {Bonfils}, {Almenara}, {Delisle}, {Bouchy}, {Delfosse}, {Forveille}, {Lovis},
  {Mayor}, {Murgas}, {Pepe}, {Santos}, {S{\'e}gransan}, {Udry}, \&
  {W{\"u}nsche}}]{Astudillo-Defru_2017}
{Astudillo-Defru}, N., {D{\'\i}az}, R.~F., {Bonfils}, X., {et~al.} 2017, \aap,
  605, L11, \dodoi{10.1051/0004-6361/201731581}

\bibitem[{{Batalha} {et~al.}(2017){Batalha}, {Mandell}, {Pontoppidan},
  {Stevenson}, {Lewis}, {Kalirai}, {Earl}, {Greene}, {Albert}, \&
  {Nielsen}}]{Batalha_2017}
{Batalha}, N.~E., {Mandell}, A., {Pontoppidan}, K., {et~al.} 2017, \pasp, 129,
  064501, \dodoi{10.1088/1538-3873/aa65b0}

\bibitem[{{Bell} {et~al.}(2021){Bell}, {Dang}, {Cowan}, {Bean}, {D{\'e}sert},
  {Fortney}, {Keating}, {Kempton}, {Kreidberg}, {Line}, {Mansfield},
  {Parmentier}, {Stevenson}, {Swain}, \& {Zellem}}]{Bell2021}
{Bell}, T.~J., {Dang}, L., {Cowan}, N.~B., {et~al.} 2021, \mnras, 504, 3316,
  \dodoi{10.1093/mnras/stab1027}

\bibitem[{{Benedict} {et~al.}(2016){Benedict}, {Henry}, {Franz}, {McArthur},
  {Wasserman}, {Jao}, {Cargile}, {Dieterich}, {Bradley}, {Nelan}, \&
  {Whipple}}]{Benedict_2016}
{Benedict}, G.~F., {Henry}, T.~J., {Franz}, O.~G., {et~al.} 2016, \aj, 152,
  141, \dodoi{10.3847/0004-6256/152/5/141}

\bibitem[{{Benneke} \& {Seager}(2013)}]{Benneke_2013}
{Benneke}, B., \& {Seager}, S. 2013, \apj, 778, 153,
  \dodoi{10.1088/0004-637X/778/2/153}

\bibitem[{{Bensby} {et~al.}(2003){Bensby}, {Feltzing}, \&
  {Lundstr{\"o}m}}]{Bensby_2003}
{Bensby}, T., {Feltzing}, S., \& {Lundstr{\"o}m}, I. 2003, \aap, 410, 527,
  \dodoi{10.1051/0004-6361:20031213}

\bibitem[{Bertaux {et~al.}(2014)Bertaux, Lallement, Ferron, Boonne, \&
  Bodichon}]{Bertaux_2014}
Bertaux, J.~L., Lallement, R., Ferron, S., Boonne, C., \& Bodichon, R. 2014,
  Astronomy \& Astrophysics, 564, A46, \dodoi{10.1051/0004-6361/201322383}

\bibitem[{{Binney} \& {Tremaine}(2008)}]{Binney_2008}
{Binney}, J., \& {Tremaine}, S. 2008, {Galactic Dynamics: Second Edition}

\bibitem[{{Blanco-Cuaresma}(2019)}]{Blanco-Cuaresma_2019}
{Blanco-Cuaresma}, S. 2019, \mnras, 486, 2075, \dodoi{10.1093/mnras/stz549}

\bibitem[{{Bland-Hawthorn} \& {Gerhard}(2016)}]{Bland-Hawthorn_2016}
{Bland-Hawthorn}, J., \& {Gerhard}, O. 2016, \araa, 54, 529,
  \dodoi{10.1146/annurev-astro-081915-023441}

\bibitem[{{B{\"o}ker} {et~al.}(2023){B{\"o}ker}, {Beck}, {Birkmann},
  {Giardino}, {Keyes}, {Kumari}, {Muzerolle}, {Rawle}, {Zeidler}, {Abul-Huda},
  {Alves de Oliveira}, {Arribas}, {Bechtold}, {Bhatawdekar}, {Bonaventura},
  {Bunker}, {Cameron}, {Carniani}, {Charlot}, {Curti}, {Espinoza}, {Ferruit},
  {Franx}, {Jakobsen}, {Karakla}, {L{\'o}pez-Caniego}, {L{\"u}tzgendorf},
  {Maiolino}, {Manjavacas}, {Marston}, {Moseley}, {Ogle}, {Perna},
  {Pe{\~n}a-Guerrero}, {Pirzkal}, {Plesha}, {Proffitt}, {Rauscher}, {Rix},
  {Rodr{\'\i}guez del Pino}, {Rustamkulov}, {Sabbi}, {Sing}, {Sirianni}, {te
  Plate}, {{\'U}beda}, {Wahlgren}, {Wislowski}, {Wu}, \&
  {Willott}}]{Boker_2023}
{B{\"o}ker}, T., {Beck}, T.~L., {Birkmann}, S.~M., {et~al.} 2023, \pasp, 135,
  038001, \dodoi{10.1088/1538-3873/acb846}

\bibitem[{{Bond} {et~al.}(2010){Bond}, {O'Brien}, \& {Lauretta}}]{Bond_2010}
{Bond}, J.~C., {O'Brien}, D.~P., \& {Lauretta}, D.~S. 2010, \apj, 715, 1050,
  \dodoi{10.1088/0004-637X/715/2/1050}

\bibitem[{{Bonsor} {et~al.}(2021){Bonsor}, {Jofr{\'e}}, {Shorttle}, {Rogers},
  {Xu(许偲艺)}, \& {Melis}}]{Bonsor_2021}
{Bonsor}, A., {Jofr{\'e}}, P., {Shorttle}, O., {et~al.} 2021, \mnras, 503,
  1877, \dodoi{10.1093/mnras/stab370}

\bibitem[{{Bouchy} {et~al.}(2001){Bouchy}, {Pepe}, \& {Queloz}}]{Bouchy_2001}
{Bouchy}, F., {Pepe}, F., \& {Queloz}, D. 2001, \aap, 374, 733,
  \dodoi{10.1051/0004-6361:20010730}

\bibitem[{{Bouchy} {et~al.}(2017){Bouchy}, {Doyon}, {Artigau}, {Melo},
  {Hernandez}, {Wildi}, {Delfosse}, {Lovis}, {Figueira}, {Canto Martins},
  {Gonz{\'a}lez Hern{\'a}ndez}, {Thibault}, {Reshetov}, {Pepe}, {Santos}, {de
  Medeiros}, {Rebolo}, {Abreu}, {Adibekyan}, {Bandy}, {Benz}, {Blind},
  {Bohlender}, {Boisse}, {Bovay}, {Broeg}, {Brousseau}, {Cabral}, {Chazelas},
  {Cloutier}, {Coelho}, {Conod}, {Cumming}, {Delabre}, {Genolet}, {Hagelberg},
  {Jayawardhana}, {K{\"a}ufl}, {Lafreni{\`e}re}, {de Castro Le{\~a}o}, {Malo},
  {de Medeiros Martins}, {Matthews}, {Metchev}, {Oshagh}, {Ouellet}, {Parro},
  {Rasilla Pi{\~n}eiro}, {Santos}, {Sarajlic}, {Segovia}, {Sordet}, {Udry},
  {Valencia}, {Vall{\'e}e}, {Venn}, {Wade}, \& {Saddlemyer}}]{Bouchy_2017}
{Bouchy}, F., {Doyon}, R., {Artigau}, {\'E}., {et~al.} 2017, The Messenger,
  169, 21, \dodoi{10.18727/0722-6691/5034}

\bibitem[{{Boutle} {et~al.}(2017){Boutle}, {Mayne}, {Drummond}, {Manners},
  {Goyal}, {Hugo Lambert}, {Acreman}, \& {Earnshaw}}]{Boutle_2017}
{Boutle}, I.~A., {Mayne}, N.~J., {Drummond}, B., {et~al.} 2017, \aap, 601,
  A120, \dodoi{10.1051/0004-6361/201630020}

\bibitem[{{Cadieux} {et~al.}(2022){Cadieux}, {Doyon}, {Plotnykov},
  {H{\'e}brard}, {Jahandar}, {Artigau}, {Valencia}, {Cook}, {Martioli},
  {Vandal}, {Donati}, {Cloutier}, {Narita}, {Fukui}, {Hirano}, {Bouchy},
  {Cowan}, {Gonzales}, {Ciardi}, {Stassun}, {Arnold}, {Benneke}, {Boisse},
  {Bonfils}, {Carmona}, {Cort{\'e}s-Zuleta}, {Delfosse}, {Forveille},
  {Fouqu{\'e}}, {Gomes da Silva}, {Jenkins}, {Kiefer}, {K{\'o}sp{\'a}l},
  {Lafreni{\`e}re}, {Martins}, {Moutou}, {do Nascimento}, {Ould-Elhkim},
  {Pelletier}, {Twicken}, {Bouma}, {Cartwright}, {Darveau-Bernier}, {Grankin},
  {Ikoma}, {Kagetani}, {Kawauchi}, {Kodama}, {Kotani}, {Latham}, {Menou},
  {Ricker}, {Seager}, {Tamura}, {Vanderspek}, \& {Watanabe}}]{Cadieux_2022}
{Cadieux}, C., {Doyon}, R., {Plotnykov}, M., {et~al.} 2022, \aj, 164, 96,
  \dodoi{10.3847/1538-3881/ac7cea}

\bibitem[{{Carter} {et~al.}(2015){Carter}, {Leinhardt}, {Elliott}, {Walter}, \&
  {Stewart}}]{Carter2015}
{Carter}, P.~J., {Leinhardt}, Z.~M., {Elliott}, T., {Walter}, M.~J., \&
  {Stewart}, S.~T. 2015, \apj, 813, 72, \dodoi{10.1088/0004-637X/813/1/72}

\bibitem[{{Charnay} {et~al.}(2021){Charnay}, {Blain}, {B{\'e}zard}, {Leconte},
  {Turbet}, \& {Falco}}]{Charnay_2021}
{Charnay}, B., {Blain}, D., {B{\'e}zard}, B., {et~al.} 2021, \aap, 646, A171,
  \dodoi{10.1051/0004-6361/202039525}

\bibitem[{{Charnay} {et~al.}(2015){Charnay}, {Meadows}, {Misra}, {Leconte}, \&
  {Arney}}]{Charnay_2015}
{Charnay}, B., {Meadows}, V., {Misra}, A., {Leconte}, J., \& {Arney}, G. 2015,
  \apjl, 813, L1, \dodoi{10.1088/2041-8205/813/1/L1}

\bibitem[{{Cherubim} {et~al.}(2023){Cherubim}, {Cloutier}, {Charbonneau},
  {Stockdale}, {Stassun}, {Schwarz}, {Safonov}, {Mortier}, {Lewin}, {Latham},
  {Horne}, {Haywood}, {Gonzales}, {Goliguzova}, {Collins}, {Ciardi}, {Bieryla},
  {Belinski}, {Wohler}, {Watson}, {Vanderspek}, {Udry}, {Sozzetti},
  {S{\'e}gransan}, {Sasselov}, {Ricker}, {Rice}, {Poretti}, {Piotto}, {Pepe},
  {Molinari}, {Micela}, {Mayor}, {Lovis}, {L{\'o}pez-Morales}, {Jenkins},
  {Essack}, {Dumusque}, {Doty}, {Col{\'o}n}, {Cameron}, \&
  {Buchhave}}]{Cherubim_2023}
{Cherubim}, C., {Cloutier}, R., {Charbonneau}, D., {et~al.} 2023, \aj, 165,
  167, \dodoi{10.3847/1538-3881/acbdfd}

\bibitem[{{Cloutier} {et~al.}(2018){Cloutier}, {Doyon}, {Bouchy}, \&
  {H{\'e}brard}}]{Cloutier_2018}
{Cloutier}, R., {Doyon}, R., {Bouchy}, F., \& {H{\'e}brard}, G. 2018, \aj, 156,
  82, \dodoi{10.3847/1538-3881/aacea9}

\bibitem[{{Cloutier} \& {Menou}(2020)}]{Cloutier-Menou_2020}
{Cloutier}, R., \& {Menou}, K. 2020, \aj, 159, 211,
  \dodoi{10.3847/1538-3881/ab8237}

\bibitem[{{Cook} {et~al.}(2022){Cook}, {Artigau}, {Doyon}, {Hobson},
  {Martioli}, {Bouchy}, {Moutou}, {Carmona}, {Usher}, {Fouqu{\'e}}, {Arnold},
  {Delfosse}, {Boisse}, {Cadieux}, {Vandal}, {Donati}, \&
  {Desli{\`e}res}}]{Cook_2022}
{Cook}, N.~J., {Artigau}, {\'E}., {Doyon}, R., {et~al.} 2022, \pasp, 134,
  114509, \dodoi{10.1088/1538-3873/ac9e74}

\bibitem[{{Coulombe} {et~al.}(2023){Coulombe}, {Benneke}, {Challener},
  {Piette}, {Wiser}, {Mansfield}, {MacDonald}, {Beltz}, {Feinstein}, {Radica},
  {Savel}, {Dos Santos}, {Bean}, {Parmentier}, {Wong}, {Rauscher}, {Komacek},
  {Kempton}, {Tan}, {Hammond}, {Lewis}, {Line}, {Lee}, {Shivkumar},
  {Crossfield}, {Nixon}, {Rackham}, {Wakeford}, {Welbanks}, {Zhang}, {Batalha},
  {Berta-Thompson}, {Changeat}, {D{\'e}sert}, {Espinoza}, {Goyal},
  {Harrington}, {Knutson}, {Kreidberg}, {L{\'o}pez-Morales}, {Shporer}, {Sing},
  {Stevenson}, {Aggarwal}, {Ahrer}, {Alam}, {Bell}, {Blecic}, {Caceres},
  {Carter}, {Casewell}, {Crouzet}, {Cubillos}, {Decin}, {Fortney}, {Gibson},
  {Heng}, {Henning}, {Iro}, {Kendrew}, {Lagage}, {Leconte}, {Lendl},
  {Lothringer}, {Mancini}, {Mikal-Evans}, {Molaverdikhani}, {Nikolov}, {Ohno},
  {Palle}, {Piaulet}, {Redfield}, {Roy}, {Tsai}, {Venot}, \&
  {Wheatley}}]{Coulombe_2023}
{Coulombe}, L.-P., {Benneke}, B., {Challener}, R., {et~al.} 2023, \nat, 620,
  292, \dodoi{10.1038/s41586-023-06230-1}

\bibitem[{{Dang} {et~al.}(2018){Dang}, {Cowan}, {Schwartz}, {Rauscher},
  {Zhang}, {Knutson}, {Line}, {Dobbs-Dixon}, {Deming}, {Sundararajan},
  {Fortney}, \& {Zhao}}]{Dang2018}
{Dang}, L., {Cowan}, N.~B., {Schwartz}, J.~C., {et~al.} 2018, Nature Astronomy,
  2, 220, \dodoi{10.1038/s41550-017-0351-6}

\bibitem[{{Dawson} \& {Johnson}(2012)}]{Dawson_2012}
{Dawson}, R.~I., \& {Johnson}, J.~A. 2012, \apj, 756, 122,
  \dodoi{10.1088/0004-637X/756/2/122}

\bibitem[{{Del Genio} {et~al.}(2018){Del Genio}, {Brain}, {Noack}, \&
  {Schaefer}}]{DelGenio_2018}
{Del Genio}, A.~D., {Brain}, D., {Noack}, L., \& {Schaefer}, L. 2018, arXiv
  e-prints, arXiv:1807.04776.
\newblock \doarXiv{1807.04776}

\bibitem[{{Del Genio} {et~al.}(2019){Del Genio}, {Way}, {Amundsen}, {Aleinov},
  {Kelley}, {Kiang}, \& {Clune}}]{DelGenio_2019}
{Del Genio}, A.~D., {Way}, M.~J., {Amundsen}, D.~S., {et~al.} 2019,
  Astrobiology, 19, 99, \dodoi{10.1089/ast.2017.1760}

\bibitem[{{Delmotte} {et~al.}(2006){Delmotte}, {Dolensky}, {Padovani},
  {Retzlaff}, {Rit{\'e}}, {Rosati}, {Slijkhuis}, {Wicenec}, {Fernique}, \&
  {Micol}}]{Delmotte_2006}
{Delmotte}, N., {Dolensky}, M., {Padovani}, P., {et~al.} 2006, in Astronomical
  Society of the Pacific Conference Series, Vol. 351, Astronomical Data
  Analysis Software and Systems XV, ed. C.~{Gabriel}, C.~{Arviset}, D.~{Ponz},
  \& S.~{Enrique}, 690

\bibitem[{{Deming} {et~al.}(2015){Deming}, {Knutson}, {Kammer}, {Fulton},
  {Ingalls}, {Carey}, {Burrows}, {Fortney}, {Todorov}, {Agol}, {Cowan},
  {Desert}, {Fraine}, {Langton}, {Morley}, \& {Showman}}]{Deming2015}
{Deming}, D., {Knutson}, H., {Kammer}, J., {et~al.} 2015, \apj, 805, 132,
  \dodoi{10.1088/0004-637X/805/2/132}

\bibitem[{{Diamond-Lowe} {et~al.}(2020){Diamond-Lowe}, {Berta-Thompson},
  {Charbonneau}, {Dittmann}, \& {Kempton}}]{Diamond-Lowe_2020}
{Diamond-Lowe}, H., {Berta-Thompson}, Z., {Charbonneau}, D., {Dittmann}, J., \&
  {Kempton}, E. M.~R. 2020, \aj, 160, 27, \dodoi{10.3847/1538-3881/ab935f}

\bibitem[{{Dittmann} {et~al.}(2017){Dittmann}, {Irwin}, {Charbonneau},
  {Bonfils}, {Astudillo-Defru}, {Haywood}, {Berta-Thompson}, {Newton},
  {Rodriguez}, {Winters}, {Tan}, {Almenara}, {Bouchy}, {Delfosse}, {Forveille},
  {Lovis}, {Murgas}, {Pepe}, {Santos}, {Udry}, {W{\"u}nsche}, {Esquerdo},
  {Latham}, \& {Dressing}}]{Dittmann_2017}
{Dittmann}, J.~A., {Irwin}, J.~M., {Charbonneau}, D., {et~al.} 2017, \nat, 544,
  333, \dodoi{10.1038/nature22055}

\bibitem[{{Donati} {et~al.}(2020){Donati}, {Kouach}, {Moutou}, {Doyon},
  {Delfosse}, {Artigau}, {Baratchart}, {Lacombe}, {Barrick}, {H{\'e}brard},
  {Bouchy}, {Saddlemyer}, {Par{\`e}s}, {Rabou}, {Micheau}, {Dolon}, {Reshetov},
  {Challita}, {Carmona}, {Striebig}, {Thibault}, {Martioli}, {Cook},
  {Fouqu{\'e}}, {Vermeulen}, {Wang}, {Arnold}, {Pepe}, {Boisse}, {Figueira},
  {Bouvier}, {Ray}, {Feugeade}, {Morin}, {Alencar}, {Hobson}, {Castilho},
  {Udry}, {Santos}, {Hernandez}, {Benedict}, {Vall{\'e}e}, {Gallou}, {Dupieux},
  {Larrieu}, {Perruchot}, {Sottile}, {Moreau}, {Usher}, {Baril}, {Wildi},
  {Chazelas}, {Malo}, {Bonfils}, {Loop}, {Kerley}, {Wevers}, {Dunn}, {Pazder},
  {Macdonald}, {Dubois}, {Carri{\'e}}, {Valentin}, {Henault}, {Yan}, \&
  {Steinmetz}}]{Donati_2020}
{Donati}, J.~F., {Kouach}, D., {Moutou}, C., {et~al.} 2020, \mnras, 498, 5684,
  \dodoi{10.1093/mnras/staa2569}

\bibitem[{{Dorn} {et~al.}(2017){Dorn}, {Hinkel}, \& {Venturini}}]{Dorn2017}
{Dorn}, C., {Hinkel}, N.~R., \& {Venturini}, J. 2017, \aap, 597, A38,
  \dodoi{10.1051/0004-6361/201628749}

\bibitem[{Doyon {et~al.}(2023)Doyon, Willott, Hutchings, Sivaramakrishnan,
  Albert, Lafreniere, Rowlands, Vila, Martel, LaMassa, Aldridge, Artigau,
  Cameron, Chayer, Cook, Cooper, Darveau-Bernier, Dupuis, Earnshaw, Espinoza,
  Filippazzo, Fullerton, Gaudreau, Gawlik, Goudfrooij, Haley, Kammerer,
  Kendall, Lambros, Ignat, Maszkiewicz, McColgan, Morishita, Ouellette,
  Pacifici, Philippi, Radica, Ravindranath, Rowe, Roy, Saad, Sohn, Talens,
  Thatte, Taylor, Vandal, Volk, Wander, Warner, Zheng, Zhou, Abraham, Beaulieu,
  Benneke, Ferrarese, Johnstone, Kaltenegger, Meyer, Pipher, Rameau, Rieke,
  Salhi, \& Sawicki}]{Doyon_2023}
Doyon, R., Willott, C.~J., Hutchings, J.~B., {et~al.} 2023, \pasp, 135, 098001,
  \dodoi{10.1088/1538-3873/acd41b}

\bibitem[{{Dumusque}(2018)}]{Dumusque_2018}
{Dumusque}, X. 2018, \aap, 620, A47, \dodoi{10.1051/0004-6361/201833795}

\bibitem[{{Edwards} {et~al.}(2021){Edwards}, {Changeat}, {Mori}, {Anisman},
  {Morvan}, {Yip}, {Tsiaras}, {Al-Refaie}, {Waldmann}, \&
  {Tinetti}}]{Edwards_2021}
{Edwards}, B., {Changeat}, Q., {Mori}, M., {et~al.} 2021, \aj, 161, 44,
  \dodoi{10.3847/1538-3881/abc6a5}

\bibitem[{{Espinoza}(2018)}]{Espinoza_2018}
{Espinoza}, N. 2018, Research Notes of the American Astronomical Society, 2,
  209, \dodoi{10.3847/2515-5172/aaef38}

\bibitem[{{Espinoza} {et~al.}(2019){Espinoza}, {Kossakowski}, \&
  {Brahm}}]{Espinoza_2019}
{Espinoza}, N., {Kossakowski}, D., \& {Brahm}, R. 2019, \mnras, 490, 2262,
  \dodoi{10.1093/mnras/stz2688}

\bibitem[{{Faria} {et~al.}(2022){Faria}, {Su{\'a}rez Mascare{\~n}o},
  {Figueira}, {Silva}, {Damasso}, {Demangeon}, {Pepe}, {Santos}, {Rebolo},
  {Cristiani}, {Adibekyan}, {Alibert}, {Allart}, {Barros}, {Cabral},
  {D'Odorico}, {Di Marcantonio}, {Dumusque}, {Ehrenreich}, {Gonz{\'a}lez
  Hern{\'a}ndez}, {Hara}, {Lillo-Box}, {Lo Curto}, {Lovis}, {Martins},
  {M{\'e}gevand}, {Mehner}, {Micela}, {Molaro}, {Nunes}, {Pall{\'e}},
  {Poretti}, {Sousa}, {Sozzetti}, {Tabernero}, {Udry}, \& {Zapatero
  Osorio}}]{Faria_2022}
{Faria}, J.~P., {Su{\'a}rez Mascare{\~n}o}, A., {Figueira}, P., {et~al.} 2022,
  \aap, 658, A115, \dodoi{10.1051/0004-6361/202142337}

\bibitem[{{Fauchez} {et~al.}(2019){Fauchez}, {Turbet}, {Villanueva}, {Wolf},
  {Arney}, {Kopparapu}, {Lincowski}, {Mandell}, {de Wit}, {Pidhorodetska},
  {Domagal-Goldman}, \& {Stevenson}}]{Fauchez_2019}
{Fauchez}, T.~J., {Turbet}, M., {Villanueva}, G.~L., {et~al.} 2019, \apj, 887,
  194, \dodoi{10.3847/1538-4357/ab5862}

\bibitem[{{Fazio} {et~al.}(2004){Fazio}, {Hora}, {Allen}, {Ashby}, {Barmby},
  {Deutsch}, {Huang}, {Kleiner}, {Marengo}, {Megeath}, {Melnick}, {Pahre},
  {Patten}, {Polizotti}, {Smith}, {Taylor}, {Wang}, {Willner}, {Hoffmann},
  {Pipher}, {Forrest}, {McMurty}, {McCreight}, {McKelvey}, {McMurray}, {Koch},
  {Moseley}, {Arendt}, {Mentzell}, {Marx}, {Losch}, {Mayman}, {Eichhorn},
  {Krebs}, {Jhabvala}, {Gezari}, {Fixsen}, {Flores}, {Shakoorzadeh}, {Jungo},
  {Hakun}, {Workman}, {Karpati}, {Kichak}, {Whitley}, {Mann}, {Tollestrup},
  {Eisenhardt}, {Stern}, {Gorjian}, {Bhattacharya}, {Carey}, {Nelson},
  {Glaccum}, {Lacy}, {Lowrance}, {Laine}, {Reach}, {Stauffer}, {Surace},
  {Wilson}, {Wright}, {Hoffman}, {Domingo}, \& {Cohen}}]{Fazio_2004}
{Fazio}, G.~G., {Hora}, J.~L., {Allen}, L.~E., {et~al.} 2004, \apjs, 154, 10,
  \dodoi{10.1086/422843}

\bibitem[{{Foreman-Mackey}(2016)}]{Foreman-Mackey_2016}
{Foreman-Mackey}, D. 2016, The Journal of Open Source Software, 1, 24,
  \dodoi{10.21105/joss.00024}

\bibitem[{{Foreman-Mackey} {et~al.}(2013){Foreman-Mackey}, {Hogg}, {Lang}, \&
  {Goodman}}]{Foreman-Mackey_2013}
{Foreman-Mackey}, D., {Hogg}, D.~W., {Lang}, D., \& {Goodman}, J. 2013, \pasp,
  125, 306, \dodoi{10.1086/670067}

\bibitem[{{Forget} \& {Leconte}(2014)}]{Forget_2014}
{Forget}, F., \& {Leconte}, J. 2014, Philosophical Transactions of the Royal
  Society of London Series A, 372, 20130084, \dodoi{10.1098/rsta.2013.0084}

\bibitem[{{Fulton} {et~al.}(2018){Fulton}, {Petigura}, {Blunt}, \&
  {Sinukoff}}]{Fulton_2018}
{Fulton}, B.~J., {Petigura}, E.~A., {Blunt}, S., \& {Sinukoff}, E. 2018, \pasp,
  130, 044504, \dodoi{10.1088/1538-3873/aaaaa8}

\bibitem[{{Fulton} {et~al.}(2017){Fulton}, {Petigura}, {Howard}, {Isaacson},
  {Marcy}, {Cargile}, {Hebb}, {Weiss}, {Johnson}, {Morton}, {Sinukoff},
  {Crossfield}, \& {Hirsch}}]{Fulton_2017}
{Fulton}, B.~J., {Petigura}, E.~A., {Howard}, A.~W., {et~al.} 2017, \aj, 154,
  109, \dodoi{10.3847/1538-3881/aa80eb}

\bibitem[{{Gaia Collaboration} {et~al.}(2021){Gaia Collaboration}, {Brown},
  {Vallenari}, {Prusti}, {de Bruijne}, {Babusiaux}, {Biermann}, {Creevey},
  {Evans}, {Eyer}, {Hutton}, {Jansen}, {Jordi}, {Klioner}, {Lammers},
  {Lindegren}, {Luri}, {Mignard}, {Panem}, {Pourbaix}, {Randich}, {Sartoretti},
  {Soubiran}, {Walton}, {Arenou}, {Bailer-Jones}, {Bastian}, {Cropper},
  {Drimmel}, {Katz}, {Lattanzi}, {van Leeuwen}, {Bakker}, {Cacciari},
  {Casta{\~n}eda}, {De Angeli}, {Ducourant}, {Fabricius}, {Fouesneau},
  {Fr{\'e}mat}, {Guerra}, {Guerrier}, {Guiraud}, {Jean-Antoine Piccolo},
  {Masana}, {Messineo}, {Mowlavi}, {Nicolas}, {Nienartowicz}, {Pailler},
  {Panuzzo}, {Riclet}, {Roux}, {Seabroke}, {Sordo}, {Tanga}, {Th{\'e}venin},
  {Gracia-Abril}, {Portell}, {Teyssier}, {Altmann}, {Andrae}, {Bellas-Velidis},
  {Benson}, {Berthier}, {Blomme}, {Brugaletta}, {Burgess}, {Busso}, {Carry},
  {Cellino}, {Cheek}, {Clementini}, {Damerdji}, {Davidson}, {Delchambre},
  {Dell'Oro}, {Fern{\'a}ndez-Hern{\'a}ndez}, {Galluccio}, {Garc{\'\i}a-Lario},
  {Garcia-Reinaldos}, {Gonz{\'a}lez-N{\'u}{\~n}ez}, {Gosset}, {Haigron},
  {Halbwachs}, {Hambly}, {Harrison}, {Hatzidimitriou}, {Heiter},
  {Hern{\'a}ndez}, {Hestroffer}, {Hodgkin}, {Holl}, {Jan{\ss}en}, {Jevardat de
  Fombelle}, {Jordan}, {Krone-Martins}, {Lanzafame}, {L{\"o}ffler}, {Lorca},
  {Manteiga}, {Marchal}, {Marrese}, {Moitinho}, {Mora}, {Muinonen}, {Osborne},
  {Pancino}, {Pauwels}, {Petit}, {Recio-Blanco}, {Richards}, {Riello},
  {Rimoldini}, {Robin}, {Roegiers}, {Rybizki}, {Sarro}, {Siopis}, {Smith},
  {Sozzetti}, {Ulla}, {Utrilla}, {van Leeuwen}, {van Reeven}, {Abbas}, {Abreu
  Aramburu}, {Accart}, {Aerts}, {Aguado}, {Ajaj}, {Altavilla}, {{\'A}lvarez},
  {{\'A}lvarez Cid-Fuentes}, {Alves}, {Anderson}, {Anglada Varela}, {Antoja},
  {Audard}, {Baines}, {Baker}, {Balaguer-N{\'u}{\~n}ez}, {Balbinot}, {Balog},
  {Barache}, {Barbato}, {Barros}, {Barstow}, {Bartolom{\'e}}, {Bassilana},
  {Bauchet}, {Baudesson-Stella}, {Becciani}, {Bellazzini}, {Bernet}, {Bertone},
  {Bianchi}, {Blanco-Cuaresma}, {Boch}, {Bombrun}, {Bossini}, {Bouquillon},
  {Bragaglia}, {Bramante}, {Breedt}, {Bressan}, {Brouillet}, {Bucciarelli},
  {Burlacu}, {Busonero}, {Butkevich}, {Buzzi}, {Caffau}, {Cancelliere},
  {C{\'a}novas}, {Cantat-Gaudin}, {Carballo}, {Carlucci}, {Carnerero},
  {Carrasco}, {Casamiquela}, {Castellani}, {Castro-Ginard}, {Castro Sampol},
  {Chaoul}, {Charlot}, {Chemin}, {Chiavassa}, {Cioni}, {Comoretto}, {Cooper},
  {Cornez}, {Cowell}, {Crifo}, {Crosta}, {Crowley}, {Dafonte}, {Dapergolas},
  {David}, {David}, {de Laverny}, {De Luise}, {De March}, {De Ridder}, {de
  Souza}, {de Teodoro}, {de Torres}, {del Peloso}, {del Pozo}, {Delbo},
  {Delgado}, {Delgado}, {Delisle}, {Di Matteo}, {Diakite}, {Diener},
  {Distefano}, {Dolding}, {Eappachen}, {Edvardsson}, {Enke}, {Esquej}, {Fabre},
  {Fabrizio}, {Faigler}, {Fedorets}, {Fernique}, {Fienga}, {Figueras},
  {Fouron}, {Fragkoudi}, {Fraile}, {Franke}, {Gai}, {Garabato},
  {Garcia-Gutierrez}, {Garc{\'\i}a-Torres}, {Garofalo}, {Gavras}, {Gerlach},
  {Geyer}, {Giacobbe}, {Gilmore}, {Girona}, {Giuffrida}, {Gomel}, {Gomez},
  {Gonzalez-Santamaria}, {Gonz{\'a}lez-Vidal}, {Granvik},
  {Guti{\'e}rrez-S{\'a}nchez}, {Guy}, {Hauser}, {Haywood}, {Helmi}, {Hidalgo},
  {Hilger}, {H{\l}adczuk}, {Hobbs}, {Holland}, {Huckle}, {Jasniewicz},
  {Jonker}, {Juaristi Campillo}, {Julbe}, {Karbevska}, {Kervella}, {Khanna},
  {Kochoska}, {Kontizas}, {Kordopatis}, {Korn}, {Kostrzewa-Rutkowska},
  {Kruszy{\'n}ska}, {Lambert}, {Lanza}, {Lasne}, {Le Campion}, {Le Fustec},
  {Lebreton}, {Lebzelter}, {Leccia}, {Leclerc}, {Lecoeur-Taibi}, {Liao},
  {Licata}, {Lindstr{\o}m}, {Lister}, {Livanou}, {Lobel}, {Madrero Pardo},
  {Managau}, {Mann}, {Marchant}, {Marconi}, {Marcos Santos}, {Marinoni},
  {Marocco}, {Marshall}, {Martin Polo}, {Mart{\'\i}n-Fleitas}, {Masip},
  {Massari}, {Mastrobuono-Battisti}, {Mazeh}, {McMillan}, {Messina},
  {Michalik}, {Millar}, {Mints}, {Molina}, {Molinaro}, {Moln{\'a}r},
  {Montegriffo}, {Mor}, {Morbidelli}, {Morel}, {Morris}, {Mulone}, {Munoz},
  {Muraveva}, {Murphy}, {Musella}, {Noval}, {Ord{\'e}novic}, {Orr{\`u}},
  {Osinde}, {Pagani}, {Pagano}, {Palaversa}, {Palicio}, {Panahi}, {Pawlak},
  {Pe{\~n}alosa Esteller}, {Penttil{\"a}}, {Piersimoni}, {Pineau}, {Plachy},
  {Plum}, {Poggio}, {Poretti}, {Poujoulet}, {Pr{\v{s}}a}, {Pulone}, {Racero},
  {Ragaini}, {Rainer}, {Raiteri}, {Rambaux}, {Ramos}, {Ramos-Lerate}, {Re
  Fiorentin}, {Regibo}, {Reyl{\'e}}, {Ripepi}, {Riva}, {Rixon}, {Robichon},
  {Robin}, {Roelens}, {Rohrbasser}, {Romero-G{\'o}mez}, {Rowell}, {Royer},
  {Rybicki}, {Sadowski}, {Sagrist{\`a} Sell{\'e}s}, {Sahlmann}, {Salgado},
  {Salguero}, {Samaras}, {Sanchez Gimenez}, {Sanna}, {Santove{\~n}a},
  {Sarasso}, {Schultheis}, {Sciacca}, {Segol}, {Segovia}, {S{\'e}gransan},
  {Semeux}, {Shahaf}, {Siddiqui}, {Siebert}, {Siltala}, {Slezak}, {Smart},
  {Solano}, {Solitro}, {Souami}, {Souchay}, {Spagna}, {Spoto}, {Steele},
  {Steidelm{\"u}ller}, {Stephenson}, {S{\"u}veges}, {Szabados}, {Szegedi-Elek},
  {Taris}, {Tauran}, {Taylor}, {Teixeira}, {Thuillot}, {Tonello}, {Torra},
  {Torra}, {Turon}, {Unger}, {Vaillant}, {van Dillen}, {Vanel}, {Vecchiato},
  {Viala}, {Vicente}, {Voutsinas}, {Weiler}, {Wevers}, {Wyrzykowski}, {Yoldas},
  {Yvard}, {Zhao}, {Zorec}, {Zucker}, {Zurbach}, \&
  {Zwitter}}]{Gaia_Collaboration_2021}
{Gaia Collaboration}, {Brown}, A.~G.~A., {Vallenari}, A., {et~al.} 2021, \aap,
  649, A1, \dodoi{10.1051/0004-6361/202039657}

\bibitem[{{Gaia Collaboration} {et~al.}(2023){Gaia Collaboration}, {Vallenari},
  {Brown}, {Prusti}, {de Bruijne}, {Arenou}, {Babusiaux}, {Biermann},
  {Creevey}, {Ducourant}, {Evans}, {Eyer}, {Guerra}, {Hutton}, {Jordi},
  {Klioner}, {Lammers}, {Lindegren}, {Luri}, {Mignard}, {Panem}, {Pourbaix},
  {Randich}, {Sartoretti}, {Soubiran}, {Tanga}, {Walton}, {Bailer-Jones},
  {Bastian}, {Drimmel}, {Jansen}, {Katz}, {Lattanzi}, {van Leeuwen}, {Bakker},
  {Cacciari}, {Casta{\~n}eda}, {De Angeli}, {Fabricius}, {Fouesneau},
  {Fr{\'e}mat}, {Galluccio}, {Guerrier}, {Heiter}, {Masana}, {Messineo},
  {Mowlavi}, {Nicolas}, {Nienartowicz}, {Pailler}, {Panuzzo}, {Riclet}, {Roux},
  {Seabroke}, {Sordo}, {Th{\'e}venin}, {Gracia-Abril}, {Portell}, {Teyssier},
  {Altmann}, {Andrae}, {Audard}, {Bellas-Velidis}, {Benson}, {Berthier},
  {Blomme}, {Burgess}, {Busonero}, {Busso}, {C{\'a}novas}, {Carry}, {Cellino},
  {Cheek}, {Clementini}, {Damerdji}, {Davidson}, {de Teodoro}, {Nu{\~n}ez
  Campos}, {Delchambre}, {Dell'Oro}, {Esquej}, {Fern{\'a}ndez-Hern{\'a}ndez},
  {Fraile}, {Garabato}, {Garc{\'\i}a-Lario}, {Gosset}, {Haigron}, {Halbwachs},
  {Hambly}, {Harrison}, {Hern{\'a}ndez}, {Hestroffer}, {Hodgkin}, {Holl},
  {Jan{\ss}en}, {Jevardat de Fombelle}, {Jordan}, {Krone-Martins}, {Lanzafame},
  {L{\"o}ffler}, {Marchal}, {Marrese}, {Moitinho}, {Muinonen}, {Osborne},
  {Pancino}, {Pauwels}, {Recio-Blanco}, {Reyl{\'e}}, {Riello}, {Rimoldini},
  {Roegiers}, {Rybizki}, {Sarro}, {Siopis}, {Smith}, {Sozzetti}, {Utrilla},
  {van Leeuwen}, {Abbas}, {{\'A}brah{\'a}m}, {Abreu Aramburu}, {Aerts},
  {Aguado}, {Ajaj}, {Aldea-Montero}, {Altavilla}, {{\'A}lvarez}, {Alves},
  {Anders}, {Anderson}, {Anglada Varela}, {Antoja}, {Baines}, {Baker},
  {Balaguer-N{\'u}{\~n}ez}, {Balbinot}, {Balog}, {Barache}, {Barbato},
  {Barros}, {Barstow}, {Bartolom{\'e}}, {Bassilana}, {Bauchet}, {Becciani},
  {Bellazzini}, {Berihuete}, {Bernet}, {Bertone}, {Bianchi}, {Binnenfeld},
  {Blanco-Cuaresma}, {Blazere}, {Boch}, {Bombrun}, {Bossini}, {Bouquillon},
  {Bragaglia}, {Bramante}, {Breedt}, {Bressan}, {Brouillet}, {Brugaletta},
  {Bucciarelli}, {Burlacu}, {Butkevich}, {Buzzi}, {Caffau}, {Cancelliere},
  {Cantat-Gaudin}, {Carballo}, {Carlucci}, {Carnerero}, {Carrasco},
  {Casamiquela}, {Castellani}, {Castro-Ginard}, {Chaoul}, {Charlot}, {Chemin},
  {Chiaramida}, {Chiavassa}, {Chornay}, {Comoretto}, {Contursi}, {Cooper},
  {Cornez}, {Cowell}, {Crifo}, {Cropper}, {Crosta}, {Crowley}, {Dafonte},
  {Dapergolas}, {David}, {David}, {de Laverny}, {De Luise}, {De March}, {De
  Ridder}, {de Souza}, {de Torres}, {del Peloso}, {del Pozo}, {Delbo},
  {Delgado}, {Delisle}, {Demouchy}, {Dharmawardena}, {Di Matteo}, {Diakite},
  {Diener}, {Distefano}, {Dolding}, {Edvardsson}, {Enke}, {Fabre}, {Fabrizio},
  {Faigler}, {Fedorets}, {Fernique}, {Fienga}, {Figueras}, {Fournier},
  {Fouron}, {Fragkoudi}, {Gai}, {Garcia-Gutierrez}, {Garcia-Reinaldos},
  {Garc{\'\i}a-Torres}, {Garofalo}, {Gavel}, {Gavras}, {Gerlach}, {Geyer},
  {Giacobbe}, {Gilmore}, {Girona}, {Giuffrida}, {Gomel}, {Gomez},
  {Gonz{\'a}lez-N{\'u}{\~n}ez}, {Gonz{\'a}lez-Santamar{\'\i}a},
  {Gonz{\'a}lez-Vidal}, {Granvik}, {Guillout}, {Guiraud},
  {Guti{\'e}rrez-S{\'a}nchez}, {Guy}, {Hatzidimitriou}, {Hauser}, {Haywood},
  {Helmer}, {Helmi}, {Sarmiento}, {Hidalgo}, {Hilger}, {H{\l}adczuk}, {Hobbs},
  {Holland}, {Huckle}, {Jardine}, {Jasniewicz}, {Jean-Antoine Piccolo},
  {Jim{\'e}nez-Arranz}, {Jorissen}, {Juaristi Campillo}, {Julbe}, {Karbevska},
  {Kervella}, {Khanna}, {Kontizas}, {Kordopatis}, {Korn}, {K{\'o}sp{\'a}l},
  {Kostrzewa-Rutkowska}, {Kruszy{\'n}ska}, {Kun}, {Laizeau}, {Lambert},
  {Lanza}, {Lasne}, {Le Campion}, {Lebreton}, {Lebzelter}, {Leccia}, {Leclerc},
  {Lecoeur-Taibi}, {Liao}, {Licata}, {Lindstr{\o}m}, {Lister}, {Livanou},
  {Lobel}, {Lorca}, {Loup}, {Madrero Pardo}, {Magdaleno Romeo}, {Managau},
  {Mann}, {Manteiga}, {Marchant}, {Marconi}, {Marcos}, {Marcos Santos},
  {Mar{\'\i}n Pina}, {Marinoni}, {Marocco}, {Marshall}, {Martin Polo},
  {Mart{\'\i}n-Fleitas}, {Marton}, {Mary}, {Masip}, {Massari},
  {Mastrobuono-Battisti}, {Mazeh}, {McMillan}, {Messina}, {Michalik}, {Millar},
  {Mints}, {Molina}, {Molinaro}, {Moln{\'a}r}, {Monari}, {Mongui{\'o}},
  {Montegriffo}, {Montero}, {Mor}, {Mora}, {Morbidelli}, {Morel}, {Morris},
  {Muraveva}, {Murphy}, {Musella}, {Nagy}, {Noval}, {Oca{\~n}a}, {Ogden},
  {Ordenovic}, {Osinde}, {Pagani}, {Pagano}, {Palaversa}, {Palicio},
  {Pallas-Quintela}, {Panahi}, {Payne-Wardenaar}, {Pe{\~n}alosa Esteller},
  {Penttil{\"a}}, {Pichon}, {Piersimoni}, {Pineau}, {Plachy}, {Plum}, {Poggio},
  {Pr{\v{s}}a}, {Pulone}, {Racero}, {Ragaini}, {Rainer}, {Raiteri}, {Rambaux},
  {Ramos}, {Ramos-Lerate}, {Re Fiorentin}, {Regibo}, {Richards}, {Rios Diaz},
  {Ripepi}, {Riva}, {Rix}, {Rixon}, {Robichon}, {Robin}, {Robin}, {Roelens},
  {Rogues}, {Rohrbasser}, {Romero-G{\'o}mez}, {Rowell}, {Royer}, {Ruz Mieres},
  {Rybicki}, {Sadowski}, {S{\'a}ez N{\'u}{\~n}ez}, {Sagrist{\`a} Sell{\'e}s},
  {Sahlmann}, {Salguero}, {Samaras}, {Sanchez Gimenez}, {Sanna},
  {Santove{\~n}a}, {Sarasso}, {Schultheis}, {Sciacca}, {Segol}, {Segovia},
  {S{\'e}gransan}, {Semeux}, {Shahaf}, {Siddiqui}, {Siebert}, {Siltala},
  {Silvelo}, {Slezak}, {Slezak}, {Smart}, {Snaith}, {Solano}, {Solitro},
  {Souami}, {Souchay}, {Spagna}, {Spina}, {Spoto}, {Steele},
  {Steidelm{\"u}ller}, {Stephenson}, {S{\"u}veges}, {Surdej}, {Szabados},
  {Szegedi-Elek}, {Taris}, {Taylor}, {Teixeira}, {Tolomei}, {Tonello}, {Torra},
  {Torra}, {Torralba Elipe}, {Trabucchi}, {Tsounis}, {Turon}, {Ulla}, {Unger},
  {Vaillant}, {van Dillen}, {van Reeven}, {Vanel}, {Vecchiato}, {Viala},
  {Vicente}, {Voutsinas}, {Weiler}, {Wevers}, {Wyrzykowski}, {Yoldas}, {Yvard},
  {Zhao}, {Zorec}, {Zucker}, \& {Zwitter}}]{Gaia_Collaboration_2023}
{Gaia Collaboration}, {Vallenari}, A., {Brown}, A.~G.~A., {et~al.} 2023, \aap,
  674, A1, \dodoi{10.1051/0004-6361/202243940}

\bibitem[{{Gan} {et~al.}(2023){Gan}, {Cadieux}, {Jahandar}, {Vazan}, {Wang},
  {Mao}, {Alvarado-Montes}, {Lin}, {Artigau}, {Cook}, {Doyon}, {Mann},
  {Stassun}, {Burgasser}, {Rackham}, {Howell}, {Collins}, {Barkaoui},
  {Shporer}, {de Leon}, {Arnold}, {Ricker}, {Vanderspek}, {Latham}, {Seager},
  {Winn}, {Jenkins}, {Burdanov}, {Charbonneau}, {Dransfield}, {Fukui},
  {Furlan}, {Gillon}, {Hooton}, {Lewis}, {Littlefield}, {Mireles}, {Narita},
  {Ormel}, {Quinn}, {Sefako}, {Timmermans}, {Vezie}, \& {de Wit}}]{Gan_2023}
{Gan}, T., {Cadieux}, C., {Jahandar}, F., {et~al.} 2023, \aj, 166, 165,
  \dodoi{10.3847/1538-3881/acf56d}

\bibitem[{{Gardner} {et~al.}(2023){Gardner}, {Mather}, {Abbott}, {Abell},
  {Abernathy}, {Abney}, {Abraham}, {Abraham}, {Abul-Huda}, {Acton}, {Adams},
  {Adams}, {Adler}, {Adriaensen}, {Aguilar}, {Ahmed}, {Ahmed}, {Ahmed},
  {Albat}, {Albert}, {Alberts}, {Aldridge}, {Allen}, {Allen}, {Altenburg},
  {Altunc}, {Alvarez}, {{\'A}lvarez-M{\'a}rquez}, {Alves de Oliveira},
  {Ambrose}, {Anandakrishnan}, {Andersen}, {Anderson}, {Anderson}, {Anderson},
  {Anderson}, {Aprea}, {Archer}, {Arenberg}, {Argyriou}, {Arribas}, {Artigau},
  {Arvai}, {Atcheson}, {Atkinson}, {Averbukh}, {Aymergen}, {Bacinski},
  {Baggett}, {Bagnasco}, {Baker}, {Balzano}, {Banks}, {Baran}, {Barker},
  {Barrett}, {Barringer}, {Barto}, {Bast}, {Baudoz}, {Baum}, {Beatty},
  {Beaulieu}, {Bechtold}, {Beck}, {Beddard}, {Beichman}, {Bellagama}, {Bely},
  {Berger}, {Bergeron}, {Bernier}, {Bertch}, {Beskow}, {Betz}, {Biagetti},
  {Birkmann}, {Bjorklund}, {Blackwood}, {Blazek}, {Blossfeld}, {Bluth},
  {Boccaletti}, {Boegner}, {Bohlin}, {Boia}, {B{\"o}ker}, {Bonaventura},
  {Bond}, {Bosley}, {Boucarut}, {Bouchet}, {Bouwman}, {Bower}, {Bowers},
  {Bowers}, {Boyce}, {Boyer}, {Boyer}, {Boyer}, {Boyer}, {Bradley}, {Brady},
  {Brandl}, {Brannen}, {Breda}, {Bremmer}, {Brennan}, {Bresnahan}, {Bright},
  {Broiles}, {Bromenschenkel}, {Brooks}, {Brooks}, {Brown}, {Brown}, {Brown},
  {Bruce}, {Bryson}, {Bujanda}, {Bullock}, {Bunker}, {Bureo}, {Burt}, {Bush},
  {Bushouse}, {Bussman}, {Cabaud}, {Cale}, {Calhoon}, {Calvani}, {Canipe},
  {Caputo}, {Cara}, {Carey}, {Case}, {Cesari}, {Cetorelli}, {Chance},
  {Chandler}, {Chaney}, {Chapman}, {Charlot}, {Chayer}, {Cheezum}, {Chen},
  {Chen}, {Cherinka}, {Chichester}, {Chilton}, {Chittiraibalan}, {Clampin},
  {Clark}, {Clark}, {Clark}, {Claybrooks}, {Cleveland}, {Cohen}, {Cohen},
  {Col{\'o}n}, {Coleman}, {Colina}, {Comber}, {Comeau}, {Comer}, {Conde Reis},
  {Connolly}, {Conroy}, {Contos}, {Contreras}, {Cook}, {Cooper}, {Cooper},
  {Correia}, {Correnti}, {Cossou}, {Costanza}, {Coulais}, {Cox}, {Coyle},
  {Cracraft}, {Crew}, {Curtis}, {Cusveller}, {Da Costa Maciel}, {Dailey},
  {Daugeron}, {Davidson}, {Davies}, {Davis}, {Davis}, {Day}, {de Chambure}, {de
  Jong}, {De Marchi}, {Dean}, {Decker}, {Delisa}, {Dell}, {Dellagatta},
  {Dembinska}, {Demosthenes}, {Dencheva}, {Deneu}, {DePriest}, {Deschenes},
  {Dethienne}, {Detre}, {Diaz}, {Dicken}, {DiFelice}, {Dillman}, {Disharoon},
  {Dixon}, {Doggett}, {Dominguez}, {Donaldson}, {Doria-Warner}, {Santos},
  {Doty}, {Douglas}, {Doyon}, {Dressler}, {Driggers}, {Driggers}, {Dunn},
  {DuPrie}, {Dupuis}, {Durning}, {Dutta}, {Earl}, {Eccleston}, {Ecobichon},
  {Egami}, {Ehrenwinkler}, {Eisenhamer}, {Eisenhower}, {Eisenstein}, {El
  Hamel}, {Elie}, {Elliott}, {Elliott}, {Engesser}, {Espinoza}, {Etienne},
  {Etxaluze}, {Evans}, {Fabreguettes}, {Falcolini}, {Falini}, {Fatig},
  {Feeney}, {Feinberg}, {Fels}, {Ferdous}, {Ferguson}, {Ferrarese}, {Ferreira},
  {Ferruit}, {Ferry}, {Filippazzo}, {Firre}, {Fix}, {Flagey}, {Flanagan},
  {Fleming}, {Florian}, {Flynn}, {Foiadelli}, {Fontaine}, {Fontanella},
  {Forshay}, {Fortner}, {Fox}, {Framarini}, {Francisco}, {Franck}, {Franx},
  {Franz}, {Friedman}, {Friend}, {Frost}, {Fu}, {Fullerton}, {Gaillard},
  {Galkin}, {Gallagher}, {Galyer}, {Garc{\'\i}a Mar{\'\i}n}, {Gardner},
  {Garland}, {Garrett}, {Gasman}, {G{\'a}sp{\'a}r}, {Gastaud}, {Gaudreau},
  {Gauthier}, {Geers}, {Geithner}, {Gennaro}, {Gerber}, {Gereau}, {Giampaoli},
  {Giardino}, {Gibbons}, {Gilbert}, {Gilman}, {Girard}, {Giuliano}, {Gkountis},
  {Glasse}, {Glassmire}, {Glauser}, {Glazer}, {Goldberg}, {Golimowski},
  {Gonzaga}, {Gordon}, {Gordon}, {Goudfrooij}, {Gough}, {Graham}, {Grau},
  {Green}, {Greene}, {Greene}, {Greenfield}, {Greenhouse}, {Greve}, {Greville},
  {Grimaldi}, {Groe}, {Groebner}, {Grumm}, {Grundy}, {G{\"u}del}, {Guillard},
  {Guldalian}, {Gunn}, {Gurule}, {Gutman}, {Guy}, {Guyot}, {Hack}, {Haderlein},
  {Hagan}, {Hagedorn}, {Hainline}, {Haley}, {Hami}, {Hamilton}, {Hammann},
  {Hammel}, {Hanley}, {Hansen}, {Hardy}, {Harnisch}, {Harr}, {Harris}, {Hart},
  {Hartig}, {Hasan}, {Hashim}, {Hashimoto}, {Haskins}, {Hawkins}, {Hayden},
  {Hayden}, {Healy}, {Hecht}, {Heeg}, {Hejal}, {Helm}, {Hengemihle}, {Henning},
  {Henry}, {Henry}, {Henshaw}, {Hernandez}, {Herrington}, {Heske}, {Hesman},
  {Hickey}, {Hilbert}, {Hines}, {Hinz}, {Hirsch}, {Hitcho}, {Hodapp}, {Hodge},
  {Hoffman}, {Holfeltz}, {Holler}, {Hoppa}, {Horner}, {Howard}, {Howard},
  {Huber}, {Hunkeler}, {Hunter}, {Hunter}, {Hurd}, {Hurst}, {Hutchings},
  {Hylan}, {Ignat}, {Illingworth}, {Irish}, {Isaacs}, {Jackson}, {Jaffe},
  {Jahic}, {Jahromi}, {Jakobsen}, {James}, {James}, {James}, {Jamieson},
  {Jandra}, {Jayawardhana}, {Jedrzejewski}, {Jeffers}, {Jensen}, {Joanne},
  {Johns}, {Johnson}, {Johnson}, {Johnson}, {Johnson}, {Johnson}, {Johnson},
  {Johnstone}, {Jollet}, {Jones}, {Jones}, {Jones}, {Jones}, {Jones}, {Jordan},
  {Jordan}, {Jue}, {Jurkowski}, {Justis}, {Justtanont}, {Kaleida}, {Kalirai},
  {Kalmanson}, {Kaltenegger}, {Kammerer}, {Kan}, {Kanarek}, {Kao}, {Karakla},
  {Karl}, {Kassin}, {Kauffman}, {Kavanagh}, {Kelley}, {Kelly}, {Kendrew},
  {Kennedy}, {Kenny}, {Keski-Kuha}, {Keyes}, {Khan}, {Kidwell}, {Kimble},
  {King}, {King}, {Kinzel}, {Kirk}, {Kirkpatrick}, {Klaassen}, {Klingemann},
  {Klintworth}, {Knapp}, {Knight}, {Knollenberg}, {Knutsen}, {Koehler},
  {Koekemoer}, {Kofler}, {Kontson}, {Kovacs}, {Kozhurina-Platais}, {Krause},
  {Kriss}, {Krist}, {Kristoffersen}, {Krogel}, {Krueger}, {Kulp}, {Kumari},
  {Kwan}, {Kyprianou}, {Labador}, {Labiano}, {Lafreni{\`e}re}, {Lagage},
  {Laidler}, {Laine}, {Laird}, {Lajoie}, {Lallo}, {Lam}, {LaMassa}, {Lambros},
  {Lampenfield}, {Lander}, {Langston}, {Larson}, {Larson}, {LaVerghetta},
  {Law}, {Lawrence}, {Lee}, {Lee}, {Lee}, {Leisenring}, {Leveille}, {Levenson},
  {Levi}, {Levine}, {Lewis}, {Lewis}, {Lewis}, {Libralato}, {Lidon},
  {Liebrecht}, {Lightsey}, {Lilly}, {Lim}, {Lim}, {Ling}, {Link}, {Link},
  {Lipinski}, {Liu}, {Lo}, {Lobmeyer}, {Logue}, {Long}, {Long}, {Long}, {Long},
  {L{\'o}pez-Caniego}, {Lotz}, {Love-Pruitt}, {Lubskiy}, {Luers}, {Luetgens},
  {Luevano}, {Lui}, {Lund}, {Lundquist}, {Lunine}, {L{\"u}tzgendorf}, {Lynch},
  {MacDonald}, {MacDonald}, {Macias}, {Macklis}, {Maghami}, {Maharaja},
  {Maiolino}, {Makrygiannis}, {Malla}, {Malumuth}, {Manjavacas}, {Marini},
  {Marrione}, {Marston}, {Martel}, {Martin}, {Martin}, {Martinez}, {Maschmann},
  {Masci}, {Masetti}, {Maszkiewicz}, {Matthews}, {Matuskey}, {McBrayer},
  {McCarthy}, {McCaughrean}, {McClare}, {McClare}, {McCloskey}, {McClurg},
  {McCoy}, {McElwain}, {McGregor}, {McGuffey}, {McKay}, {McKenzie}, {McLean},
  {McMaster}, {McNeil}, {De Meester}, {Mehalick}, {Meixner}, {Mel{\'e}ndez},
  {Menzel}, {Menzel}, {Merz}, {Mesterharm}, {Meyer}, {Meyett}, {Meza},
  {Midwinter}, {Milam}, {Miller}, {Miller}, {Miskey}, {Misselt}, {Mitchell},
  {Mohan}, {Montoya}, {Moran}, {Morishita}, {Moro-Mart{\'\i}n}, {Morrison},
  {Morrison}, {Morse}, {Moschos}, {Moseley}, {Mosier}, {Mosner}, {Mountain},
  {Muckenthaler}, {Mueller}, {Mueller}, {Muhiem}, {M{\"u}hlmann}, {Mullally},
  {Mullen}, {Munger}, {Murphy}, {Murray}, {Muzerolle}, {Mycroft}, {Myers},
  {Myers}, {Myers}, {Myers}, {Myrick}, {Nagle}, {Nayak}, {Naylor}, {Neff},
  {Nelan}, {Nella}, {Nguyen}, {Nguyen}, {Nickson}, {Nidhiry}, {Niedner},
  {Nieto-Santisteban}, {Nikolov}, {Nishisaka}, {Noriega-Crespo}, {Nota},
  {O'Mara}, {Oboryshko}, {O'Brien}, {Ochs}, {Offenberg}, {Ogle}, {Ohl},
  {Olmsted}, {Osborne}, {O'Shaughnessy}, {{\"O}stlin}, {O'Sullivan}, {Otor},
  {Ottens}, {Ouellette}, {Outlaw}, {Owens}, {Pacifici}, {Page}, {Paranilam},
  {Park}, {Parrish}, {Paschal}, {Patapis}, {Patel}, {Patrick}, {Pattishall},
  {Paul}, {Paul}, {Pauly}, {Pavlovsky}, {Pe{\~n}a-Guerrero}, {Pedder}, {Peek},
  {Pelham}, {Penanen}, {Perriello}, {Perrin}, {Perrine}, {Perrygo}, {Peslier},
  {Petach}, {Peterson}, {Pfarr}, {Pierson}, {Pietraszkiewicz}, {Pilchen},
  {Pipher}, {Pirzkal}, {Pitman}, {Player}, {Plesha}, {Plitzke}, {Pohner},
  {Poletis}, {Pollizzi}, {Polster}, {Pontius}, {Pontoppidan}, {Porges},
  {Potter}, {Prescott}, {Proffitt}, {Pueyo}, {Quispe Neira}, {Radich}, {Rager},
  {Rameau}, {Ramey}, {Ramos Alarcon}, {Rampini}, {Rapp}, {Rashford},
  {Rauscher}, {Ravindranath}, {Rawle}, {Rawlings}, {Ray}, {Regan}, {Rehm},
  {Rehm}, {Reid}, {Reis}, {Renk}, {Reoch}, {Ressler}, {Rest}, {Reynolds},
  {Richon}, {Richon}, {Ridgaway}, {Riedel}, {Rieke}, {Rieke}, {Rifelli},
  {Rigby}, {Riggs}, {Ringel}, {Ritchie}, {Rix}, {Robberto}, {Robinson},
  {Robinson}, {Robinson}, {Rock}, {Rodriguez}, {Rodr{\'\i}guez del Pino},
  {Roellig}, {Rohrbach}, {Roman}, {Romelfanger}, {Romo}, {Rosales}, {Rose},
  {Roteliuk}, {Roth}, {Rothwell}, {Rouzaud}, {Rowe}, {Rowlands}, {Roy},
  {Royer}, {Rui}, {Rumler}, {Rumpl}, {Russ}, {Ryan}, {Ryan}, {Saad}, {Sabata},
  {Sabatino}, {Sabbi}, {Sabelhaus}, {Sabia}, {Sahu}, {Saif}, {Salvignol},
  {Samara-Ratna}, {Samuelson}, {Sanders}, {Sappington}, {Sargent}, {Sauer},
  {Savadkin}, {Sawicki}, {Schappell}, {Scheffer}, {Scheithauer}, {Scherer},
  {Schiff}, {Schlawin}, {Schmeitzky}, {Schmitz}, {Schmude}, {Schneider},
  {Schreiber}, {Schroeven-Deceuninck}, {Schultz}, {Schwab}, {Schwartz},
  {Scoccimarro}, {Scott}, {Scott}, {Seaton}, {Seely}, {Seery}, {Seidleck},
  {Sembach}, {Shanahan}, {Shaughnessy}, {Shaw}, {Shay}, {Sheehan}, {Sheth},
  {Shih}, {Shivaei}, {Siegel}, {Sienkiewicz}, {Simmons}, {Simon}, {Sirianni},
  {Sivaramakrishnan}, {Slade}, {Sloan}, {Slocum}, {Slowinski}, {Smith},
  {Smith}, {Smith}, {Smith}, {Smith}, {Smith}, {Smolik}, {Soderblom}, {Sohn},
  {Sokol}, {Sonneborn}, {Sontag}, {Sooy}, {Soummer}, {Southwood}, {Spain},
  {Sparmo}, {Speer}, {Spencer}, {Sprofera}, {Stallcup}, {Stanley},
  {Stansberry}, {Stark}, {Starr}, {Stassi}, {Steck}, {Steeley}, {Stephens},
  {Stephenson}, {Stewart}, {Stiavelli}, {}, {Strada}, {Straughn}, {Streetman},
  {Strickland}, {Strobele}, {Stuhlinger}, {Stys}, {Such}, {Sukhatme},
  {Sullivan}, {Sullivan}, {Sumner}, {Sun}, {Sunnquist}, {Swade}, {Swam},
  {Swenton}, {Swoish}, {Tam Litten}, {Tamas}, {Tao}, {Taylor}, {Taylor}, {te
  Plate}, {Van Tea}, {Teague}, {Telfer}, {Temim}, {Texter}, {Thatte},
  {Thompson}, {Thompson}, {Thomson}, {Thronson}, {Tierney}, {Tikkanen},
  {Tinnin}, {Tippet}, {Todd}, {Tran}, {Trauger}, {Trejo}, {Vinh Truong},
  {Tsukamoto}, {Tufail}, {Tumlinson}, {Tustain}, {Tyra}, {Ubeda}, {Underwood},
  {Uzzo}, {Vaclavik}, {Valenduc}, {Valenti}, {Van Campen}, {van de Wetering},
  {Van Der Marel}, {van Haarlem}, {Vandenbussche}, {van Dishoeck},
  {Vanterpool}, {Vernoy}, {Vila Costas}, {Volk}, {Voorzaat}, {Voyton}, {Vydra},
  {Waddy}, {Waelkens}, {Wahlgren}, {Walker}, {Wander}, {Warfield}, {Warner},
  {Wasiak}, {Wasiak}, {Wehner}, {Weiler}, {Weilert}, {Weiss}, {Wells}, {Welty},
  {Wheate}, {Wheeler}, {White}, {Whitehouse}, {Whiteleather}, {Whitman},
  {Williams}, {Willmer}, {Willott}, {Willoughby}, {Wilson}, {Wilson}, {Wilson},
  {Windhorst}, {Wislowski}, {Wolfe}, {Wolfe}, {Wolff}, {Wondel}, {Woo},
  {Woods}, {Worden}, {Workman}, {Wright}, {Wu}, {Wu}, {Wun}, {Wymer},
  {Yadetie}, {Yan}, {Yang}, {Yates}, {Yeager}, {Yerger}, {Young}, {Young},
  {Yu}, {Yu}, {Zak}, {Zeidler}, {Zepp}, {Zhou}, {Zincke}, {Zonak}, \&
  {Zondag}}]{Gardner_2023}
{Gardner}, J.~P., {Mather}, J.~C., {Abbott}, R., {et~al.} 2023, \pasp, 135,
  068001, \dodoi{10.1088/1538-3873/acd1b5}

\bibitem[{{Gillon} {et~al.}(2017){Gillon}, {Triaud}, {Demory}, {Jehin}, {Agol},
  {Deck}, {Lederer}, {de Wit}, {Burdanov}, {Ingalls}, {Bolmont}, {Leconte},
  {Raymond}, {Selsis}, {Turbet}, {Barkaoui}, {Burgasser}, {Burleigh}, {Carey},
  {Chaushev}, {Copperwheat}, {Delrez}, {Fernandes}, {Holdsworth}, {Kotze}, {Van
  Grootel}, {Almleaky}, {Benkhaldoun}, {Magain}, \& {Queloz}}]{Gillon_2017}
{Gillon}, M., {Triaud}, A. H.~M.~J., {Demory}, B.-O., {et~al.} 2017, \nat, 542,
  456, \dodoi{10.1038/nature21360}

\bibitem[{{Ginzburg} {et~al.}(2016){Ginzburg}, {Schlichting}, \&
  {Sari}}]{Ginzburg_2016}
{Ginzburg}, S., {Schlichting}, H.~E., \& {Sari}, R. 2016, \apj, 825, 29,
  \dodoi{10.3847/0004-637X/825/1/29}

\bibitem[{{Ginzburg} {et~al.}(2018){Ginzburg}, {Schlichting}, \&
  {Sari}}]{Ginzburg_2018}
---. 2018, \mnras, 476, 759, \dodoi{10.1093/mnras/sty290}

\bibitem[{{Guillot} \& {Morel}(1995)}]{Guillot_1995}
{Guillot}, T., \& {Morel}, P. 1995, \aaps, 109, 109

\bibitem[{{Hallatt} \& {Wiegert}(2020)}]{Hallatt_2020}
{Hallatt}, T., \& {Wiegert}, P. 2020, \aj, 159, 147,
  \dodoi{10.3847/1538-3881/ab7336}

\bibitem[{{Harris} {et~al.}(2020){Harris}, {Millman}, {van der Walt},
  {Gommers}, {Virtanen}, {Cournapeau}, {Wieser}, {Taylor}, {Berg}, {Smith},
  {Kern}, {Picus}, {Hoyer}, {van Kerkwijk}, {Brett}, {Haldane}, {del R{\'\i}o},
  {Wiebe}, {Peterson}, {G{\'e}rard-Marchant}, {Sheppard}, {Reddy}, {Weckesser},
  {Abbasi}, {Gohlke}, \& {Oliphant}}]{Harris_2020}
{Harris}, C.~R., {Millman}, K.~J., {van der Walt}, S.~J., {et~al.} 2020, \nat,
  585, 357, \dodoi{10.1038/s41586-020-2649-2}

\bibitem[{{Hawkins} {et~al.}(2015){Hawkins}, {Jofr{\'e}}, {Masseron}, \&
  {Gilmore}}]{Hawkins_2015}
{Hawkins}, K., {Jofr{\'e}}, P., {Masseron}, T., \& {Gilmore}, G. 2015, \mnras,
  453, 758, \dodoi{10.1093/mnras/stv1586}

\bibitem[{{Haywood} {et~al.}(2014){Haywood}, {Collier Cameron}, {Queloz},
  {Barros}, {Deleuil}, {Fares}, {Gillon}, {Lanza}, {Lovis}, {Moutou}, {Pepe},
  {Pollacco}, {Santerne}, {S{\'e}gransan}, \& {Unruh}}]{Haywood_2014}
{Haywood}, R.~D., {Collier Cameron}, A., {Queloz}, D., {et~al.} 2014, \mnras,
  443, 2517, \dodoi{10.1093/mnras/stu1320}

\bibitem[{{Hemley} {et~al.}(1987){Hemley}, {Jephcoat}, {Mao}, {Zha}, {Finger},
  \& {Cox}}]{Hemley_1987}
{Hemley}, R.~J., {Jephcoat}, A.~P., {Mao}, H.~K., {et~al.} 1987, \nat, 330,
  737, \dodoi{10.1038/330737a0}

\bibitem[{{Higson} {et~al.}(2019){Higson}, {Handley}, {Hobson}, \&
  {Lasenby}}]{Higson_2019}
{Higson}, E., {Handley}, W., {Hobson}, M., \& {Lasenby}, A. 2019, Statistics
  and Computing, 29, 891, \dodoi{10.1007/s11222-018-9844-0}

\bibitem[{{Hunter}(2007)}]{Hunter_2007}
{Hunter}, J.~D. 2007, Computing in Science and Engineering, 9, 90,
  \dodoi{10.1109/MCSE.2007.55}

\bibitem[{{Husser} {et~al.}(2013){Husser}, {Wende-von Berg}, {Dreizler},
  {Homeier}, {Reiners}, {Barman}, \& {Hauschildt}}]{husser2013new}
{Husser}, T.~O., {Wende-von Berg}, S., {Dreizler}, S., {et~al.} 2013, \aap,
  553, A6, \dodoi{10.1051/0004-6361/201219058}

\bibitem[{{Inamdar} \& {Schlichting}(2016)}]{Inamdar_2016}
{Inamdar}, N.~K., \& {Schlichting}, H.~E. 2016, \apjl, 817, L13,
  \dodoi{10.3847/2041-8205/817/2/L13}

\bibitem[{{Irwin} {et~al.}(2009){Irwin}, {Charbonneau}, {Nutzman}, \&
  {Falco}}]{Irwin_2009}
{Irwin}, J., {Charbonneau}, D., {Nutzman}, P., \& {Falco}, E. 2009, in
  Transiting Planets, ed. F.~{Pont}, D.~{Sasselov}, \& M.~J. {Holman}, Vol.
  253, 37--43, \dodoi{10.1017/S1743921308026215}

\bibitem[{{Jahandar} {et~al.}(2023){Jahandar}, {Doyon}, {Artigau}, {Cook},
  {Cadieux}, {Lafreni{\`e}re}, {Forveille}, {Donati}, {Fouqu{\'e}}, {Carmona},
  {Cloutier}, {Cristofari}, {Gaidos}, {Gomes da Silva}, {Malo}, {Martioli}, {do
  Nascimento}, {Pelletier}, {Vandal}, \& {Venn}}]{Jahandar_2023}
{Jahandar}, F., {Doyon}, R., {Artigau}, {\'E}., {et~al.} 2023, arXiv e-prints,
  arXiv:2310.12125, \dodoi{10.48550/arXiv.2310.12125}

\bibitem[{{Jenkins} {et~al.}(2016){Jenkins}, {Twicken}, {McCauliff},
  {Campbell}, {Sanderfer}, {Lung}, {Mansouri-Samani}, {Girouard}, {Tenenbaum},
  {Klaus}, {Smith}, {Caldwell}, {Chacon}, {Henze}, {Heiges}, {Latham},
  {Morgan}, {Swade}, {Rinehart}, \& {Vanderspek}}]{Jenkins_2016}
{Jenkins}, J.~M., {Twicken}, J.~D., {McCauliff}, S., {et~al.} 2016, in Society
  of Photo-Optical Instrumentation Engineers (SPIE) Conference Series, Vol.
  9913, Software and Cyberinfrastructure for Astronomy IV, ed. G.~{Chiozzi} \&
  J.~C. {Guzman}, 99133E, \dodoi{10.1117/12.2233418}

\bibitem[{{Karamanis} {et~al.}(2021){Karamanis}, {Beutler}, \&
  {Peacock}}]{Karamanis_2021}
{Karamanis}, M., {Beutler}, F., \& {Peacock}, J.~A. 2021, \mnras, 508, 3589,
  \dodoi{10.1093/mnras/stab2867}

\bibitem[{{Kipping}(2010)}]{Kipping_2010}
{Kipping}, D.~M. 2010, \mnras, 407, 301,
  \dodoi{10.1111/j.1365-2966.2010.16894.x}

\bibitem[{{Kipping}(2013)}]{Kipping_2013}
---. 2013, \mnras, 435, 2152, \dodoi{10.1093/mnras/stt1435}

\bibitem[{{Kipping}(2014)}]{Kipping_2014}
---. 2014, \mnras, 440, 2164, \dodoi{10.1093/mnras/stu318}

\bibitem[{{Kite} \& {Ford}(2018)}]{Kite_2018}
{Kite}, E.~S., \& {Ford}, E.~B. 2018, \apj, 864, 75,
  \dodoi{10.3847/1538-4357/aad6e0}

\bibitem[{{Kordopatis} {et~al.}(2023){Kordopatis}, {Schultheis}, {McMillan},
  {Palicio}, {de Laverny}, {Recio-Blanco}, {Creevey}, {{\'A}lvarez}, {Andrae},
  {Poggio}, {Spitoni}, {Contursi}, {Zhao}, {Oreshina-Slezak}, {Ordenovic}, \&
  {Bijaoui}}]{Kordopatis_2023}
{Kordopatis}, G., {Schultheis}, M., {McMillan}, P.~J., {et~al.} 2023, \aap,
  669, A104, \dodoi{10.1051/0004-6361/202244283}

\bibitem[{{Kreidberg}(2015)}]{Kreidberg_2015}
{Kreidberg}, L. 2015, \pasp, 127, 1161, \dodoi{10.1086/683602}

\bibitem[{{Leconte} {et~al.}(2015){Leconte}, {Wu}, {Menou}, \&
  {Murray}}]{Leconte_2015}
{Leconte}, J., {Wu}, H., {Menou}, K., \& {Murray}, N. 2015, Science, 347, 632,
  \dodoi{10.1126/science.1258686}

\bibitem[{{Lee} \& {Connors}(2021)}]{Lee_2021}
{Lee}, E.~J., \& {Connors}, N.~J. 2021, \apj, 908, 32,
  \dodoi{10.3847/1538-4357/abd6c7}

\bibitem[{{Li} \& {Zhao}(2017)}]{Li_2017}
{Li}, C., \& {Zhao}, G. 2017, \apj, 850, 25, \dodoi{10.3847/1538-4357/aa93f4}

\bibitem[{{Lillo-Box} {et~al.}(2020){Lillo-Box}, {Figueira}, {Leleu},
  {Acu{\~n}a}, {Faria}, {Hara}, {Santos}, {Correia}, {Robutel}, {Deleuil},
  {Barrado}, {Sousa}, {Bonfils}, {Mousis}, {Almenara}, {Astudillo-Defru},
  {Marcq}, {Udry}, {Lovis}, \& {Pepe}}]{Lillo-Box_2020}
{Lillo-Box}, J., {Figueira}, P., {Leleu}, A., {et~al.} 2020, \aap, 642, A121,
  \dodoi{10.1051/0004-6361/202038922}

\bibitem[{Lindegren(2018)}]{Lindegren_2018}
Lindegren, L. 2018.
\newblock \url{http://www.rssd.esa.int/doc_fetch.php?id=3757412}

\bibitem[{{Luque} \& {Pall{\'e}}(2022)}]{Luque_2022}
{Luque}, R., \& {Pall{\'e}}, E. 2022, Science, 377, 1211,
  \dodoi{10.1126/science.abl7164}

\bibitem[{{Madhusudhan} {et~al.}(2021){Madhusudhan}, {Piette}, \&
  {Constantinou}}]{Madhusudhan_2021}
{Madhusudhan}, N., {Piette}, A. A.~A., \& {Constantinou}, S. 2021, \apj, 918,
  1, \dodoi{10.3847/1538-4357/abfd9c}

\bibitem[{{Madhusudhan} {et~al.}(2023){Madhusudhan}, {Sarkar}, {Constantinou},
  {Holmberg}, {Piette}, \& {Moses}}]{Madhusudhan_2023}
{Madhusudhan}, N., {Sarkar}, S., {Constantinou}, S., {et~al.} 2023, \apjl, 956,
  L13, \dodoi{10.3847/2041-8213/acf577}

\bibitem[{{Majewski} {et~al.}(2016){Majewski}, {APOGEE Team}, \& {APOGEE-2
  Team}}]{Majewski_2016}
{Majewski}, S.~R., {APOGEE Team}, \& {APOGEE-2 Team}. 2016, Astronomische
  Nachrichten, 337, 863, \dodoi{10.1002/asna.201612387}

\bibitem[{{Mann} {et~al.}(2015){Mann}, {Feiden}, {Gaidos}, {Boyajian}, \& {von
  Braun}}]{Mann_2015}
{Mann}, A.~W., {Feiden}, G.~A., {Gaidos}, E., {Boyajian}, T., \& {von Braun},
  K. 2015, \apj, 804, 64, \dodoi{10.1088/0004-637X/804/1/64}

\bibitem[{{Mann} {et~al.}(2019){Mann}, {Dupuy}, {Kraus}, {Gaidos}, {Ansdell},
  {Ireland}, {Rizzuto}, {Hung}, {Dittmann}, {Factor}, {Feiden}, {Martinez},
  {Ru{\'\i}z-Rodr{\'\i}guez}, \& {Thao}}]{Mann_2019}
{Mann}, A.~W., {Dupuy}, T., {Kraus}, A.~L., {et~al.} 2019, \apj, 871, 63,
  \dodoi{10.3847/1538-4357/aaf3bc}

\bibitem[{{Marounina} \& {Rogers}(2020)}]{Marounina_2022}
{Marounina}, N., \& {Rogers}, L.~A. 2020, \apj, 890, 107,
  \dodoi{10.3847/1538-4357/ab68e4}

\bibitem[{{Ment} {et~al.}(2019){Ment}, {Dittmann}, {Astudillo-Defru},
  {Charbonneau}, {Irwin}, {Bonfils}, {Murgas}, {Almenara}, {Forveille}, {Agol},
  {Ballard}, {Berta-Thompson}, {Bouchy}, {Cloutier}, {Delfosse}, {Doyon},
  {Dressing}, {Esquerdo}, {Haywood}, {Kipping}, {Latham}, {Lovis}, {Newton},
  {Pepe}, {Rodriguez}, {Santos}, {Tan}, {Udry}, {Winters}, \&
  {W{\"u}nsche}}]{Ment_2019}
{Ment}, K., {Dittmann}, J.~A., {Astudillo-Defru}, N., {et~al.} 2019, \aj, 157,
  32, \dodoi{10.3847/1538-3881/aaf1b1}

\bibitem[{{Morrison} {et~al.}(2018){Morrison}, {Jackson}, {Sturhahn}, {Zhang},
  \& {Greenberg}}]{Morrison2018}
{Morrison}, R.~A., {Jackson}, J.~M., {Sturhahn}, W., {Zhang}, D., \&
  {Greenberg}, E. 2018, Journal of Geophysical Research (Solid Earth), 123,
  4647, \dodoi{10.1029/2017JB015343}

\bibitem[{{Owen} \& {Wu}(2017)}]{Owen_2017}
{Owen}, J.~E., \& {Wu}, Y. 2017, \apj, 847, 29,
  \dodoi{10.3847/1538-4357/aa890a}

\bibitem[{{Patel} \& {Espinoza}(2022)}]{Patel_2022}
{Patel}, J.~A., \& {Espinoza}, N. 2022, \aj, 163, 228,
  \dodoi{10.3847/1538-3881/ac5f55}

\bibitem[{Pepe {et~al.}(2002)Pepe, Mayor, Rupprecht, Avila, Ballester, Beckers,
  Benz, Bertaux, Bouchy, Buzzoni, Cavadore, Deiries, Dekker, Delabre,
  D'Odorico, Eckert, Fischer, Fleury, George, Gilliotte, Gojak, Guzman, Koch,
  Kohler, Kotzlowski, Lacroix, Le~Merrer, Lizon, Lo~Curto, Longinotti,
  Megevand, Pasquini, Petitpas, Pichard, Queloz, Reyes, Richaud, Sivan,
  Sosnowska, Soto, Udry, Ureta, van Kesteren, Weber, Weilenmann, Wicenec,
  Wieland, Christensen-Dalsgaard, Dravins, Hatzes, Kürster, Paresce, \&
  Penny}]{Pepe_2002}
Pepe, F., Mayor, M., Rupprecht, G., {et~al.} 2002, The Messenger, 110, 9.
\newblock \url{https://ui.adsabs.harvard.edu/2002Msngr.110....9P/abstract}

\bibitem[{{Pepe} {et~al.}(2021){Pepe}, {Cristiani}, {Rebolo}, {Santos},
  {Dekker}, {Cabral}, {Di Marcantonio}, {Figueira}, {Lo Curto}, {Lovis},
  {Mayor}, {M{\'e}gevand}, {Molaro}, {Riva}, {Zapatero Osorio}, {Amate},
  {Manescau}, {Pasquini}, {Zerbi}, {Adibekyan}, {Abreu}, {Affolter}, {Alibert},
  {Aliverti}, {Allart}, {Allende Prieto}, {{\'A}lvarez}, {Alves}, {Avila},
  {Baldini}, {Bandy}, {Barros}, {Benz}, {Bianco}, {Borsa}, {Bourrier},
  {Bouchy}, {Broeg}, {Calderone}, {Cirami}, {Coelho}, {Conconi}, {Coretti},
  {Cumani}, {Cupani}, {D'Odorico}, {Damasso}, {Deiries}, {Delabre},
  {Demangeon}, {Dumusque}, {Ehrenreich}, {Faria}, {Fragoso}, {Genolet},
  {Genoni}, {G{\'e}nova Santos}, {Gonz{\'a}lez Hern{\'a}ndez}, {Hughes},
  {Iwert}, {Kerber}, {Knudstrup}, {Landoni}, {Lavie}, {Lillo-Box}, {Lizon},
  {Maire}, {Martins}, {Mehner}, {Micela}, {Modigliani}, {Monteiro}, {Monteiro},
  {Moschetti}, {Murphy}, {Nunes}, {Oggioni}, {Oliveira}, {Oshagh}, {Pall{\'e}},
  {Pariani}, {Poretti}, {Rasilla}, {Rebord{\~a}o}, {Redaelli}, {Santana
  Tschudi}, {Santin}, {Santos}, {S{\'e}gransan}, {Schmidt}, {Segovia},
  {Sosnowska}, {Sozzetti}, {Sousa}, {Span{\`o}}, {Su{\'a}rez Mascare{\~n}o},
  {Tabernero}, {Tenegi}, {Udry}, \& {Zanutta}}]{Pepe_2021}
{Pepe}, F., {Cristiani}, S., {Rebolo}, R., {et~al.} 2021, \aap, 645, A96,
  \dodoi{10.1051/0004-6361/202038306}

\bibitem[{{Piaulet} {et~al.}(2023){Piaulet}, {Benneke}, {Almenara}, {Dragomir},
  {Knutson}, {Thorngren}, {Peterson}, {Crossfield}, {Kempton}, {Kubyshkina},
  {Howard}, {Angus}, {Isaacson}, {Weiss}, {Beichman}, {Fortney}, {Fossati},
  {Lammer}, {McCullough}, {Morley}, \& {Wong}}]{Piaulet_2023}
{Piaulet}, C., {Benneke}, B., {Almenara}, J.~M., {et~al.} 2023, Nature
  Astronomy, 7, 206, \dodoi{10.1038/s41550-022-01835-4}

\bibitem[{{Plotnykov} \& Valencia(2020)}]{Plotnykov_2020}
{Plotnykov}, A., \& Valencia, D. 2020, MNRAS, 499, 932

\bibitem[{{Rajpaul} {et~al.}(2015){Rajpaul}, {Aigrain}, {Osborne}, {Reece}, \&
  {Roberts}}]{Rajpaul_2015}
{Rajpaul}, V., {Aigrain}, S., {Osborne}, M.~A., {Reece}, S., \& {Roberts}, S.
  2015, \mnras, 452, 2269, \dodoi{10.1093/mnras/stv1428}

\bibitem[{{Reddy} {et~al.}(2006){Reddy}, {Lambert}, \& {Allende
  Prieto}}]{Reddy_2006}
{Reddy}, B.~E., {Lambert}, D.~L., \& {Allende Prieto}, C. 2006, \mnras, 367,
  1329, \dodoi{10.1111/j.1365-2966.2006.10148.x}

\bibitem[{{Reyl{\'e}} {et~al.}(2021){Reyl{\'e}}, {Jardine}, {Fouqu{\'e}},
  {Caballero}, {Smart}, \& {Sozzetti}}]{Reyle_2021}
{Reyl{\'e}}, C., {Jardine}, K., {Fouqu{\'e}}, P., {et~al.} 2021, \aap, 650,
  A201, \dodoi{10.1051/0004-6361/202140985}

\bibitem[{{Reyl{\'e}} {et~al.}(2022){Reyl{\'e}}, {Jardine}, {Fouqu{\'e}},
  {Caballero}, {Smart}, \& {Sozzetti}}]{Reyle_2022}
{Reyl{\'e}}, C., {Jardine}, K., {Fouqu{\'e}}, P., {et~al.} 2022, in Cambridge
  Workshop on Cool Stars, Stellar Systems, and the Sun, Cambridge Workshop on
  Cool Stars, Stellar Systems, and the Sun, 218, \dodoi{10.5281/zenodo.7669746}

\bibitem[{{Ribas} {et~al.}(2016){Ribas}, {Bolmont}, {Selsis}, {Reiners},
  {Leconte}, {Raymond}, {Engle}, {Guinan}, {Morin}, {Turbet}, {Forget}, \&
  {Anglada-Escud{\'e}}}]{Ribas_2016}
{Ribas}, I., {Bolmont}, E., {Selsis}, F., {et~al.} 2016, \aap, 596, A111,
  \dodoi{10.1051/0004-6361/201629576}

\bibitem[{{Ricker} {et~al.}(2015){Ricker}, {Winn}, {Vanderspek}, {Latham},
  {Bakos}, {Bean}, {Berta-Thompson}, {Brown}, {Buchhave}, {Butler}, {Butler},
  {Chaplin}, {Charbonneau}, {Christensen-Dalsgaard}, {Clampin}, {Deming},
  {Doty}, {De Lee}, {Dressing}, {Dunham}, {Endl}, {Fressin}, {Ge}, {Henning},
  {Holman}, {Howard}, {Ida}, {Jenkins}, {Jernigan}, {Johnson}, {Kaltenegger},
  {Kawai}, {Kjeldsen}, {Laughlin}, {Levine}, {Lin}, {Lissauer}, {MacQueen},
  {Marcy}, {McCullough}, {Morton}, {Narita}, {Paegert}, {Palle}, {Pepe},
  {Pepper}, {Quirrenbach}, {Rinehart}, {Sasselov}, {Sato}, {Seager},
  {Sozzetti}, {Stassun}, {Sullivan}, {Szentgyorgyi}, {Torres}, {Udry}, \&
  {Villasenor}}]{Ricker_2015}
{Ricker}, G.~R., {Winn}, J.~N., {Vanderspek}, R., {et~al.} 2015, Journal of
  Astronomical Telescopes, Instruments, and Systems, 1, 014003,
  \dodoi{10.1117/1.JATIS.1.1.014003}

\bibitem[{{Rogers} {et~al.}(2023){Rogers}, {Schlichting}, \&
  {Owen}}]{Rogers_2023}
{Rogers}, J.~G., {Schlichting}, H.~E., \& {Owen}, J.~E. 2023, \apjl, 947, L19,
  \dodoi{10.3847/2041-8213/acc86f}

\bibitem[{{Rogers}(2015)}]{Rogers_2015}
{Rogers}, L.~A. 2015, \apj, 801, 41, \dodoi{10.1088/0004-637X/801/1/41}

\bibitem[{{Sabotta} {et~al.}(2021){Sabotta}, {Schlecker}, {Chaturvedi},
  {Guenther}, {Mu{\~n}oz Rodr{\'\i}guez}, {Mu{\~n}oz S{\'a}nchez}, {Caballero},
  {Shan}, {Reffert}, {Ribas}, {Reiners}, {Hatzes}, {Amado}, {Klahr}, {Morales},
  {Quirrenbach}, {Henning}, {Dreizler}, {Pall{\'e}}, {Perger}, {Azzaro},
  {Jeffers}, {Kaminski}, {K{\"u}rster}, {Lafarga}, {Montes}, {Passegger}, \&
  {Zechmeister}}]{Sabotta_2021}
{Sabotta}, S., {Schlecker}, M., {Chaturvedi}, P., {et~al.} 2021, \aap, 653,
  A114, \dodoi{10.1051/0004-6361/202140968}

\bibitem[{{Scora} {et~al.}(2020){Scora}, {Valencia}, {Morbidelli}, \&
  {Jacobson}}]{Scora2020}
{Scora}, J., {Valencia}, D., {Morbidelli}, A., \& {Jacobson}, S. 2020, \mnras,
  493, 4910, \dodoi{10.1093/mnras/staa568}

\bibitem[{{Seager} \& {Mall{\'e}n-Ornelas}(2003)}]{Seager_2003}
{Seager}, S., \& {Mall{\'e}n-Ornelas}, G. 2003, \apj, 585, 1038,
  \dodoi{10.1086/346105}

\bibitem[{{Silva} {et~al.}(2022){Silva}, {Faria}, {Santos}, {Sousa}, {Viana},
  {Martins}, {Figueira}, {Lovis}, {Pepe}, {Cristiani}, {Rebolo}, {Allart},
  {Cabral}, {Mehner}, {Sozzetti}, {Su{\'a}rez Mascare{\~n}o}, {Martins},
  {Ehrenreich}, {M{\'e}gevand}, {Palle}, {Lo Curto}, {Tabernero}, {Lillo-Box},
  {Gonz{\'a}lez Hern{\'a}ndez}, {Zapatero Osorio}, {Hara}, {Nunes}, {Di
  Marcantonio}, {Udry}, {Adibekyan}, \& {Dumusque}}]{Silva_2022}
{Silva}, A.~M., {Faria}, J.~P., {Santos}, N.~C., {et~al.} 2022, \aap, 663,
  A143, \dodoi{10.1051/0004-6361/202142262}

\bibitem[{{Skrutskie} {et~al.}(2006){Skrutskie}, {Cutri}, {Stiening},
  {Weinberg}, {Schneider}, {Carpenter}, {Beichman}, {Capps}, {Chester},
  {Elias}, {Huchra}, {Liebert}, {Lonsdale}, {Monet}, {Price}, {Seitzer},
  {Jarrett}, {Kirkpatrick}, {Gizis}, {Howard}, {Evans}, {Fowler}, {Fullmer},
  {Hurt}, {Light}, {Kopan}, {Marsh}, {McCallon}, {Tam}, {Van Dyk}, \&
  {Wheelock}}]{Skrutskie_2006}
{Skrutskie}, M.~F., {Cutri}, R.~M., {Stiening}, R., {et~al.} 2006, \aj, 131,
  1163, \dodoi{10.1086/498708}

\bibitem[{{Smith} {et~al.}(2012){Smith}, {Stumpe}, {Van Cleve}, {Jenkins},
  {Barclay}, {Fanelli}, {Girouard}, {Kolodziejczak}, {McCauliff}, {Morris}, \&
  {Twicken}}]{Smith_2012}
{Smith}, J.~C., {Stumpe}, M.~C., {Van Cleve}, J.~E., {et~al.} 2012, \pasp, 124,
  1000, \dodoi{10.1086/667697}

\bibitem[{{Spaargaren} {et~al.}(2023){Spaargaren}, {Wang}, {Mojzsis},
  {Ballmer}, \& {Tackley}}]{Spaargaren2023}
{Spaargaren}, R.~J., {Wang}, H.~S., {Mojzsis}, S.~J., {Ballmer}, M.~D., \&
  {Tackley}, P.~J. 2023, \apj, 948, 53, \dodoi{10.3847/1538-4357/acac7d}

\bibitem[{{Speagle}(2020)}]{Speagle_2020}
{Speagle}, J.~S. 2020, \mnras, 493, 3132, \dodoi{10.1093/mnras/staa278}

\bibitem[{{Stewart} \& {Ahrens}(2005)}]{Stewart_2005}
{Stewart}, S.~T., \& {Ahrens}, T.~J. 2005, Journal of Geophysical Research
  (Planets), 110, E03005, \dodoi{10.1029/2004JE002305}

\bibitem[{{Stixrude} \& {Lithgow-Bertelloni}(2011)}]{Stixrude2011}
{Stixrude}, L., \& {Lithgow-Bertelloni}, C. 2011, Geophysical Journal
  International, 184, 1180, \dodoi{10.1111/j.1365-246X.2010.04890.x}

\bibitem[{{Stock} {et~al.}(2023){Stock}, {Kemmer}, {Kossakowski}, {Sabotta},
  {Reffert}, \& {Quirrenbach}}]{Stock_2023}
{Stock}, S., {Kemmer}, J., {Kossakowski}, D., {et~al.} 2023, \aap, 674, A108,
  \dodoi{10.1051/0004-6361/202244629}

\bibitem[{{Stumpe} {et~al.}(2014){Stumpe}, {Smith}, {Catanzarite}, {Van Cleve},
  {Jenkins}, {Twicken}, \& {Girouard}}]{Stumpe_2014}
{Stumpe}, M.~C., {Smith}, J.~C., {Catanzarite}, J.~H., {et~al.} 2014, \pasp,
  126, 100, \dodoi{10.1086/674989}

\bibitem[{{Stumpe} {et~al.}(2012){Stumpe}, {Smith}, {Van Cleve}, {Twicken},
  {Barclay}, {Fanelli}, {Girouard}, {Jenkins}, {Kolodziejczak}, {McCauliff}, \&
  {Morris}}]{Stumpe_2012}
{Stumpe}, M.~C., {Smith}, J.~C., {Van Cleve}, J.~E., {et~al.} 2012, \pasp, 124,
  985, \dodoi{10.1086/667698}

\bibitem[{{Su{\'a}rez Mascare{\~n}o} {et~al.}(2023){Su{\'a}rez Mascare{\~n}o},
  {Gonz{\'a}lez-{\'A}lvarez}, {Zapatero Osorio}, {Lillo-Box}, {Faria},
  {Passegger}, {Gonz{\'a}lez Hern{\'a}ndez}, {Figueira}, {Sozzetti}, {Rebolo},
  {Pepe}, {Santos}, {Cristiani}, {Lovis}, {Silva}, {Ribas}, {Amado},
  {Caballero}, {Quirrenbach}, {Reiners}, {Zechmeister}, {Adibekyan}, {Alibert},
  {B{\'e}jar}, {Benatti}, {D'Odorico}, {Damasso}, {Delisle}, {Di Marcantonio},
  {Dreizler}, {Ehrenreich}, {Hatzes}, {Hara}, {Henning}, {Kaminski},
  {L{\'o}pez-Gonz{\'a}lez}, {Martins}, {Micela}, {Montes}, {Pall{\'e}},
  {Pedraz}, {Rodr{\'\i}guez}, {Rodr{\'\i}guez-L{\'o}pez}, {Tal-Or}, {Sousa}, \&
  {Udry}}]{Suarez_2023}
{Su{\'a}rez Mascare{\~n}o}, A., {Gonz{\'a}lez-{\'A}lvarez}, E., {Zapatero
  Osorio}, M.~R., {et~al.} 2023, \aap, 670, A5,
  \dodoi{10.1051/0004-6361/202244991}

\bibitem[{{Thiabaud} {et~al.}(2015){Thiabaud}, {Marboeuf}, {Alibert}, {Leya},
  \& {Mezger}}]{Thiabaud_2015}
{Thiabaud}, A., {Marboeuf}, U., {Alibert}, Y., {Leya}, I., \& {Mezger}, K.
  2015, \aap, 580, A30, \dodoi{10.1051/0004-6361/201525963}

\bibitem[{{Trotta}(2008)}]{Trotta_2008}
{Trotta}, R. 2008, Contemporary Physics, 49, 71,
  \dodoi{10.1080/00107510802066753}

\bibitem[{{Tsiaras} {et~al.}(2016){Tsiaras}, {Waldmann}, {Rocchetto}, {Varley},
  {Morello}, {Damiano}, \& {Tinetti}}]{Tsiaras_2016}
{Tsiaras}, A., {Waldmann}, I.~P., {Rocchetto}, M., {et~al.} 2016, \apj, 832,
  202, \dodoi{10.3847/0004-637X/832/2/202}

\bibitem[{{Turbet} {et~al.}(2020){Turbet}, {Bolmont}, {Ehrenreich}, {Gratier},
  {Leconte}, {Selsis}, {Hara}, \& {Lovis}}]{Turbet_2020}
{Turbet}, M., {Bolmont}, E., {Ehrenreich}, D., {et~al.} 2020, \aap, 638, A41,
  \dodoi{10.1051/0004-6361/201937151}

\bibitem[{{Turbet} {et~al.}(2016){Turbet}, {Leconte}, {Selsis}, {Bolmont},
  {Forget}, {Ribas}, {Raymond}, \& {Anglada-Escud{\'e}}}]{Turbet_2016}
{Turbet}, M., {Leconte}, J., {Selsis}, F., {et~al.} 2016, \aap, 596, A112,
  \dodoi{10.1051/0004-6361/201629577}

\bibitem[{{Turbet} {et~al.}(2018){Turbet}, {Bolmont}, {Leconte}, {Forget},
  {Selsis}, {Tobie}, {Caldas}, {Naar}, \& {Gillon}}]{Turbet_2018}
{Turbet}, M., {Bolmont}, E., {Leconte}, J., {et~al.} 2018, \aap, 612, A86,
  \dodoi{10.1051/0004-6361/201731620}

\bibitem[{{Turbet} {et~al.}(2023){Turbet}, {Fauchez}, {Leconte}, {Bolmont},
  {Chaverot}, {Forget}, {Millour}, {Selsis}, {Charnay}, {Ducrot}, {Gillon},
  {Maurel}, \& {Villanueva}}]{Turbet_2023}
{Turbet}, M., {Fauchez}, T.~J., {Leconte}, J., {et~al.} 2023, arXiv e-prints,
  arXiv:2308.15110, \dodoi{10.48550/arXiv.2308.15110}

\bibitem[{{Unterborn} {et~al.}(2016){Unterborn}, {Dismukes}, \&
  {Panero}}]{Unterborn_2016}
{Unterborn}, C.~T., {Dismukes}, E.~E., \& {Panero}, W.~R. 2016, \apj, 819, 32,
  \dodoi{10.3847/0004-637X/819/1/32}

\bibitem[{{Valencia} {et~al.}(2013){Valencia}, {Guillot}, {Parmentier}, \&
  {Freedman}}]{Valencia_2013}
{Valencia}, D., {Guillot}, T., {Parmentier}, V., \& {Freedman}, R.~S. 2013,
  \apj, 775, 10, \dodoi{10.1088/0004-637X/775/1/10}

\bibitem[{{Valencia} {et~al.}(2007){Valencia}, {Sasselov}, \&
  {O'Connell}}]{Valencia_2007}
{Valencia}, D., {Sasselov}, D.~D., \& {O'Connell}, R.~J. 2007, \apj, 656, 545,
  \dodoi{10.1086/509800}

\bibitem[{{Van Eylen} \& {Albrecht}(2015)}]{Van-Eylen_2015}
{Van Eylen}, V., \& {Albrecht}, S. 2015, \apj, 808, 126,
  \dodoi{10.1088/0004-637X/808/2/126}

\bibitem[{{Villanueva} {et~al.}(2022){Villanueva}, {Liuzzi}, {Faggi},
  {Protopapa}, {Kofman}, {Fauchez}, {Stone}, \& {Mandell}}]{Villanueva_2022}
{Villanueva}, G.~L., {Liuzzi}, G., {Faggi}, S., {et~al.} 2022, {Fundamentals of
  the Planetary Spectrum Generator}

\bibitem[{{Villanueva} {et~al.}(2018){Villanueva}, {Smith}, {Protopapa},
  {Faggi}, \& {Mandell}}]{Villanueva_2018}
{Villanueva}, G.~L., {Smith}, M.~D., {Protopapa}, S., {Faggi}, S., \&
  {Mandell}, A.~M. 2018, \jqsrt, 217, 86, \dodoi{10.1016/j.jqsrt.2018.05.023}

\bibitem[{{Virtanen} {et~al.}(2020){Virtanen}, {Gommers}, {Oliphant},
  {Haberland}, {Reddy}, {Cournapeau}, {Burovski}, {Peterson}, {Weckesser},
  {Bright}, {van der Walt}, {Brett}, {Wilson}, {Millman}, {Mayorov}, {Nelson},
  {Jones}, {Kern}, {Larson}, {Carey}, {Polat}, {Feng}, {Moore}, {VanderPlas},
  {Laxalde}, {Perktold}, {Cimrman}, {Henriksen}, {Quintero}, {Harris},
  {Archibald}, {Ribeiro}, {Pedregosa}, {van Mulbregt}, \& {SciPy 1. 0
  Contributors}}]{Virtanen_2020}
{Virtanen}, P., {Gommers}, R., {Oliphant}, T.~E., {et~al.} 2020, Nature
  Methods, 17, 261, \dodoi{10.1038/s41592-019-0686-2}

\bibitem[{{Wagner} \& {Pru{\ss}}(2002)}]{Wagner_2002}
{Wagner}, W., \& {Pru{\ss}}, A. 2002, Journal of Physical and Chemical
  Reference Data, 31, 387, \dodoi{10.1063/1.1461829}

\bibitem[{{Waskom}(2021)}]{Waskom_2021}
{Waskom}, M. 2021, The Journal of Open Source Software, 6, 3021,
  \dodoi{10.21105/joss.03021}

\bibitem[{{Wildi} {et~al.}(2022){Wildi}, {Bouchy}, {Doyon}, {Blind}, {Genolet},
  {Sordet}, {Segovia}, {Grieves}, {Malo}, {Artigau}, {St-Antoine},
  {Vall{\'e}e}, {Rasilla}, {Gracia Temich}, {Poulin-Girard}, {Brousseau},
  {Sosnowska}, {Reshetov}, {Baron}, {Thibault}, {Bovay}, {Frensch}, {Lo Curto},
  {Hubin}, {Zins}, {Peroux}, \& {Cabral}}]{Wildi_2022}
{Wildi}, F., {Bouchy}, F., {Doyon}, R., {et~al.} 2022, in Society of
  Photo-Optical Instrumentation Engineers (SPIE) Conference Series, Vol. 12184,
  Ground-based and Airborne Instrumentation for Astronomy IX, ed. C.~J.
  {Evans}, J.~J. {Bryant}, \& K.~{Motohara}, 121841H,
  \dodoi{10.1117/12.2630016}

\bibitem[{{Wolf} {et~al.}(2017){Wolf}, {Shields}, {Kopparapu}, {Haqq-Misra}, \&
  {Toon}}]{Wolf2017}
{Wolf}, E.~T., {Shields}, A.~L., {Kopparapu}, R.~K., {Haqq-Misra}, J., \&
  {Toon}, O.~B. 2017, \apj, 837, 107, \dodoi{10.3847/1538-4357/aa5ffc}

\bibitem[{{Wordsworth} {et~al.}(2011){Wordsworth}, {Forget}, {Selsis},
  {Millour}, {Charnay}, \& {Madeleine}}]{Wordsworth_2011}
{Wordsworth}, R.~D., {Forget}, F., {Selsis}, F., {et~al.} 2011, \apjl, 733,
  L48, \dodoi{10.1088/2041-8205/733/2/L48}

\bibitem[{{Yang} {et~al.}(2020){Yang}, {Ji}, \& {Zeng}}]{Yang_2020}
{Yang}, J., {Ji}, W., \& {Zeng}, Y. 2020, Nature Astronomy, 4, 58,
  \dodoi{10.1038/s41550-019-0883-z}

\bibitem[{{Zechmeister} {et~al.}(2018){Zechmeister}, {Reiners}, {Amado},
  {Azzaro}, {Bauer}, {B{\'e}jar}, {Caballero}, {Guenther}, {Hagen}, {Jeffers},
  {Kaminski}, {K{\"u}rster}, {Launhardt}, {Montes}, {Morales}, {Quirrenbach},
  {Reffert}, {Ribas}, {Seifert}, {Tal-Or}, \& {Wolthoff}}]{Zechmeister_2018}
{Zechmeister}, M., {Reiners}, A., {Amado}, P.~J., {et~al.} 2018, \aap, 609,
  A12, \dodoi{10.1051/0004-6361/201731483}

\end{thebibliography}

\appendix
\counterwithin{table}{section}
\counterwithin{figure}{section}

\section{Light curves}
\label{appendix_A}
\setcounter{figure}{0}
\renewcommand{\thefigure}{A\arabic{figure}}
\setcounter{table}{0}
\renewcommand{\thetable}{A\arabic{table}}

\begin{figure*}[h]
\centering
\includegraphics[width=1\linewidth]{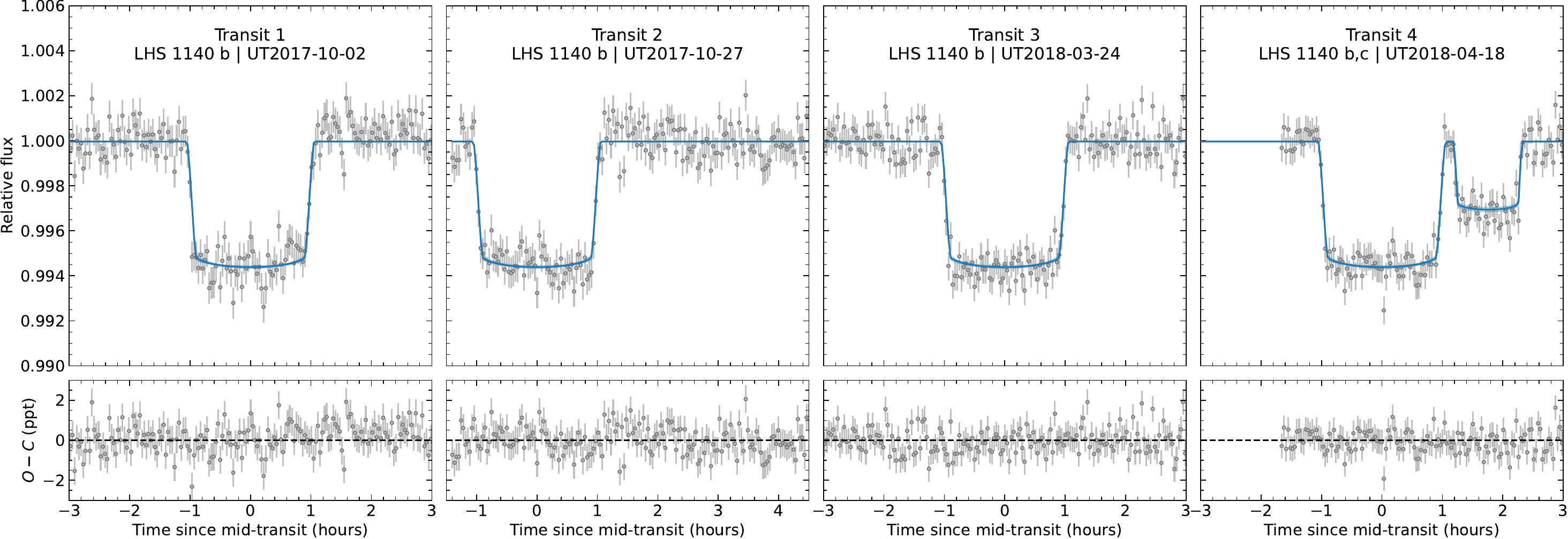}
  \caption{Individual \textit{Spitzer} transits of LHS~1140 b (Transit 1--4) and c (Transit 4 only). The best-fit transit models of the planets are shown in blue with the residual of this fit shown below.}
  \label{fig:spitzer_individual}
\end{figure*}

\begin{figure}[h!]
\centering
\includegraphics[width=0.45\linewidth]{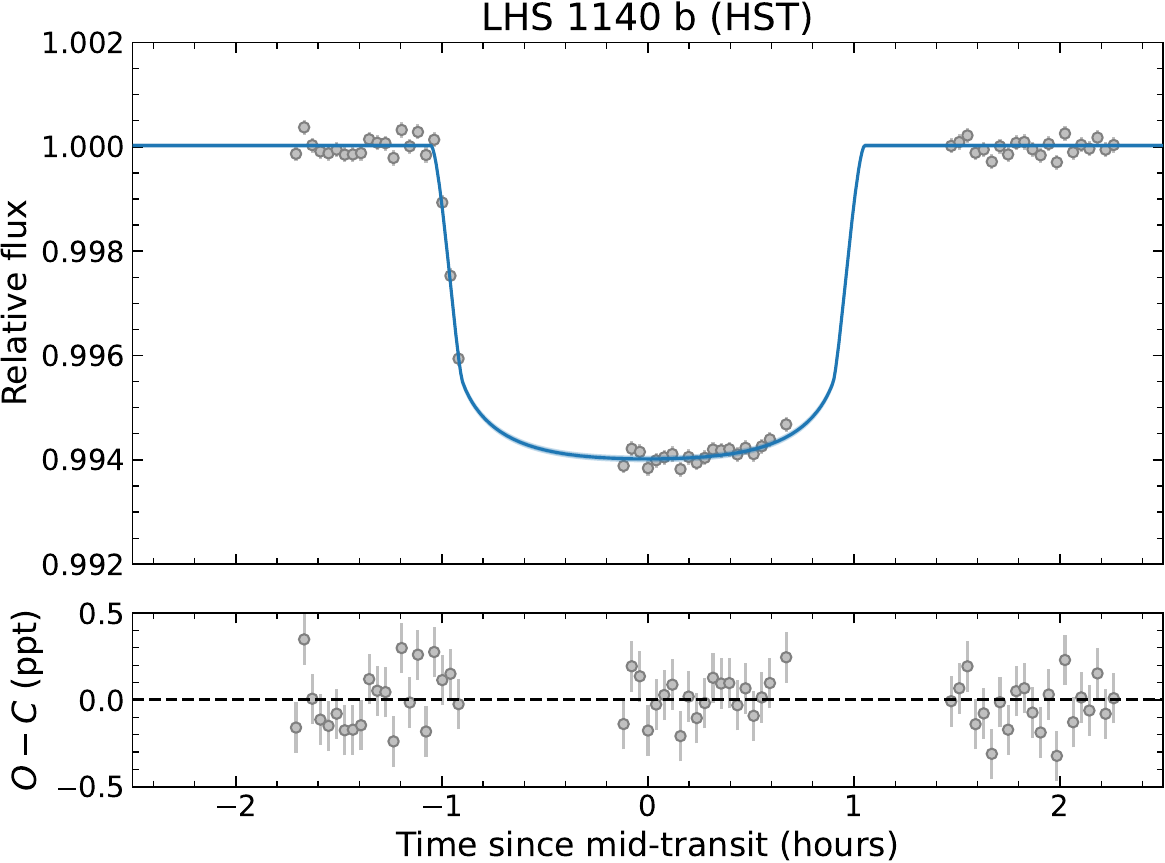}
  \caption{White light curve transit of LHS~1140\,b with the Wide Field Camera 3 on HST. The best-fit transit model is shown in blue with the residuals of this fit shown on the bottom panel.}
  \label{fig:hst_transit}
\end{figure}

\begin{figure*}[t!]
\centering
\includegraphics[width=1\linewidth]{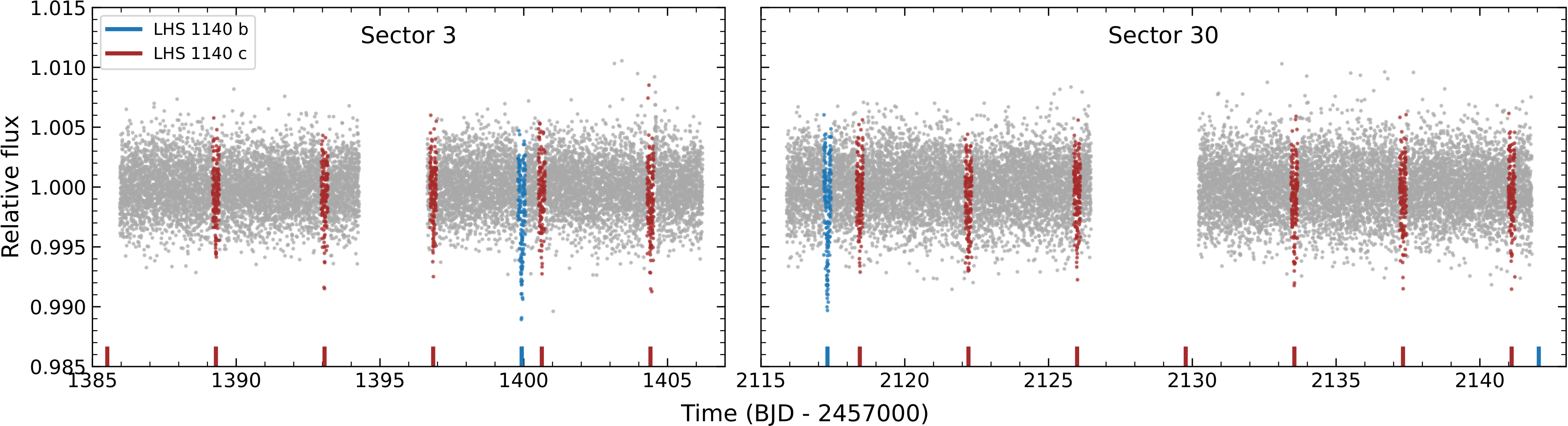}
\minipage{0.45\textwidth}
  \includegraphics[width=0.94\linewidth]{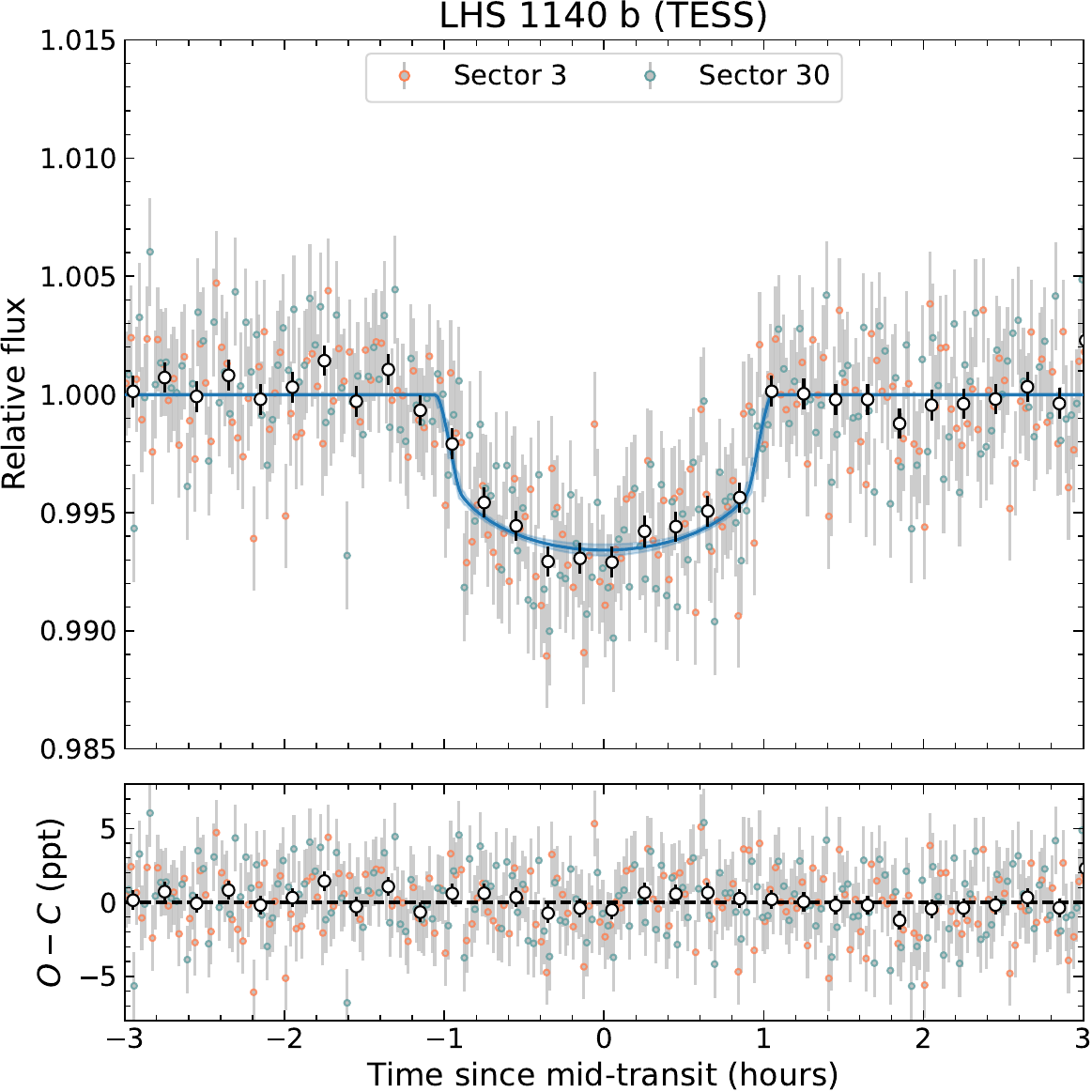}
\endminipage
  \minipage{0.45\textwidth}
  \includegraphics[width=0.94\linewidth]{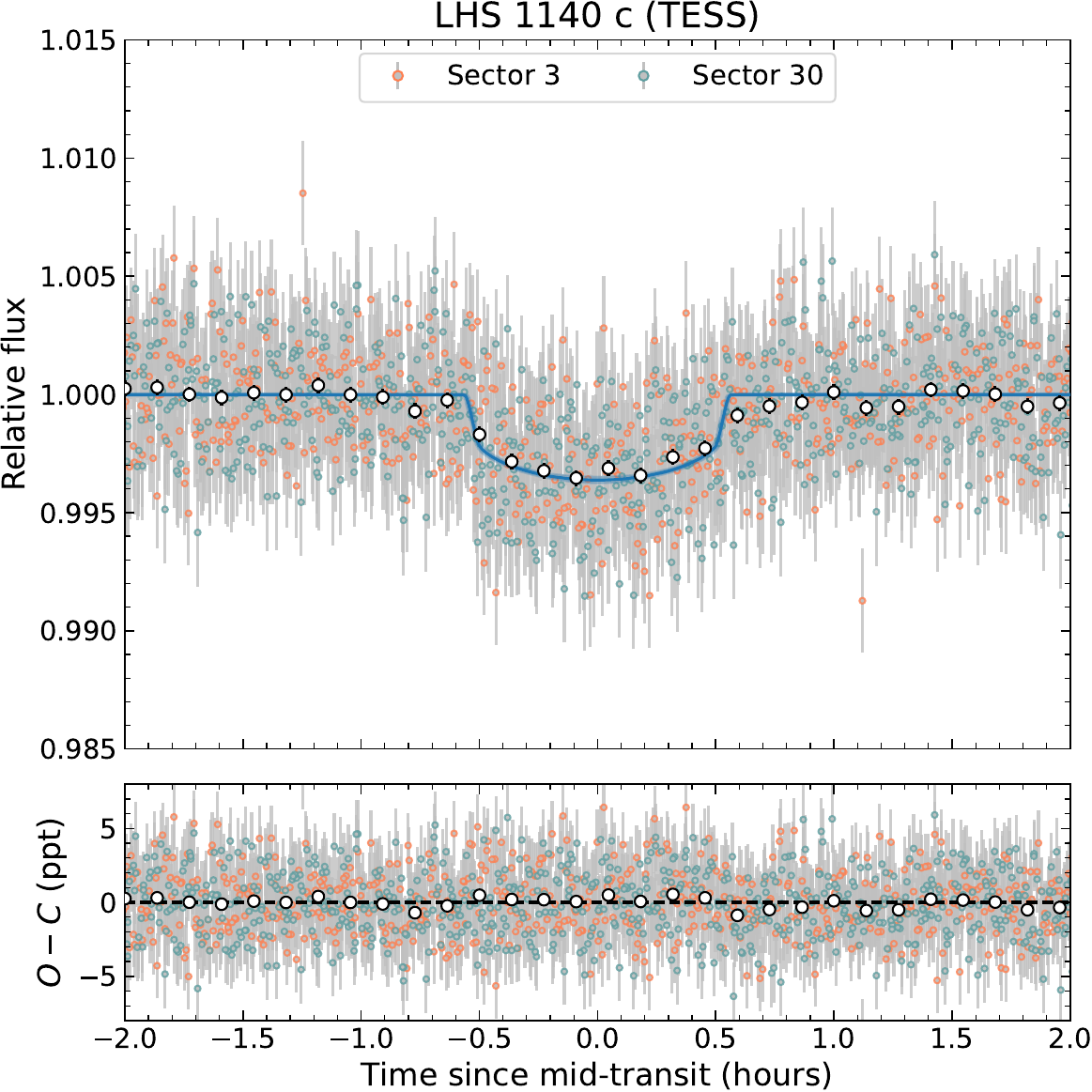}
\endminipage
  \caption{\textit{Top panels}: Normalized \texttt{PDCSAP} light curve of LHS~1140 from TESS Sectors 3 and 30. The blue and red vertical ticks and data points highlight the transits of planet b and c. \textit{Bottom panels}: Phase-folded transit of LHS~1140 b (left) and c (right). The open circles represent the binned photometry, respectively 12- and 8-minute bins for b and c. The best-fit transit models are depicted with blue curves. The residuals of the fits are shown below their respective transit.}
  \label{fig:tess_transit}
\end{figure*}

In this appendix, we present the light curves of LHS~1140 from \textit{Spitzer}, HST, and TESS. The four individual transits of LHS~1140\,b and one of LHS~1140\,c acquired with \textit{Spitzer} are presented in Figure~\ref{fig:spitzer_individual}. The single transit visit from HST with the Wide Field Camera 3 (white light curve) is shown in Figure~\ref{fig:hst_transit}. Lastly, the full TESS light curves from Sectors 3 and 30 are presented in Figure~\ref{fig:tess_transit}.

\section{Radial velocity measurements}
\label{appendix_B}
\setcounter{figure}{0}
\renewcommand{\thefigure}{B\arabic{figure}}
\setcounter{table}{0}
\renewcommand{\thetable}{B\arabic{table}}

\begin{figure}[ht!]
  \includegraphics[width=1\linewidth]{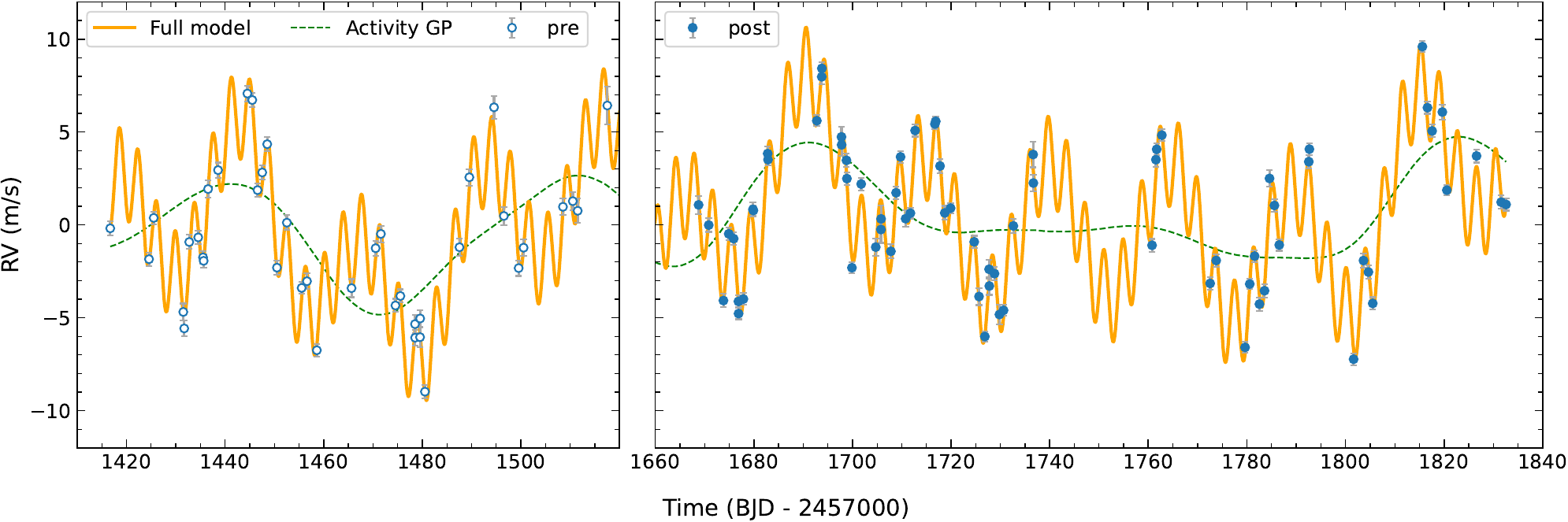}
  \caption{Radial velocity time series of LHS~1140 with ESPRESSO produced with the line-by-line method. The data before (pre) and after (post) the fiber upgrade of ESPRESSO (June 2019) are distinguished with open and solid circles, respectively. The full RV model (planet b, planet c, and activity GP) is represented by an orange curve, with the activity GP component highlighted with a green dashed curve (details in Appendix~\ref{sec:jointfit}).}
  \label{fig:RV_raw}
\end{figure}

The ESPRESSO radial velocity of LHS~1140 extracted with the line-by-line method are shown in Figure~\ref{fig:RV_raw} and listed in Table~\ref{table:rv} fully available online.

\begin{deluxetable}{ccc}
\tablecaption{ESPRESSO radial velocity of LHS~1140 obtained with the line-by-line method \label{table:rv}}
\tablehead{
\colhead{BJD - 2\,400\,000} & \colhead{RV (m\,s$^{-1}$)} & \colhead{$\sigma_{\rm RV}$ (m\,s$^{-1}$)}
}
\startdata
58416.711656 & $-13457.505$ & 0.364 \\
58424.576752 & $-13459.164$ & 0.360 \\
58425.528804 & $-13456.943$ & 0.343 \\
58431.527205 & $-13462.007$ & 0.453 \\
58431.714392 & $-13462.893$ & 0.423 \\
58432.733559 & $-13458.241$ & 0.383 \\
58434.555946 & $-13457.993$ & 0.346 \\
\ldots & \ldots & \ldots \\
\enddata
\tablecomments{Table~\ref{table:rv} is published in its entirety in machine-readable format.}
\end{deluxetable}

\section{Supplementary material of the stellar characterization}
\label{appendix_C}
\setcounter{figure}{0}
\renewcommand{\thefigure}{C\arabic{figure}}
\setcounter{table}{0}
\renewcommand{\thetable}{C\arabic{table}}

\subsection{Stellar age}\label{sec:age}

We analysed the kinematics of LHS~1140 to characterize its age. This was done by assessing whether the star belongs to the Galaxy's thin or thick disk stellar populations; thick disk stars are older ($\sim$10 Gyr) than their counterparts in the thin disk, have different chemistry ([Fe/H]$\sim$-0.5, [$\alpha$/Fe]$\sim$+0.3; \citealt{Reddy_2006}), and are kinematically hotter, with larger velocity dispersions relative to the Local Standard of Rest (LSR), and larger orbital excursions from the Galactic midplane \citep{Binney_2008}. We employed the astrometric solution from Gaia DR3 \citep{Gaia_Collaboration_2023}. We note that the solution's renormalized unit weight error \texttt{ruwe}=1.53 indicates that there is some uncertainty in the astrometry \citep[\texttt{ruwe}$\lesssim$1.4 for well-behaved solutions;][]{Lindegren_2018}. The orbit of LHS~1140 was integrated forward in time 60 Myr using the Monte Carlo model outlined in \cite{Hallatt_2020}.

These calculations produce a velocity with respect to the Local Standard of Rest $(U_{\rm LSR}, V_{\rm LSR}, W_{\rm LSR})=(14.38\pm 0.041,-38.52\pm 0.09,11.18\pm 0.41)$ km\,s$^{-1}$, yielding $V_{\rm tot}=\sqrt{U^{2}_{\rm LSR}+V^{2}_{\rm LSR}+W^{2}_{\rm LSR}}=42.61 \pm 0.20 $ km\,s$^{-1}$ (adopting the Local Standard of Rest from \citealt{Bland-Hawthorn_2016}). This places LHS~1140 in the thin disk, where $V_{\rm tot, thin}\lesssim 50$ km\,s$^{-1}$ \citep[e.g.,][]{Bensby_2003,Hawkins_2015}. Its orbital oscillation amplitude above/below the Galactic midplane is $131\pm 5$ pc, significantly smaller than that of thick disk stars ($\sim$1 kpc; e.g., \citealt{Li_2017}) and consistent with that of thin disk stars $\sim$a few Gyr old \citep[see Fig.~20 of][]{Kordopatis_2023}. This result for the age of LHS~1140 is consistent with \citetalias{Dittmann_2017} ($>$5\,Gyr) estimated from its slow rotation period and absence of H$\alpha$ emission. We thus adopt that LHS~1140 has a relatively old age $>$5\,Gyr and is a thin disk star.

\subsection{Bayesian inference of stellar mass and radius from transits}
\label{sec:bayes_stellar}

\begin{figure*}[t!]
\centering
\minipage{0.5\textwidth}
\includegraphics[width=0.95\linewidth]{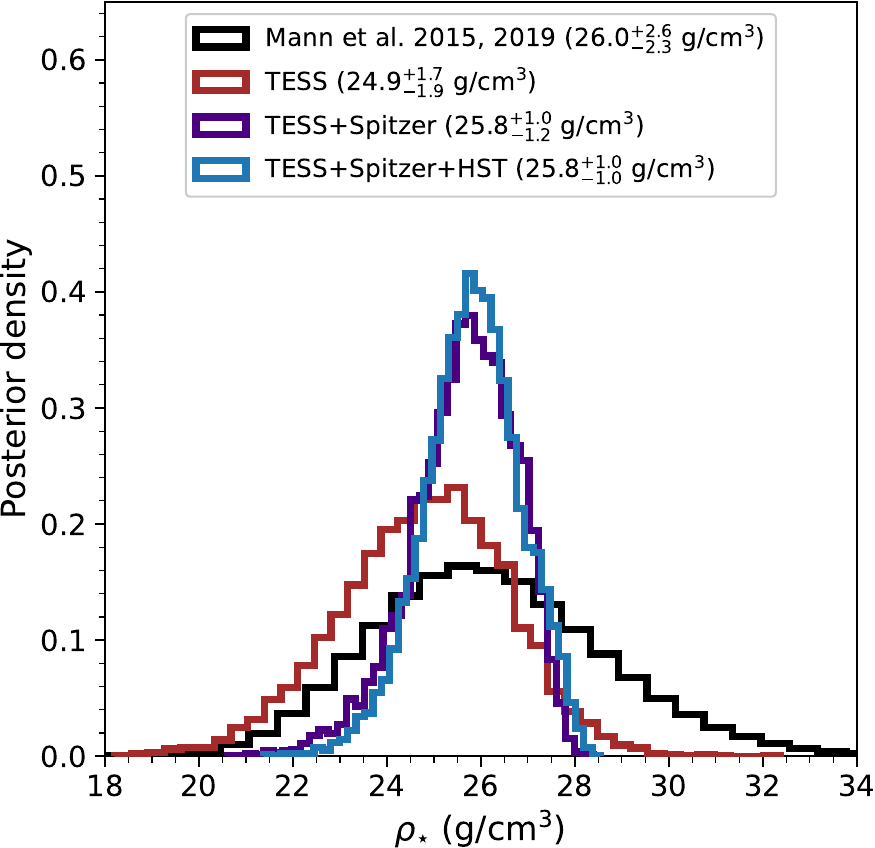}
\endminipage
\minipage{0.5\textwidth}
\includegraphics[width=1\linewidth]{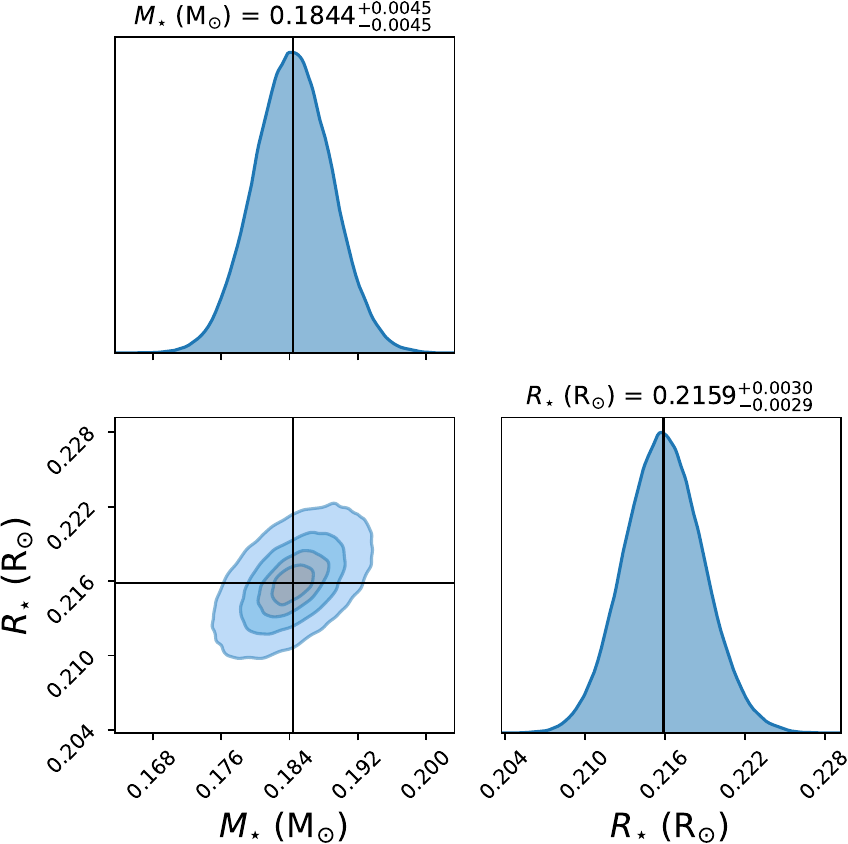}
\endminipage
  \caption{\textit{Left panel}: Stellar density ($\rho_{\star}$) distribution of LHS~1140 using mass and radius from \citealt{Mann_2015, Mann_2019} (black) and posterior distributions of $\rho_{\star}$ inferred from TESS (red), TESS+\textit{Spitzer} (purple), and TESS+\textit{Spitzer}+HST (blue) transits. \textit{Right panel}: Bayesian inference of the stellar mass and radius of LHS~1140 using the TESS+\textit{Spitzer}+HST constraints on $\rho_{\star}$ as a measurement and \cite{Mann_2015, Mann_2019} as a prior. The positive covariance in $M_{\star}$--$R_{\star}$ space indicates that both parameters are likely to vary in the same direction to produce a constant $\rho_{\star}$ constrained by the transits.}
  \label{fig:rhos}
\end{figure*}

Measured directly from transit light curves, the orbital period ($P$) and the scaled semi-major axis ($a/R_{\star}$) of an exoplanet allow the determination of the density of its host star \citep{Seager_2003}:
\begin{equation}
    \rho_{\star\rm{,transit}} = \frac{3 \pi}{G P^2} \left(\frac{a}{R_{\star}}\right)^3
    \label{eq:rho_transit}
\end{equation}
While Equation~\ref{eq:rho_transit} is fundamentally true for all orbits, as it is essentially a reformulation of Kepler's Third Law (with $G$ the gravitational constant), the $a/R_{\star}$ inferred from transit can be significantly biased when assuming a circular orbit (e.g., \citealt{Kipping_2010}, \citealt{Dawson_2012}, \citealt{Kipping_2014}). Following the notation of \cite{Van-Eylen_2015}, the true stellar density ($\rho_{\star}$) when photo-eccentric effects are considered is given by:
\begin{equation}
    \frac{\rho_{\star}}{\rho_{\star\rm{,transit}}} = \frac{\left(1 - e^2\right)^{3/2}}{\left(1 + e \sin \omega \right)^3}
    \label{eq:rho_real}
\end{equation}
where $e$ and $\omega$ are respectively the orbital eccentricity and argument of periastron of the transiting planet. Unaccounted eccentricity as small as $e \sim 0.1$ can potentially induce a 30\% error in $\rho_{\star}$. From our joint transit RV analysis (Appendix~\ref{sec:jointfit}), we measure $\rho_{\star}/\rho_{\star\rm{,transit}}$ of $1.00 \pm 0.03$ for LHS~1140\,b and $1.02 \pm 0.04$ for LHS~1140\,c consistent with perfectly circular orbits ($e_{\rm b} < 0.043$, $e_{\rm c} < 0.050$ with 95\% confidence). As both orbital solutions satisfy $\rho_{\star} \approx \rho_{\star\rm{,transit}}$, we hereafter drop the transit subscript when referring to stellar density obtained from our transit light curves. As summarized in Figure~\ref{fig:rhos}, our measurement of $\rho_{\star}$ constrained by \textit{Spitzer}, HST, and TESS has resulted in new posteriors for the mass and radius of the star LHS~1140, namely $M_{\star} = 0.1844 \pm 0.0045$\,M$_{\odot}$ and $R_{\star} = 0.2159 \pm 0.0030$\,R$_{\odot}$. This Bayesian approach is taken to improve the precision on $M_{\star}$ and $R_{\star}$, otherwise the dominant sources of uncertainty for the inferred planetary mass and radius.

\subsection{Additional Tables and Figure}

\begin{figure}[t!]
\centering
\includegraphics[width=0.5\linewidth]{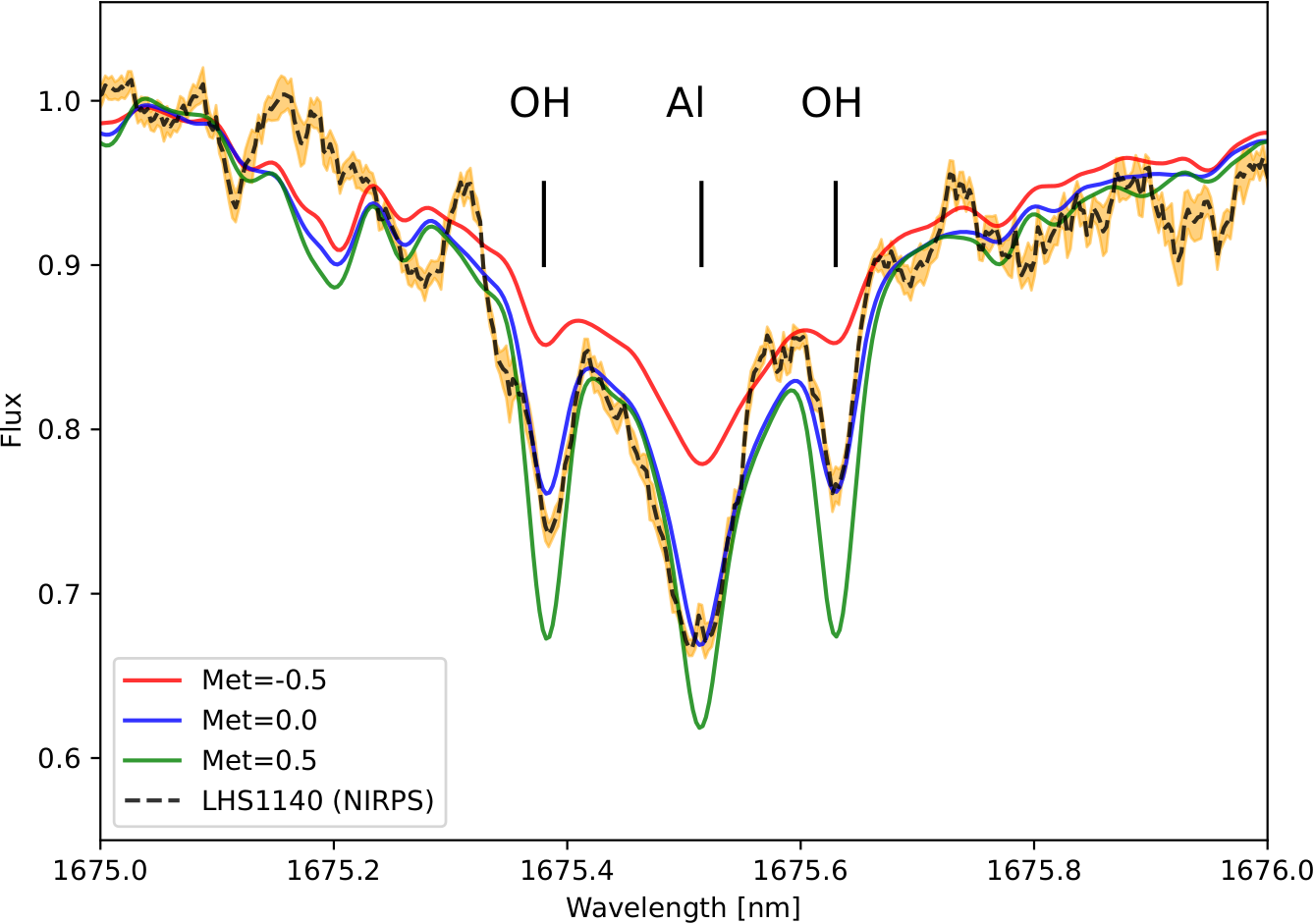}
  \caption{NIRPS template spectrum of LHS~1140 (black dashed line) around the Al I line (1675.514\,nm). The orange envelope depicts the flux dispersion measured on 29 individual observations. Three ACES models with a fixed $T_{\rm eff} = 3100$\,K and metallicities of $-0.5$ (red line), 0.0 (blue line) and 0.5\,dex (green line) are shown. For this single line, we measure [Al/H]~=~$0.0\pm0.1$\,dex. This figure illustrates the precision of the NIRPS data to constrain elemental abundances often different from that of iron ([Fe/H]~=~$-0.15 \pm 0.09$\,dex).}
  \label{fig:nirps_met}
\end{figure}

We recapitulate the stellar parameters of LHS~1140 in Table~\ref{table:stellarparams}. After, we present the stellar abundance determination from NIRPS (Sect.~\ref{sec:abundances}) in Table~\ref{table:abundances} and give corresponding chemical weight ratios in Table~\ref{table:ratios}. An example of this chemical spectroscopy analysis for the Al I line (1675.514\,nm) is presented in Figure~\ref{fig:nirps_met}.

\begin{deluxetable}{lcr}
\tablecaption{LHS~1140 stellar parameters}
\tablehead{
\colhead{Parameter} & \colhead{Value} & \colhead{Ref.}
}
\startdata
\multicolumn{3}{c}{\textit{Astrometry and kinematics}}\\
RA (J2016.0) & 00:44:59.33 & 1\\
DEC (J2016.0) & -15:16:17.54 & 1\\
$\mu_{\alpha} \cos \delta$ (mas\,$\cdot$\,yr$^{-1}$) & 318.152 $\pm$ 0.049 & 1\\
$\mu_{\delta}$ (mas\,$\cdot$\,yr$^{-1}$) & -596.623 $\pm$ 0.054 & 1\\
$\pi$ (mas) & 66.8287 $\pm$ 0.0479 & 1\\
$d$ (pc) & 14.9636 $\pm$ 0.0107 & 1\\
$U$ (km\,s$^{-1}$) & 14.38 $\pm$ 0.041 & 2\\
$V$ (km\,s$^{-1}$) & -38.52 $\pm$ 0.09 & 2\\
$W$ (km\,s$^{-1}$) & 11.18 $\pm$ 0.41 & 2\\
\hline
\multicolumn{3}{c}{\textit{Physical parameters}}\\
$M_{\star}$ (M$_{\odot}$) & 0.1844 $\pm$ 0.0045 & 2\\
$R_{\star}$ (R$_{\odot}$) & 0.2159 $\pm$ 0.0030 & 2\\
$\rho_{\star}$ (g\,$\cdot$\,cm$^{-3}$) & 25.8 $\pm$ 1.0 & 3\\
$T_{\rm eff}$ (K) & 3096 $\pm$ 48 & 2\\
$L_{\star}$ (L$_{\odot}$) & 0.0038 $\pm$ 0.0003 & 3\\
SpT & M4.5V & 4\\
$\left[ {\rm Fe/H} \right]$ (dex) & $-0.15$ $\pm$ 0.09 & 2\\
log $g$ (cgs) & 5.041 $\pm$ 0.016 & 3\\
Age & $>5$\,Gyr & 4\\
$P_{\rm rot}$ (days) & 131 $\pm$ 5 & 4
\enddata
\tablenocomments{}
\tablerefs{(1) Gaia DR3 \citep{Gaia_Collaboration_2021}. (2) This work. (3) This work derived from $M_{\star}$, $R_{\star}$, and $T_{\rm eff}$ (4) \cite{Dittmann_2017}.}
\label{table:stellarparams}
\end{deluxetable}

\begin{deluxetable}{lccc}
\setlength{\tabcolsep}{16pt}
\tablecaption{LHS~1140 stellar abundances measured with NIRPS for various chemical species with important features in the near-infrared}
\tablehead{
\colhead{Element} & \colhead{[X/H]} & \colhead{$\sigma$} & \colhead{\# of lines}}
\startdata
Fe I      &  -0.15   &  0.09   & 3           \\
Al I      &  0.00   &  0.10   & 2           \\
Mg I      &  0.11   &  0.10   & 2        \\
Si I      &  -0.20   &  0.10   & 1         \\
Ca I      &  0.20   &  0.10   & 1        \\
O I$^*$   &  0.00   &  0.01   & 82         \\
C I       &  0.10   &  0.10   & 3           \\
$<>^{\dag}$&  0.01   &  0.04   & -- 
\enddata
\tablecomments{$^*$The oxygen abundance is inferred from OH lines.\\ $^{\dag}$Average abundance of all elements.}
\label{table:abundances}
\end{deluxetable}

\begin{deluxetable}{lccc}
\setlength{\tabcolsep}{16pt}
\tablecaption{LHS~1140 stellar abundance weight ratios$^{*}$}
\tablehead{
\colhead{Ratios} & \colhead{LHS~1140} & \colhead{Sun$^{\dag}$} & \colhead{M dwarf$^{\ddag}$}}
\startdata
Fe/Mg [w] & 1.03$^{+0.40}_{-0.29}$ & 1.87 $\pm$ 0.22 & [0.89, 2.92]\\
Mg/Si [w] & 1.94$^{+0.79}_{-0.56}$ & 0.95 $\pm$ 0.09 & [0.69, 1.66]\\
Fe/O [w] & 0.15$^{+0.04}_{-0.03}$  &  0.21 $\pm$ 0.03 & [0.08, 0.28]\\
C/O [w] & 0.56$^{+0.17}_{-0.13}$  & 0.44 $\pm$ 0.06 & [0.21, 0.60]
\enddata
\tablecomments{$^{*}$Weight ratios calculated using $\textrm{X/Y [w]} = 10^{A\left(X\right)} / 10^{A\left(Y\right)} \times (m_{\rm X}/m_{\rm Y})$ with $A\left(\textrm{X}\right) = \left[ \textrm{X/H}\right] + A_{\rm solar}\left(\textrm{X}\right)$ the absolute logarithmic abundance, $A_{\rm solar}\left(\textrm{X}\right)$ taken from Table~2 of \cite{Asplund_2021}, and $m_{\rm X}$ the atomic mass of element X.\\$^{\dag}$Solar weight ratios from \cite{Asplund_2021}.\\$^{\ddag}$95\% confidence interval of the M dwarf population ($\sim$1000) of APOGEE DR16 (\citealt{Majewski_2016}; \citealt{Ahumada_2020})}
\label{table:ratios}
\end{deluxetable}

\section{Data analysis}
\label{appendix_D}
\setcounter{figure}{0}
\renewcommand{\thefigure}{D\arabic{figure}}
\setcounter{table}{0}
\renewcommand{\thetable}{D\arabic{table}}

\subsection{Joint transit RV fit} \label{sec:jointfit}

The joint analysis of the photometric (\textit{Spitzer}, HST, and TESS) and RV (ESPRESSO) data is done with \texttt{juliet} \citep{Espinoza_2018}, an all-in-one package that combines transit and RV modeling using \texttt{batman} \citep{Kreidberg_2015} and \texttt{radvel} \citep{Fulton_2018} with multiple sampling options (e.g., Markov Chain Monte Carlo, nested sampling). Here, we select the \texttt{dynesty} \citep{Speagle_2020} sampler in \texttt{juliet} for parameter estimations and Bayesian log-evidence ($\ln Z$) calculations relevant for model comparisons. The \texttt{dynesty} package implements dynamic nested sampling algorithms \citep{Higson_2019} designed for more efficient and robust estimations of complex posterior distributions. We follow the \texttt{dynesty} documentation\footnote{\href{https://dynesty.readthedocs.io/en/stable/index.html}{\texttt{dynesty.readthedocs.io/en/stable/index.html}}} and choose the random slice sampling option since the number of free parameters exceeds 20.

The orbit of planet $k$ ($k$: `b', `c', `d') is described by four parameters, the orbital period $P_{k}$, the time of inferior conjunction $t_{0,k}$, the eccentricity $e_{k}$, and the argument of periastron $\omega_{k}$, and one systemic parameter, the stellar density $\rho_{\star}$, common for all planets. For multi-planetary systems, a single $\rho_{\star}$ exists, which eliminates the need to fit semi-major axes ($a_{k}/R_{\star}$) for each planet (Equation~\ref{eq:rho_transit}). We define a Gaussian prior on the stellar density of LHS~1140 based on \cite{Mann_2015, Mann_2019}: $\rho_{\star} \sim \mathcal{N}\left(26.0, 2.6^2\right)$\,g\,cm$^{-3}$. For the transiting planets b and c, we follow the \cite{Espinoza_2019} transformation of the transit impact parameter $b$ and planet-to-star radius ratio $p = R_{\rm p} / R_{\star}$ into $r_1$ and $r_2$ parameters to only sample physically plausible regions in $b$--$p$ space. We model the baseline flux of the \textit{Spitzer}, HST, and TESS light curves with the parameter $M$ described in \cite{Espinoza_2018} and include per-instrument extra jitter terms ($\sigma_{\rm Spitzer}$, $\sigma_{\rm HST}$, and $\sigma_{\rm TESS}$).

Synthetic spectra of M dwarfs often show significant discrepancy with the observations \citep{Blanco-Cuaresma_2019}, implying that theoretical limb-darkening (LD) predictions may be unreliable. \cite{Patel_2022} have found systematic offsets between empirical quadratic LD coefficients ($u_1$,$u_2$) and theoretical predictions in the TESS bandpass of the order $\Delta u_1$,$ \Delta u_2 \approx 0.2$ for cool stars similar to LHS~1140. Differing from previous transit analyses (\citetalias{Dittmann_2017}; \citetalias{Ment_2019}; \citetalias{Lillo-Box_2020}; \citealt{Edwards_2021}), we do not fix or apply Gaussian priors on the LD coefficients, but let them vary freely (uniform priors). The stellar LD effects in the \textit{Spitzer}, HST, and TESS transits are modeled using per-instrument quadratic $q_1$ and $q_2$ parameters \citep{Kipping_2013} constructed to only allow physical solutions for values between 0 and 1. Note fixing the LD parameters to those measured by previous studies for the same instrument does not change the median of our $R_{\rm p} / R_{\star}$ posteriors.

For the Keplerian component, a semi-amplitude $K_k$ for each planet is fitted, as well as instrumental RV offsets ($\gamma_{\rm pre}$, $\gamma_{\rm post}$) and extra white noise terms ($\sigma_{\rm pre}$, $\sigma_{\rm post}$) for ESPRESSO pre- and post-fiber upgrade. When testing for possible eccentric orbits, we sample uniformly $\sqrt{e_k} \cos \omega_k$ and $\sqrt{e_k} \sin \omega_k$ between -1 and 1. We include in \texttt{juliet} a Gaussian Process (GP) to model stellar activity in the ESPRESSO RVs. Our GP implementation in \texttt{juliet} runs \texttt{george} \citep{Ambikasaran_2015} with a quasi-periodic covariance kernel (\citealt{Haywood_2014}; \citealt{Rajpaul_2015}):
\begin{equation}
k_{i,j} = A^2 \exp \left[ -\frac{|t_i - t_j|^2}{2 \ell^2} - \Gamma \sin^2 \left( \frac{\pi | t_i - t_j|}{P_{\rm rot}} \right) \right]
\end{equation}
where $|t_i - t_j|$ is the time interval between data $i$ and $j$, $A$ is the amplitude of the GP, $\ell$ is the coherence timescale, $\Gamma$ scales the periodic component of the GP, and $P_{\rm rot}$ is the stellar rotation period. We adopt a Gaussian prior on the known rotation period of the star $P_{\rm rot} \sim \mathcal{N}\left(131, 5^2\right)$\,days. The priors for the other GP hyperparameters are listed in Table~\ref{table:modelparams} and mostly follow the recommendation of \citealt{Stock_2023} (GP Prior III) when the rotation period is already constrained.

We inspected three different joint models ($\mathcal{M}$):
\begin{itemize}
\setlength\itemsep{0em}
\item Two planets, LHS~1140 b and c, on circular orbits ($\mathcal{M}_{\rm 2cp}$; $e_{\rm b} = e_{\rm c} = 0$, $\omega_{\rm b} = \omega_{\rm c} = 90^{\circ}$)
\item Two planets, LHS~1140 b and c, on eccentric orbits ($\mathcal{M}_{\rm 2ep}$)
\item Three planets, LHS~1140 b, c, and the candidate planet d reported by \citetalias{Lillo-Box_2020}, on circular orbits ($\mathcal{M}_{\rm 3cp}$; $e_{\rm b} = e_{\rm c} = e_{\rm d} =0$, $\omega_{\rm b} = \omega_{\rm c} = \omega_{\rm d} = 90^{\circ}$)
\end{itemize}
The difference in Bayesian log-evidence ($\Delta \ln Z$) yields the probability that one model describes better the observations over another. The empirical scale of \citealt{Trotta_2008} (see Table 1 therein) serves to interpret the significance of $\Delta \ln Z$ and to select the ``best" model. A $\Delta \ln Z > 5$ constitutes ``strong" evidence in favour of the model with the highest $\ln Z$. A $2.5 < \Delta \ln Z < 5$ corresponds to ``moderate'' evidence, but a $\Delta \ln Z \leq 2.5$ means that neither model should be favoured. Multiple runs of each $\mathcal{M}$ were carried with \texttt{dynesty} to verify the consistency of $\ln Z$.

For the two planet models $\mathcal{M}_{\rm 2cp}$ and $\mathcal{M}_{\rm 2ep}$, we obtain a $\Delta \ln Z = 0.8 \pm 0.9$ in favour of the circular orbit solutions (inconclusive). We report from model $\mathcal{M}_{\rm 2ep}$ upper limits on $e_{\rm b}$ and $e_{\rm c}$ (95\% confidence) of 0.043 and 0.050, respectively, implying that in all likelihood, the orbit of LHS~1140 b and c are de facto circular. For this reason, we select the simpler $\mathcal{M}_{\rm 2cp}$ as the preferred model. We present the relevant planetary parameters derived from the joint transit RV fit for model $\mathcal{M}_{\rm 2cp}$ in Table~\ref{table:derivedparams}. The priors and posteriors (16$^{\rm th}$, 50$^{\rm th}$, and 84$^{\rm th}$ percentiles) of the free parameters of model $\mathcal{M}_{\rm 2cp}$ are reported in Table~\ref{table:modelparams}. The best-fit transit models of the \textit{Spitzer}, HST, and TESS light curves are respectively shown in Figures~\ref{fig:spitzer_transit}, \ref{fig:hst_transit}, and \ref{fig:tess_transit}. The phase-folded RVs with the best-fit orbital solutions of LHS~1140~b~and~c is shown in Figure~\ref{fig:RV} with the full RV model (Keplerian + activity GP) presented in Figure~\ref{fig:RV_raw}. This full RV model yields a residual RMS of 41\,cm\,s$^{-1}$ for the ``pre" data consistent with the median RV errors of 42\,cm\,s$^{-1}$. For the ``post" data, the residual dispersion of 54\,cm\,s$^{-1}$ is larger than the typical RV errors of 34\,cm\,s$^{-1}$ so that a $\sigma_{\rm post} = 36\pm6$\,cm\,s$^{-1}$ jitter term is needed to fully describe the scatter.

The reanalysis of the ESPRESSO data with the LBL framework is an opportunity to test the presence of the candidate LHS~1140\,d on a 78.9-day orbit reported by \citetalias{Lillo-Box_2020}. For model $\mathcal{M}_{\rm 3cp}$, we chose the same priors on $P_{\rm d}$, $t_{\rm 0,d}$, and $K_{\rm d}$ as in \citetalias{Lillo-Box_2020}, namely $\mathcal{U}\left(70, 120\right)$\,days, $\mathcal{U}\left(2458350, 2458400\right)$\,BJD, and $\mathcal{U}\left(0, 10\right)$\,m\,s$^{-1}$. This fit converges to a small semi-amplitude of $0.9^{+0.8}_{-0.6}$\,m\,s$^{-1}$ and an undefined period of $87^{+14}_{-9}$\,days for a planet d. The Bayesian log-evidence does not increase when adding a third planet with $\Delta \ln Z = -1.0 \pm 0.9$ between $\mathcal{M}_{\rm 3cp}$ and $\mathcal{M}_{\rm 2cp}$. We also reject a $K_{\rm d}$ larger than 2.21\,m\,s$^{-1}$ at 2$\sigma$, corresponding to the median signal detected in
\citetalias{Lillo-Box_2020}. LHS~1140 could realistically have other planets, but given the precision of our RV measurements, we see no evidence of an additional companion sharing the parameters of candidate LHS~1140\,d. In Appendix~\ref{sec:validating}, we further demonstrate that an 80-day RV signal is most likely of stellar origin.

\subsection{Transit depth discrepancy for LHS~1140\,c} \label{sec:tension}

\begin{figure}[h]
\centering
\minipage{0.324\textwidth}
  \includegraphics[width=1\linewidth]{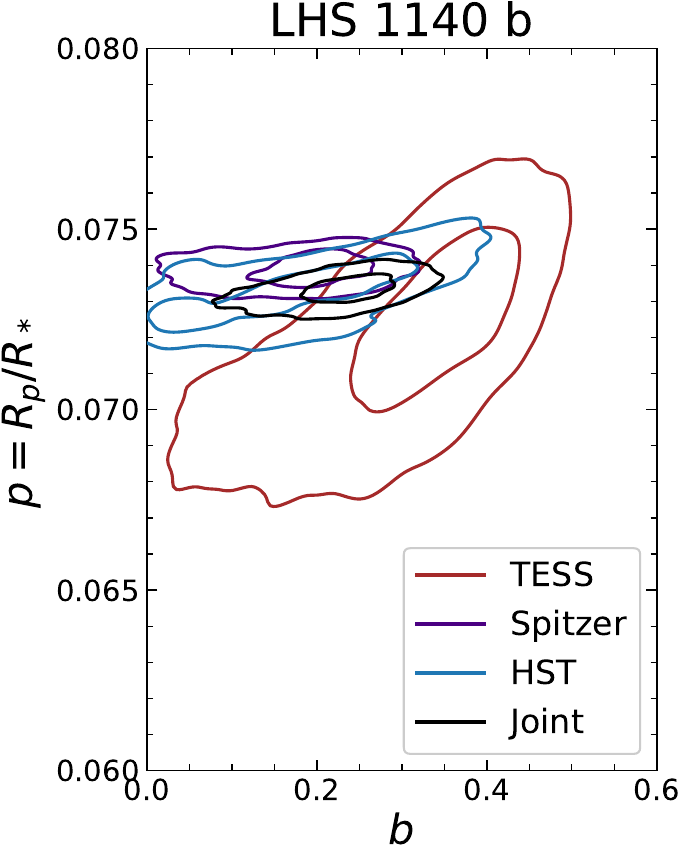}
\endminipage
\minipage{0.3\textwidth}
 \includegraphics[width=1\linewidth]{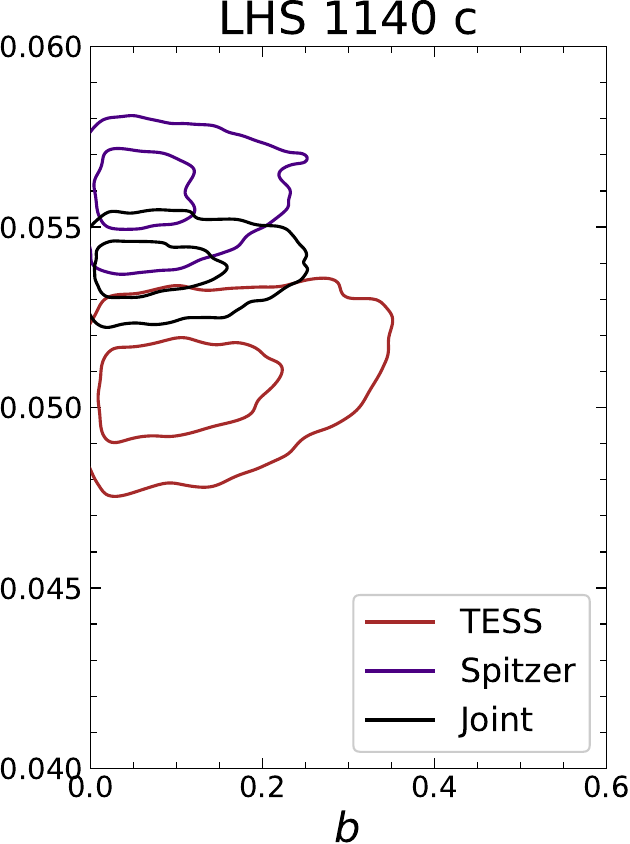}
\endminipage
  \caption{Posterior distributions (1 and 2$\sigma$ contours) for the transit impact parameter ($b$) and scaled radius ($p = R_{\rm p}/R_{\star}$) of LHS~1140~b~and~c. Different colors refer to fit on TESS (red), \textit{Spitzer} (purple), and HST (blue) data only. The black contours are the final joint posteriors on all available instruments. A $\sim$4$\sigma$ tension in $p$ exists between \textit{Spitzer} and TESS for LHS~1140\,c.}
  \label{fig:rpr*}
\end{figure}

In \citetalias{Lillo-Box_2020}, the radius of LHS~1140~b~and~c measured by TESS was slightly smaller (by 1.5$\sigma$ and 2$\sigma$ respectively) compared with previous results by \citetalias{Ment_2019} obtained with \textit{Spitzer} (see Fig.~\ref{fig:MR}). Here, our joint analysis of \textit{Spitzer}, HST (for LHS~1140\,b only), and TESS data has resulted in a similar discrepancy for planet c, but not for b. In Figure~\ref{fig:rpr*}, we show the transit impact parameter ($b$) and scaled radius ($p = R_{\rm p}/R_{\star}$) of LHS~1140~b~and~c derived from fitting each instrument independently. The $b$--$p$ posteriors of LHS~1140\,b agree well for all available instruments, but we detect a 4$\sigma$ tension for the $p$ of LHS~1140\,c measured by \textit{Spitzer} and TESS.

We remain cautious before interpreting the transit depth discrepancy of LHS~1140\,c as real because we only have a single visit with \textit{Spitzer}. It is possible that data reduction systematics affected the depth measurement. Nonetheless, it is worth mentioning that \textit{Spitzer} Channel 2 at 4.5\,$\mu$m covers a strong CO$_2$ feature. We detect a difference of 500\,ppm between the \textit{Spitzer} and TESS bandpasses, meaning that if excess atmospheric absorption is causing this discrepancy, it would be readily detectable with JWST. This work calls for better radius determination for LHS~1140\,c, particularly obtaining the full near-infrared transmission spectrum of this planet with JWST to reveal its true radius that we currently report as an average between \textit{Spitzer} and TESS (see Joint in Fig.~\ref{fig:rpr*}) and test whether its atmosphere is CO$_2$-rich.

\begin{deluxetable}{lccr}
\tablecaption{Planetary parameters derived from the joint transit RV fit}
\tablehead{\colhead{Parameter}
& \colhead{LHS~1140\,b} & \colhead{LHS~1140\,c} & \colhead{Description}}
\startdata
\multicolumn{4}{c}{\textit{Orbital parameters}}\\
$P$ (days) &  24.73723 $\pm$ 0.00002 & 3.777940 $\pm$ 0.000002 & Period\\
$t_{\rm 0}$ (BJD - 2\,457\,000) & 1399.9300 $\pm$ 0.0003 & 1389.2939 $\pm$ 0.0002 & Time of inferior conjunction\\
$a$ (au) & 0.0946 $\pm$ 0.0017 & 0.0270 $\pm$ 0.0005 & Semi-major axis\\
$i$ ($^{\circ}$) & 89.86 $\pm$ 0.04 & 89.80$^{+0.14}_{-0.19}$ & Inclination\\
$e$ & $<0.043$ (95\%) & $<0.050$ (95\%) & Eccentricity\\
\multicolumn{4}{c}{\textit{Transit parameters}}\\
$b$ & 0.23$^{+0.05}_{-0.07}$ & 0.09$^{+0.09}_{-0.06}$ & Impact parameter\\
$\delta$ (ppt) & 5.38 $\pm$ 0.06 & 2.90 $\pm$ 0.09 & Depth\\
$t_{14}$ (hours) & 2.15 $\pm$ 0.05 & 1.13 $\pm$ 0.02 & Duration\\
\multicolumn{4}{c}{\textit{Physical parameters}}\\
$R_{\rm p}$ (R$_{\oplus}$) & 1.730 $\pm$ 0.025 & 1.272 $\pm$ 0.026 & Radius\\
$M_{\rm p}$ (M$_{\oplus}$) & 5.60 $\pm$ 0.19 & 1.91 $\pm$ 0.06 & Mass\\
$\rho$ (g\,$\cdot$\,cm$^{-3}$) & 5.9 $\pm$ 0.3 & 5.1 $\pm$ 0.4 & Bulk density\\
$S$ (S$_{\oplus}$) & 0.43 $\pm$ 0.03 & 5.3 $\pm$ 0.4 & Insolation\\
$T_{\rm eq}$ $\left[A_{\rm B} = 0\right]$ (K) & 226 $\pm$ 4 & 422 $\pm$ 7 & Equilibrium temperature\\
\enddata
\tablenocomments{}
\label{table:derivedparams}
\end{deluxetable}

\begin{deluxetable}{lccr}
\tablecaption{Prior and posterior distributions of the joint transit RV fit}
\tablehead{
\colhead{Parameter} & \colhead{Prior$^{\rm 1}$} & \colhead{Posterior} & \colhead{Description}
}
\startdata
\textit{Stellar parameter} & & &\\
$\rho_{\star}$ (g\,$\cdot$\,cm$^{-3}$) & $\mathcal{N}\left(26.0, 2.6^2\right)$ & 25.8 $\pm$ 1.0 & Stellar density\\
\\
\textit{LHS~1140 b} & & &\\
$P_{\rm b}$ (days) & $\mathcal{U}\left(24.7,24.8\right)$ & 24.73723 $\pm$ 0.00002 & Orbital period\\
$t_{\rm 0,b}$ (BJD - 2\,457\,000) & $\mathcal{U}\left(1399.9, 1400.0\right)$ & 1399.9300 $\pm$ 0.0003 & Time of inferior conjunction\\
$r_{\rm 1,b}$ & $\mathcal{U}\left(0, 1\right)$ & 0.49 $\pm$ 0.03 & Parameterization$^2$\,for $b$ and $p$\\
$r_{\rm 2,b}$ & $\mathcal{U}\left(0, 1\right)$ & 0.0733 $\pm$ 0.0004 & Parameterization$^2$\,for $b$ and $p$\\
$K_{\rm b}$ (m\,s$^{-1}$) & $\mathcal{U}\left(0, 10\right)$ & 3.80 $\pm$ 0.11 & RV semi-amplitude\\
\\
\textit{LHS~1140 c} & & &\\
$P_{\rm c}$ (days) & $\mathcal{U}\left(3.7, 3.8\right)$ & 3.777940 $\pm$ 0.000002 & Orbital period\\
$t_{\rm 0,c}$ (BJD - 2\,457\,000) & $\mathcal{U}\left(1389.25, 1389.35\right)$ & 1389.2939 $\pm$ 0.0002 & Time of inferior conjunction\\
$r_{\rm 1,c}$ & $\mathcal{U}\left(0, 1\right)$ & 0.39 $\pm$ 0.05 & Parameterization$^2$\,for $b$ and $p$\\
$r_{\rm 2,c}$ & $\mathcal{U}\left(0, 1\right)$ & 0.0539 $\pm$ 0.0008 & Parameterization$^2$\,for $b$ and $p$\\
$K_{\rm c}$ (m\,s$^{-1}$) & $\mathcal{U}\left(0, 10\right)$ & 2.42 $\pm$ 0.07 & RV semi-amplitude\\
\\
\textit{Photometric parameters} & & &\\
$q_{\rm 1,Spitzer}$ & $\mathcal{U}\left(0, 1\right)$ & 0.016$^{+0.018}_{-0.009}$ & Limb-darkening parameter$^3$\\
$q_{\rm 2,Spitzer}$ & $\mathcal{U}\left(0, 1\right)$ & 0.42$^{+0.36}_{-0.28}$ & Limb-darkening parameter$^3$\\
$M_{\rm Spitzer}$ (ppm) & $\mathcal{N}\left(0, 1\,000^2\right)$ & 34 $\pm$ 23 & Baseline flux\\
$\sigma_{\rm Spitzer}$ (ppm) & $\mathcal{LU}\left(1, 1\,000\right)$ & 15$^{+57}_{-12}$ & Extra white noise\\
$q_{\rm 1,HST}$ & $\mathcal{U}\left(0, 1\right)$ & 0.28 $\pm$ 0.08 & Limb-darkening parameter$^3$\\
$q_{\rm 2,HST}$ & $\mathcal{U}\left(0, 1\right)$ & 0.12$^{+0.11}_{-0.08}$ & Limb-darkening parameter$^3$\\
$M_{\rm HST}$ (ppm) & $\mathcal{N}\left(0, 1\,000^2\right)$ & $-25$ $\pm$ 22 & Baseline flux\\
$\sigma_{\rm HST}$ (ppm) & $\mathcal{LU}\left(1, 1\,000\right)$ & 12$^{+38}_{-10}$ & Extra white noise\\
$q_{\rm 1,TESS}$ & $\mathcal{U}\left(0, 1\right)$ & 0.33$^{+0.20}_{-0.12}$ & Limb-darkening parameter$^3$\\
$q_{\rm 2,TESS}$ & $\mathcal{U}\left(0, 1\right)$ & 0.56 $\pm$ 0.27 & Limb-darkening parameter$^3$\\
$M_{\rm TESS}$ (ppm) & $\mathcal{N}\left(0, 1000^2\right)$ & 9 $\pm$ 48 & Baseline flux\\
$\sigma_{\rm TESS}$ (ppm) & $\mathcal{LU}\left(1, 1\,000\right)$ & 17$^{+116}_{-14}$ & Extra white noise\\
\\
\textit{RV parameters} & & &\\
$\gamma_{\rm pre}$ (m\,s$^{-1}$) & $\mathcal{U}\left(-10, 10\right)$ & 1.3 $\pm$ 1.5 & RV offset$^4$\\
$\gamma_{\rm post}$ (m\,s$^{-1}$) & $\mathcal{U}\left(-10, 10\right)$ & -0.6 $\pm$ 1.4 & RV offset$^4$\\
$\sigma_{\rm pre}$ (m\,s$^{-1}$) & $\mathcal{LU}\left(10^{-3}, 10\right)$ & 0.04$^{+0.18}_{-0.03}$ & Extra white noise\\
$\sigma_{\rm post}$ (m\,s$^{-1}$) & $\mathcal{LU}\left(10^{-3}, 10\right)$ & 0.36 $\pm$ 0.06 & Extra white noise\\
\\
\textit{RV activity GP} & & &\\
$A$ (m\,s$^{-1}$) & $\mathcal{LU}\left(10^{-3}, 10\right)$ & 2.8$^{+1.0}_{-0.6}$ & Amplitude of the GP\\
$\ell$ (days) & $\mathcal{LU}\left(100, 1000\right)$ & 164$^{+62}_{-42}$ & Timescale of the GP\\
$\Gamma$ & $\mathcal{LU}\left(0.1, 10\right)$ & 3.7$^{+1.5}_{-1.2}$ & Periodic scale of the GP\\
$P_{\rm rot}$ (days) & $\mathcal{N}\left(131, 5^2\right)$ & 133 $\pm$ 3 & Rotation period\\
\hline
\enddata
\tablecomments{$^1\mathcal{U}\left(a,b \right)$ is the uniform distribution between value $a$ and $b$, $\mathcal{LU}\left(a,b \right)$ is the log-uniform (Jeffreys) distribution between value $a$ and $b$, $\mathcal{N}\left(\mu, \sigma^2 \right)$ is the normal distribution with mean $\mu$ and variance $\sigma^2$.\\
$^2$Parameterization of the transit impact parameter ($b$) and the planet-to-star radius ratio ($p = R_{\rm p}/R_{\star}$) outlined in \cite{Espinoza_2018}.\\
$^3q_1, q_2$ are related to the quadratic limb-darkening coefficients $u_1, u_2$ by the transformations described in \cite{Kipping_2013}.\\
$^4$Relative to the median RV (-13\,458.5 and -13\,450.5 m\,s$^{-1}$ for pre- and post-fiber change data).}
\label{table:modelparams}
\end{deluxetable}

\subsection{Validating genuine Keplerian signals with line-by-line radial velocity}
\label{sec:validating}

\begin{figure*}[t!]
\centering
\includegraphics[width=1\linewidth]{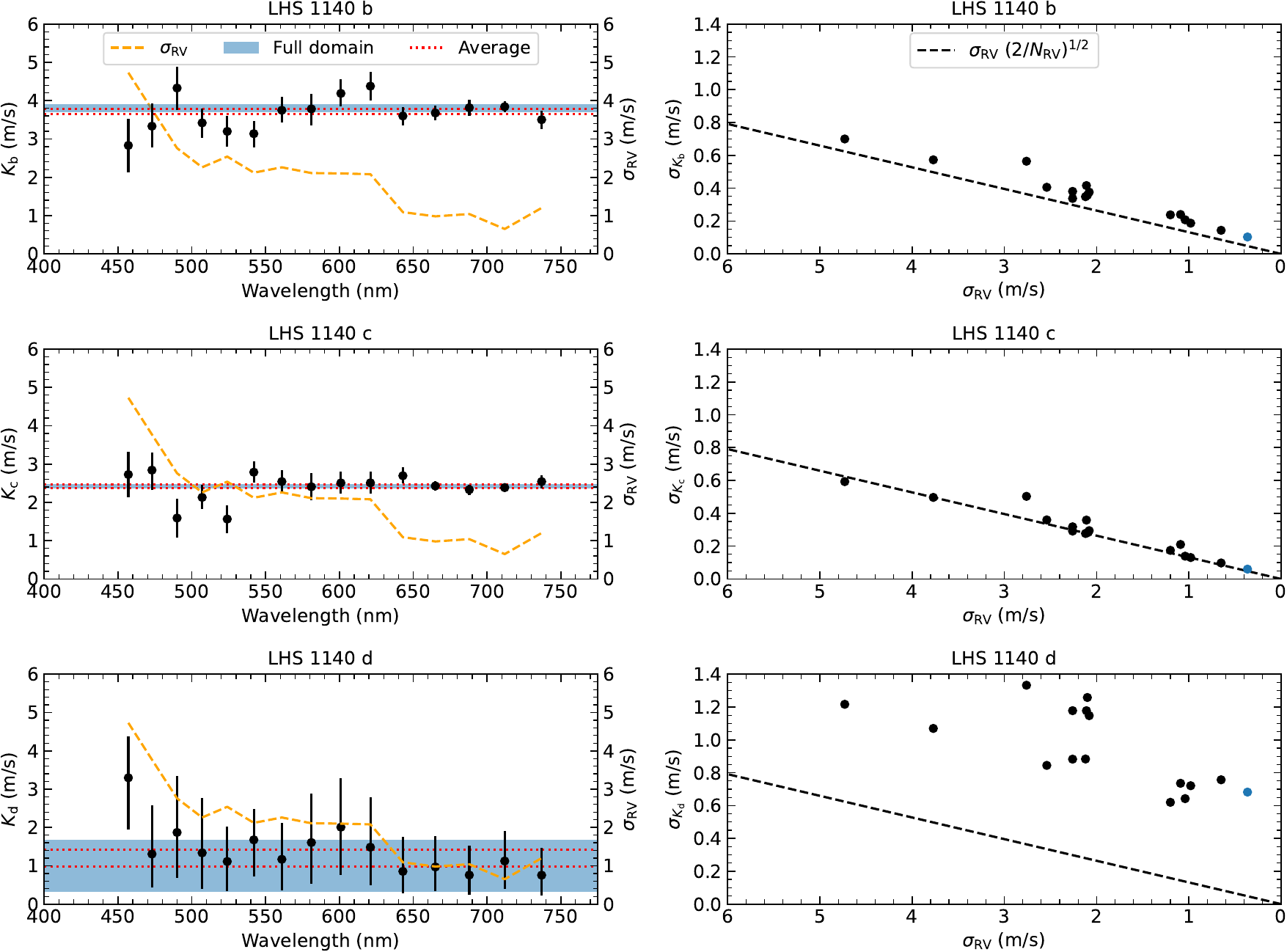}
  \caption{Validating Keplerian signals in the ESPRESSO spectral domain of LHS~1140. \textit{Left panels}: Semi-amplitudes $K_{\lambda}$ obtained with model $\mathcal{M}_{\rm 3cp}$ from LBL radial velocities extracted for different bands ($\Delta \lambda \approx 20$\,nm). The per-band velocities have uncertainties ($\sigma_{\rm RV}$) varying from 4.73 to 0.65 m\,s$^{-1}$ (orange dashed lines). A genuine Keplerian signal is achromatic ($dK_{\lambda}/d\lambda = 0$). We measure no chromatic slopes for LHS~1140~b~and~c, but the signal of candidate LHS~1140\,d is marginally larger in the blue wavelengths of ESPRESSO ($dK_{\lambda}/d\lambda = -0.42 \pm 0.26$ m\,s$^{-1}$ per 100 nm). A coherent signal such as a Doppler shift affecting all wavelengths should produce consistent $K$ between the full spectrum constraints (1$\sigma$ confidence regions in blue) and the weighted average of the $K_{\lambda}$ (1$\sigma$ confidence interval between the red dotted lines). Assuming a coherent signal, the $K_{\rm d}$ inferred from the full ESPRESSO domain should have an uncertainty $\sigma_{K_{\rm d}} \sim 0.18$\,m\,s$^{-1}$, which is inconsistent with observations. \textit{Right panels}: Semi-amplitude uncertainties $\sigma_{K}$ with increasing RV precision (smaller $\sigma_{\rm RV}$). A coherent signal in wavelengths should follow the \cite{Cloutier_2018} formalism ($\sigma_{K} = \sigma_{\rm RV} \sqrt{2 / N_{\rm RV}}$), which is not seen for LHS~1140\,d. The LBL data set shows no evidence of an 80-day signal associated with the candidate planet LHS~1140\,d.}
  \label{fig:spectral_test}
\end{figure*}

The LBL method allows to verify the achromaticity of the inferred semi-amplitudes of the LHS~1140 planets. The three-planet model $\mathcal{M}_{\rm 3cp}$ can be applied on velocities derived per spectral bins of $\Delta \lambda \approx 20$\,nm in the entire ESPRESSO domain. For genuine Keplerian signals:
\begin{itemize}
    \setlength\itemsep{0em}
    \item (\textbf{chromaticity test}) the $K_{\lambda}$ calculated for different bands are achromatic, i.e., $dK_{\lambda}/d\lambda = 0$.
    \item (\textbf{coherence test}) the $K_{\lambda}$ add constructively (coherent period and phase), yielding a consistent weighted average (value and uncertainty) with $K$ resulting from the full spectral range.
    \item (\textbf{noise test}) the semi-amplitude uncertainties ($\sigma_{K}$) should scale with RV precision ($\sigma_{\rm RV}$) according to the \cite{Cloutier_2018} formalism, i.e., $\sigma_{K} = \sigma_{\rm RV} \sqrt{2 / N_{\rm RV}}$ when no important extra jitter is detected. 
\end{itemize}
These three criteria can be tested for LHS~1140 b, c, and the non-transiting candidate d. As summarized in Figure~\ref{fig:spectral_test}, we measure no chromaticity for the Keplerian signals of LHS~1140~b~and~c, respectively with $dK_{\lambda}/d\lambda = 0.11 \pm 0.10$ and $0.03 \pm 0.08$ m\,s$^{-1}$ per 100 nm. For the candidate LHS~1140\,d, the signal is marginally larger in the blue wavelengths of ESPRESSO ($dK_{\lambda}/d\lambda = -0.42 \pm 0.26$ m\,s$^{-1}$ per 100 nm). Similarly, since the Doppler signals of LHS~1140\,b and c are perfectly coherent, the weighted average of the $K_{\lambda}$ are consistent with the $K$ obtained from the full spectral range (see Fig.~\ref{fig:spectral_test}). This is not observed for the 80-day signal associated with candidate d for which fitting a Keplerian model on RVs derived from the full ESPRESSO wavelength range is three times less accurate. Finally, given the RV precision of this data set varying from 4.73 to 0.65 m\,s$^{-1}$ in the 450--750\,nm interval, we find that LHS~1140~b~and~c follow the \cite{Cloutier_2018} formalism, but not d (see Fig.~\ref{fig:spectral_test}, right). Interestingly, the spectral interval 700--725\,nm covering only 8\% of the ESPRESSO domain contains 30\% of all the RV content of LHS~1140. The tests presented in this section provide evidence against the existence of LHS~1140\,d first announced by \citetalias{Lillo-Box_2020} as a candidate non-transiting planet.

\section{Supplementary material of the internal structure analysis}
\label{appendix_E}
\setcounter{figure}{0}
\renewcommand{\thefigure}{E\arabic{figure}}
\setcounter{table}{0}
\renewcommand{\thetable}{E\arabic{table}}

\subsection{Interior modeling} \label{sec:interior_modeling}

In this appendix, we summarize the method of \cite{Plotnykov_2020} adapted to constrain the detailed internal structure of LHS~1140\,b.

We used a Bayesian analysis with Markov Chain Monte Carlo (MCMC) sampling with \texttt{emcee} \citep{Foreman-Mackey_2013} coupled to the internal structure model of \cite{Valencia_2007}. The model assumes three layers for the interior of the planet: a condensed water (liquid/ice) layer at the surface, a mantle (Mg-Si rock) and a core (Fe-Ni-Si alloy). A fixed atmospheric layer is also considered for models in which the atmosphere could significantly contribute to the planet radius. The water layer is described by the equation of state (EOS) of \citealt{Hemley_1987, Wagner_2002, Stewart_2005} and we adopt the density mixing model with EOS parameters  of \citet{Stixrude2011} and \citet{Morrison2018} for the mantle and the core, respectively. The lower mantle is assumed to be composed of bridgmanite/post-perovskite and wustite (at different proportions to that of Earth), while the upper mantle is modeled as pure olivine ($\mathrm{Mg_2SiO_4}$). This choice was made in light of the possible low Si content of the star that may indicate olivine formation is favoured compared to pyroxenes ($\mathrm{Mg_2Si_2O_6}$).

Important ingredients to the model are the abundance ratios (by weight) of refractory elements, namely Fe/Mg and Mg/Si. We explored a no prior (Fe/Mg unconstrained) and a stellar prior (Fe/Mg following the star) models for the water world case of LHS~1140\,b (Sect.~\ref{sec:interior_waterworld}). For the pure rocky case (Sect.~\ref{sec:interior_barerock}), the planet Fe/Mg ratio must be lower than that of the host star to produce the observed density. Consequently, only the unconstrained Fe/Mg analysis was done. Note that because we model three layers for the interior, the no prior case cannot fully constrain both the CMF and WMF, i.e., planetary interior models are inherently degenerate in this case. For all models, the Mg/Si ratio in the mantle is assumed to be that of the host star: $\mathrm{Mg/Si_{planet}} \sim \mathcal{N}(\mathrm{Mg/Si_{star}},\sigma_{\rm star}^2$). Despite LHS~1140 having a Mg/Si approximately twice that of the Sun/Earth (see Table \ref{table:ratios}), this ratio only weakly influences the final radius of the planet meaning that our results do not depend on the exact mineralogy of the rocks.

We report the posterior distributions of our interior analysis in Appendix~\ref{posterior_dist} for all interior models considered. Additionally, we compare the Fe/Mg distributions of LHS~1140\,b for different scenarios in Figure~\ref{fig:LHS1140b_Fe2Mg}.

\subsection{Posterior distributions} \label{posterior_dist}
\begin{figure}[ht!]
\centering
\includegraphics[width=0.4\linewidth]{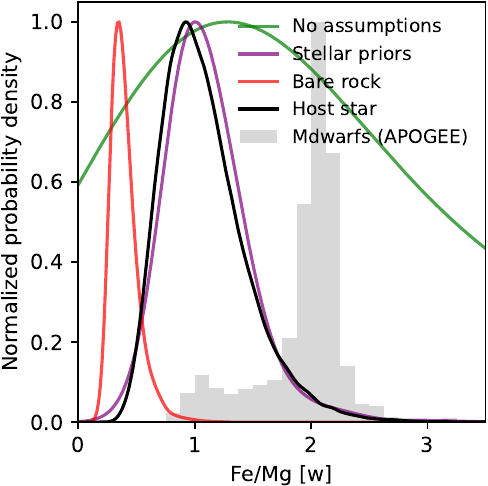}
  \caption{Fe/Mg distribution for LHS~1140\,b according to different assumptions (no prior or stellar prior) for the bare rock (red) and water world (green and purple) models compared to the host star (black). We plot the kernel density estimate of the interior model posterior, which is normalized to have mode of 1. We include the M dwarfs distribution of Fe/Mg from the APOGEE DR16 (gray histogram, sample size of $\sim$1000; \citealt{Majewski_2016}; \citealt{Ahumada_2020}) for comparison.}  
  \label{fig:LHS1140b_Fe2Mg}
\end{figure}

\begin{figure}[ht!]
\centering
  \includegraphics[width=0.8\linewidth]{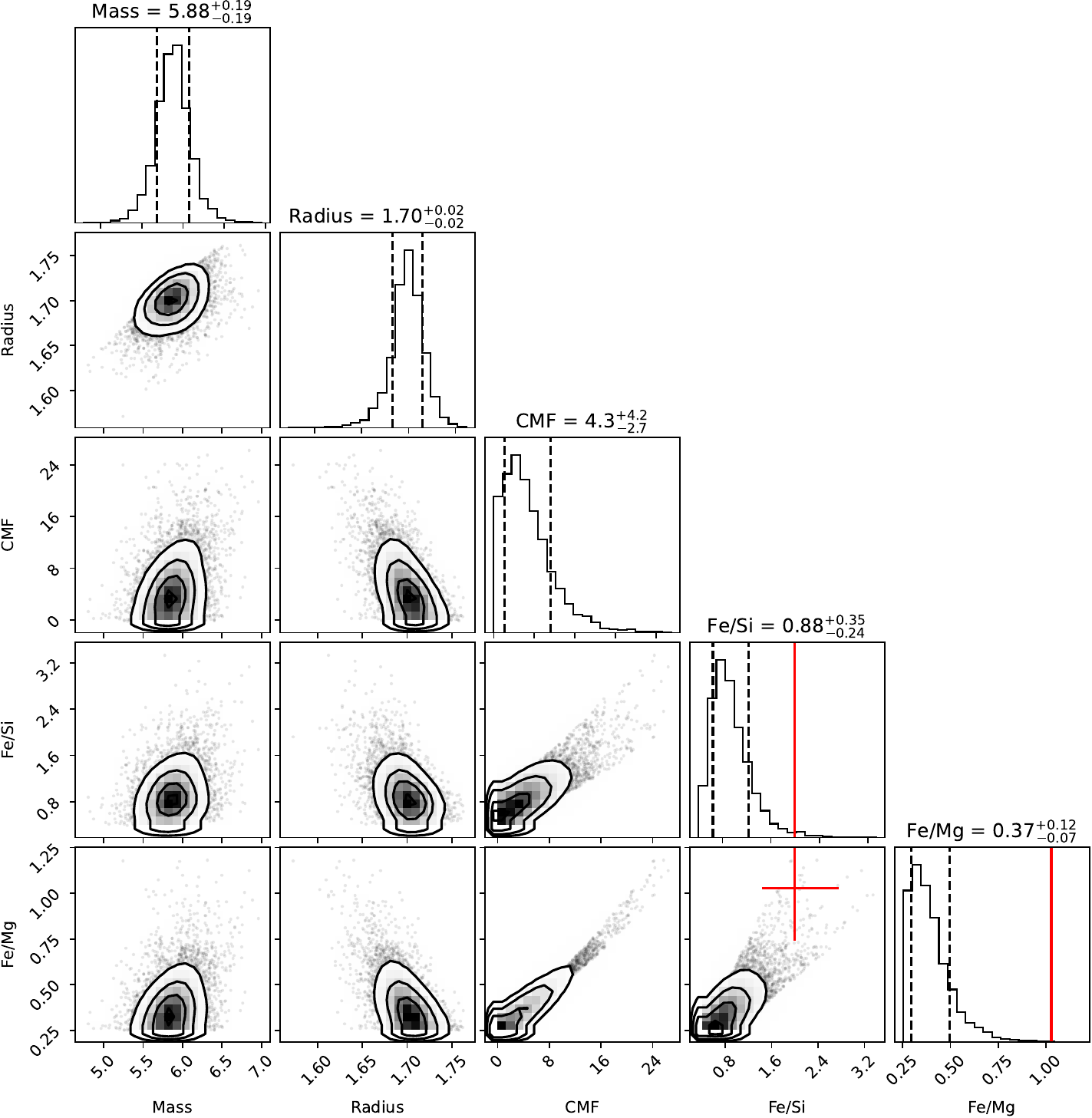}
  \linespread{1}
  \caption{Posterior distribution for LHS~1140\,b interior parameters for the bare rock scenario (WMF~=~0). Chemical weight ratios Fe/Si and Fe/Mg are derived quantities. The red lines (and red cross) in Fe/Si and Fe/Mg space correspond to the stellar measurements. The vertical black dashed lines represent 16$^{\rm th}$ and 84$^{\rm th}$ percentiles of the posterior.}
  \label{fig:cornerb_rocky}
\end{figure}

\begin{figure}[ht!]
\centering
  \includegraphics[width=0.9\linewidth]{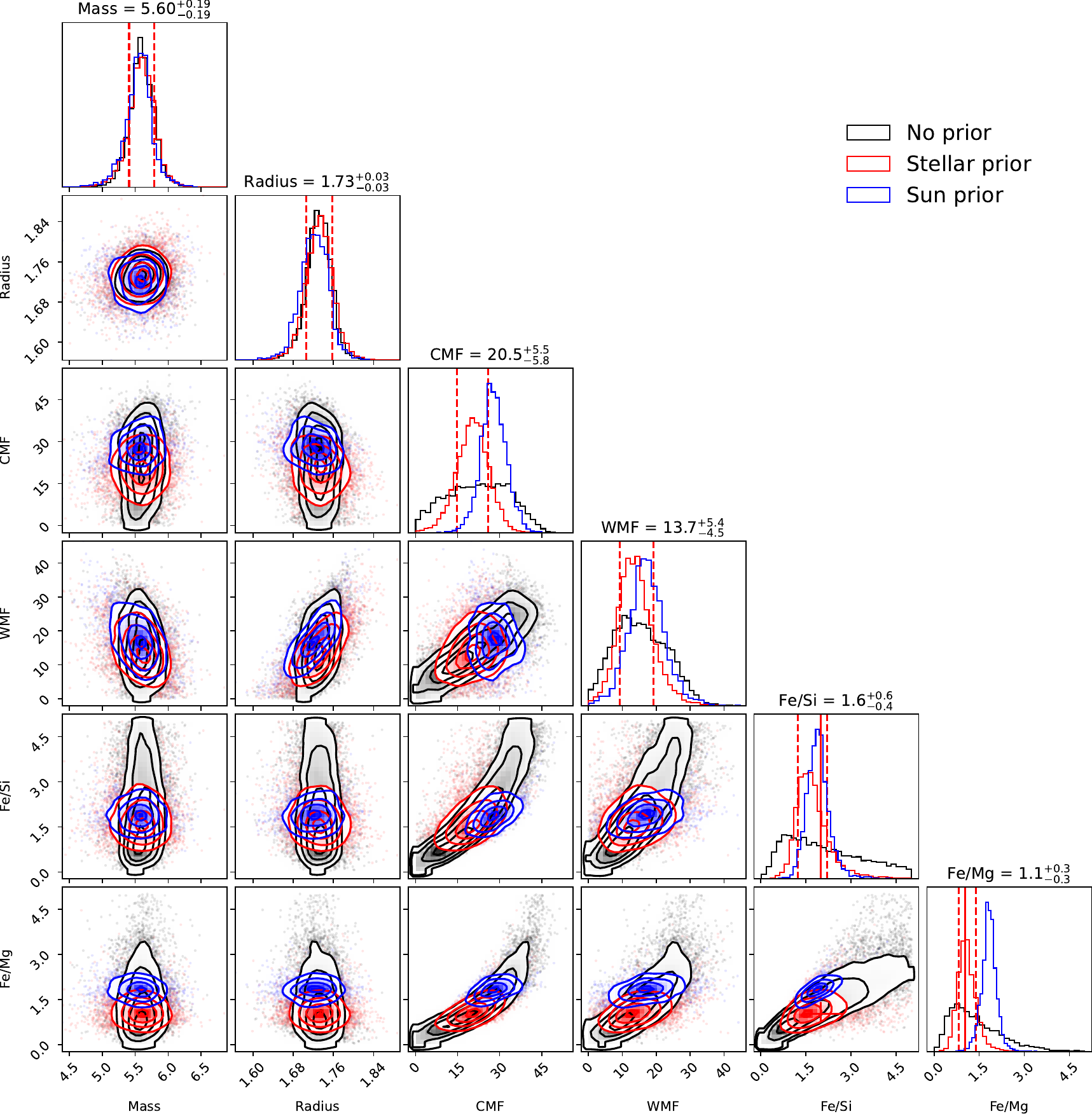}
  \linespread{1}
  \caption{Posterior distribution for LHS~1140\,b interior parameters for the water world scenario (WMF~$>$~0) for different Fe/Mg assumptions: no prior (black), stellar prior (red) and Sun prior (blue). For the stellar and Sun prior cases, the log-probability function is modified to include the restriction posed by Fe/Mg and Mg/Si ratios measured on the host star or the Sun (Table~\ref{table:ratios}). The red lines (and red cross) in Fe/Si and Fe/Mg space correspond to the stellar measurements. The vertical red dashed lines represent 16$^{\rm th}$ and 84$^{\rm th}$ percentiles of the posterior for the stellar prior case. Note the values in the parameter titles are for the stellar prior case only.}
  \label{fig:cornerb_np}
\end{figure}

\begin{figure}[ht!]
\centering
  \includegraphics[width=0.9\linewidth]{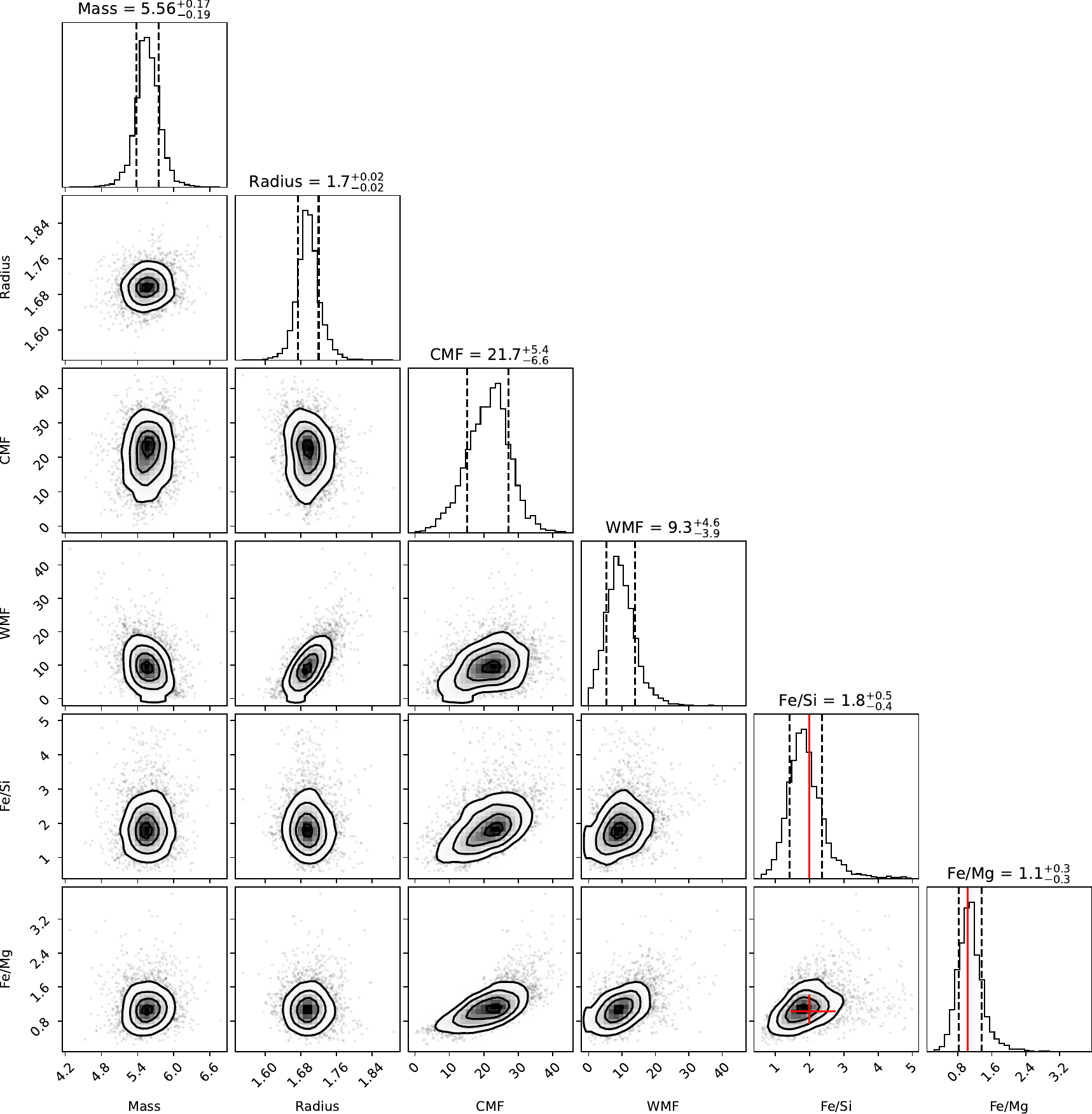}
  \linespread{1}
  \caption{Posterior distribution for LHS~1140\,b interior parameters for the Hycean world scenario (icy planet with 250\,km H$_2$ atmosphere on top). Here, we use the same assumptions as the stellar prior water world model. The red lines (and red cross) represent the stellar ratios, while the vertical black dashed lines are the 16$^{\rm th}$ and 84$^{\rm th}$ percentiles of the posterior. Note the simulation radius is smaller ($\Delta R_{\rm p} \sim 0.03\mathrm{R_\oplus}$) due to H$_2$ atmosphere which is not included in our interior model.}
  \label{fig:cornerb_wp}
\end{figure}

We summarize the posterior distributions of Fe/Mg for LHS~1140\,b in Figure~\ref{fig:LHS1140b_Fe2Mg} for the pure rocky (Sect.~\ref{sec:interior_barerock}) and water world (Sect.~\ref{sec:interior_waterworld}) scenarios. The posteriors from our MCMC sampling are also presented in the following corner plots of Figures \ref{fig:cornerb_rocky}, \ref{fig:cornerb_np}, and \ref{fig:cornerb_wp} showing the results for a bare rock, a water world (no prior, stellar prior, and solar prior cases), and a Hycean world, respectively.

\bibliographystyle{aasjournal}

\end{document}